\documentclass[11pt,twocolumn,preprintnumbers,amsmath,amssymb,nofootinbib,superscriptaddress,aps,prd,numberedappendix]{openjournal}
\usepackage{natbib}
\usepackage{newtxtext,newtxmath}
\usepackage{amsmath,amssymb}
\usepackage{graphicx}
\usepackage{dcolumn}
\usepackage{bm}
\usepackage[dvipsnames]{xcolor}
\usepackage{xspace}
\usepackage{ulem}
\usepackage{lipsum}
\usepackage[colorlinks,linkcolor=red,citecolor=blue,urlcolor=blue ]{hyperref}
\usepackage{graphicx} 
\usepackage{multirow}
\usepackage{cancel}
\newcommand{\nv}{\hat{\bf n}}

\newcommand{\wtj}[6]{\left(\begin{array}{ccc} #1 & #2 & #3\\#4 & #5 & #6\end{array} \right)}
\newcommand{\gaunt}[6]{{\cal G}_{#1 #2 #3}^{#4 #5 #6}}
\newcommand{\dfk}[1]{{\cal D}^N #1}
\newcommand{\avg}[1]{\left\langle #1 \right\rangle}
\newcommand{\nmt}{{\tt NaMaster}\xspace}

\begin{document}
\title{Fast projected bispectra: the filter-square approach}

\author{Lea Harscouet$^{1,*}$}
\author{Jessica A. Cowell$^{1,2}$}
\author{Julia Ereza$^{3}$}
\author{David Alonso$^{1}$}
\author{Hugo Camacho$^{4}$}
\author{Andrina Nicola$^{5}$}
\author{An\v ze Slosar$^{4}$}
\email{$^*$lea.harscouet@physics.ox.ac.uk}
\affiliation{$^1$Department of Physics, University of Oxford, Denys Wilkinson Building, Keble Road, Oxford OX1 3RH, United Kingdom}
\affiliation{$^2$Kavli Institute for the Physics and Mathematics of the Universe (Kavli IPMU, WPI), UTIAS, The University of Tokyo, Kashiwa, Chiba 277-8583, Japan}
\affiliation{$^3$Instituto de Astrof\'isica de Andaluc\'ia (CSIC), Glorieta de la Astronom\'ia, E-18080 Granada, Spain}
\affiliation{$^4$Brookhaven National Laboratory, Physics Department, Upton, NY 11973, USA}
\affiliation{$^5$Argelander Institut f\"ur Astronomie, Universit\"at Bonn, Auf dem H\"ugel 71, 53121 Bonn, Germany}
\date{\today}

\begin{abstract}
  The study of third-order statistics in large-scale structure analyses has been hampered by the increased complexity of bispectrum estimators (compared to power spectra), the large dimensionality of the data vector, and the difficulty in estimating its covariance matrix. In this paper we present the filtered-squared bispectrum (FSB), an estimator of the projected bispectrum effectively consisting of the cross-correlation between the square of a field filtered on a range of scales and the original field. Within this formalism, we are able to recycle much of the infrastructure built around power spectrum measurement to construct an estimator that is both fast and robust against mode-coupling effects caused by incomplete sky observations. Furthermore, we demonstrate that the existing techniques for the estimation of analytical power spectrum covariances can be used within this formalism to calculate the bispectrum covariance at very high accuracy, naturally accounting for the most relevant Gaussian and non-Gaussian contributions in a model-independent manner.
\end{abstract}

\maketitle

\section{Introduction}\label{sec:intro}
  Studying the large-scale structure (LSS) of the Universe is central to cosmology, with many cosmological surveys aiming to investigate the properties of the underlying matter density field. The traditional tool for cosmologists is the power spectrum, or `2-point correlation function', which measures correlations between pairs of points in the Universe across spatial scales. The power spectrum is known to capture all information for a Gaussian field. However, while the Universe evolved from an initially random field, at late times, the matter density field is highly non-Gaussian due to nonlinear gravitational collapse. Indeed, it is possible to generate many topologically distinct non-Gaussian fields with an identical power spectrum. To extract all possible information, it becomes necessary to use alternative summary statistics, able to capture this non-Gaussian information. Unlike in the case of Gaussian fields, however, no single summary statistic exists that is guaranteed to be lossless. Thus, various different strategies have been developed to try to extract this additional information in an efficient manner. These include wavelet scattering transforms \citep{simbig_wavelet,eickenberg2022waveletmomentscosmologicalparameter}, marked power spectra \citep{Massara_2023,2409.05695}, persistent homology \citep{yip2024persistenthomology}, Minkowski functionals \citep{Grewal_2022}, phase correlation functions \citep{1211.5213,2206.11005}, density-split statistics \citep{1710.05162}, and others.

  The natural extension of power spectrum-based analyses, however, is the measurement and exploitation of field correlators of increasingly higher order, starting with the three-point function or its Fourier-space equivalent, the bispectrum. Although there is no guarantee that this approach is optimally efficient, in terms of information extracted in relation to the data vector size, correlator-based studies have particular advantages over potentially more efficient ad-hoc non-linear transformations. Most notably a theoretical prediction can be derived that connects the correlators estimated in real data to fundamental quantities (e.g. correlators of the matter overdensities) without requiring the construction of expensive emulators that must reproduce all the details of the specific dataset under study \citep[e.g. sky mask, source selection, survey inhomogeneity][]{2018MNRAS.474..712M,2204.13717}. Moreover, these fundamental quantities can be accurately predicted as a function of a small number of cosmological parameters, and hence can be more easily emulated. Alternatively, in particular regimes they can be predicted directly in terms of fundamental principles using perturbation theory \citep{Bernardeau_Colombi_Gaztanaga_Scoccimarro_2002}. Predictions for the density correlators are direct products of the perturbation theory approaches to structure formation and offer unique theoretical control over errors in the predictions, allowing for a highly robust interpretation of the scientific results.

  The bispectrum has been exploited in several areas of cosmological science. Most notably, measurements of the primary CMB bispectrum have led to the most stringent constraints to date on the level of primordial non-Gaussianity, a powerful discriminator of inflationary scenarios. Measurements of the angular bispectrum in CMB experiments have also been used to characterise the most relevant extragalactic foreground sources (e.g. thermal Sunyaev-Zel'dovich, cosmic infrared background, radio point sources) \citep{1711.07879,1901.04515}, and there are prospects to measure and exploit the CMB lensing bispectrum \citep{1604.08578,1812.10635,2210.16203}. The other area where the bispectrum has played a central role is in the study of spectroscopic galaxy clustering in three dimensions \citep{1407.5668,1606.00439,2206.02800, sugiyamaCompleteFFTbasedDecomposition2019,2010.06179}. Its main role in this area is its ability to break key degeneracies between the galaxy bias relation and cosmological parameters, including the amplitude of matter fluctuations, improving constraints on the latter.
 
  Surprisingly, however, the study of the angular bispectrum in projected probes of the late-time large-scale structure (e.g. photometric galaxy clustering and cosmic shear) has lagged significantly behind its three-dimensional counterpart. This is in spite of its potential to significantly increase the scientific yield of the now traditional $3\times2$-point analyses combining weak lensing and galaxy clustering (again, by breaking degeneracies with galaxy bias), and to improve constraints on primordial non-Gaussianity from galaxy clustering studies \citep{2103.01229}. These are key targets of current and near future surveys covering large swathes of the observable Universe, such as the Vera C. Rubin Observatory's Legacy Survey of Space and Time (LSST) \citep{1809.01669}, the Euclid satellite \citep{1606.00180}, the Nancy Grace Roman space telescope \citep{1503.03757}, and the Square Kilometre Array \citep{1501.03825}. Note, however, that third-order statistics in weak lensing analyses have been studied in the recent past, in the form of third-order aperture mass statistics \citep{astro-ph/0505581,2309.08602} and moments \citep{2110.10141}. In part, the reason for this relatively slow progress is the absence of estimators able to measure the bispectrum in realistic datasets achieving the same level of accuracy and robustness as angular power spectra (e.g. in the handling of inhomogeneous sky coverage, or the estimation of their statistical uncertainties), although significant progress has recently been made in this area \citep{2107.06287,2306.03915, carronSphericalBispectrumExpansion2024}. Additionally, standard bispectrum estimators are significantly more costly to deploy on realistic data than the fast power spectrum estimators currently on the market, which hampers their use in analyses where the data vector and its covariance matrix must be estimated multiple times as new cuts are applied to the data, or to test for different systematics. Finally, the number of potentially independent bispectrum configurations, or ``triangles'', is significantly larger than that of power spectra, and data compression schemes are often necessary in order to make their analysis practical.

One of the simplest bispectrum estimators is the skew spectrum, which consists in taking the power spectrum of a field and its squared version, in what is essentially a compressed form of the bispectrum. While related estimators for the 3D bispectrum are prevalent in the literature \citep{schmittfullOptimalBispectrumEstimators2015, Skew_2020, chakrabortySkewingCMB$times$LSSFast2022, giriConstraining$f_NL$Using2023}, the concept of skew spectrum was also adapted for its angular counterpart \citep[in particular, the Sunyaev-Zel'dovich-CMB temperature bispectrum in][]{cooraySquaredTemperaturetemperaturePower2001}, with recent applications to the detection of the galaxy-galaxy-CMB lensing bispectrum \citep{farrenDetectionCMBLensing2023}.

Reducing the number of mode configurations contributing to the estimator has also been explored as a means to simplify and make bispectrum estimators more interpretable. In particular, the squeezed-limit bispectrum -- the limit in which only triangle configurations where one mode (side) is much smaller than the other two contribute -- is thought to capture much of the non-Gaussianity of LSS fields, and is therefore the focus of many bispectrum analyses \citep{chiangPositiondependentPowerSpectrum2014, barreiraSqueezedMatterBispectrum2019, biagettiCovarianceSqueezedBispectrum2022, giriConstraining$f_NL$Using2023, salvalaggioBispectrumNonGaussianCovariance2024a}. To probe such configurations, \cite{chiangPositiondependentPowerSpectrum2014} introduced an additional selection of modes in their bispectrum estimator, effectively setting some constraints on the shapes that contribute to what they call the integrated bispectrum. They first compute position-dependent power spectra, defined on large patches of characteristic size $r_L$ in configuration space, then correlate them to the average value of the field in these same patches, thus restricting the range of possible values for the longer legs of the triangle to modes $> k_L$ (where $k_L = 1/r_L$ is the mode which corresponds to the characteristic patch size). A similar estimator for the angular bispectrum was implemented in \cite{Jung_2020}, where the task of selecting patches on a spherical surface becomes convoluted if one wants to keep an analytically tractable estimator. 

Both the skew-spectrum and integrated bispectrum strive to reduce the complexity of the bispectrum data-vector: the skew spectrum effectively collapses the information down to a single dimension -- not unlike the power spectrum -- while the integrated bispectrum only probes the so-called squeezed triangle configurations. The skew spectrum however loses some of its interpretability by erasing individual contributions, while the integrated bispectrum faces challenges when it comes to real-space patch selection (especially in the presence of a mask). An alternative and simpler way of probing said squeezed configurations -- chosen in this paper -- is to implement this selection of scales in the skew spectrum estimator via the use of filters in frequency space, rather than configuration space \citep[this was implemented for the 3D bispectrum in e.g.][]{schmittfullOptimalBispectrumEstimators2015, giriConstraining$f_NL$Using2023}. For the angular bispectrum, this means we keep track of the exact mode combinations contributing to the estimator, while avoiding unwanted mode-coupling effects originating in configuration-space filtering; this compromise between both approaches also greatly improves the flexibility and interpretability of the angular skew-spectrum estimator and its covariance.

The estimator we introduce in this paper -- the ``filtered-squared bispectrum'' or FSB -- aims to solve aforementioned issues with previous bispectrum estimators; it is in essence an angular skew spectrum with an additional filtering step in harmonic space. The FSB therefore reinterprets the bispectrum as a power spectrum between a filtered and squared field, and a second field. Although, as we will show, the FSB is mathematically equivalent to standard bispectrum estimators in the full sky, the power spectrum interpretation allows us to make use of the significant infrastructure built around the estimation of second-order statistics to construct an estimator that is both relatively fast, and robust against survey geometry effects. Moreover, and perhaps more importantly, we are able to make use of existing techniques for the estimation of power spectrum covariances to build a highly-reliable estimate of the bispectrum covariance in a completely data-driven manner, without the need for expensive simulations, while accounting for all relevant non-Gaussian contributions. 

This paper is structured as follows. In Section \ref{sec:meth} we review the theory behind power spectra and bispectra, and the techniques used in power spectrum estimation. We then introduce the FSB, compare it to existing estimators, and describe the methods used to calculate its covariance matrix. Section \ref{sec:res} describes the simulations used to validate the FSB estimator and its covariance, and presents the results of this validation. We summarise these results and conclude in Section \ref{sec:conclusion}. A number of technical results are then provided in the appendices. \vspace{10mm}

\section{Methods}\label{sec:meth}

In this section we outline the basic method in mathematical detail. We work directly with the fields defined on the sphere motivated by our ultimate application which is to study projected cosmological fields. The method, however, is more general than that and can be applied also to flat two-dimensional fields, which corresponds to the flat-sky limit of the formalism described, or even to three-dimensional fields. 

  \subsection{Power spectra}\label{ssec:meth.cl}
    \subsubsection{Generalities}\label{sssec:meth.cl.gen}
      We will limit much of our discussion to the power spectrum and bispectrum of a single scalar field, although the methods and formalism described later in this section are easily applicable to the more general multi-field case. The general treatment of power- and bispectra for spin-$s$ fields has been described in the literature \citep{astro-ph/0612571,astro-ph/0303414,alonsoUnifiedPseudoC_2019}, and the generalisation of our FSB estimator to these cases will be the subject of future work. In what follows, spherical harmonic transforms (SHTs) and their inverse are given by
      \begin{equation}\label{eq:sht}
        a_{\ell m}\equiv\int d\nv\,a(\nv)\,Y^*_{\ell m}(\nv),\hspace{6pt}a(\nv)=\sum_{\ell m}a_{\ell m} Y_{\ell m}(\nv),
      \end{equation}
      where $Y_{\ell m}(\nv)$ are the spherical harmonic functions for the angles corresponding to a unit vector $\nv$. For the sake of conciseness, we will label the action of the spherical harmonic transform and its inverse transform, $\mathcal S$ and $\mathcal S^{-1}$ respectively.

      Let $a(\nv)$ be a random field. The power spectrum $C_\ell$ and reduced bispectrum of the field are defined in terms of the harmonic-space 2-point and 3-point correlators:
      \begin{align}
        \langle a_{\ell m}a_{\ell'm'}\rangle=(-1)^m\delta^K_{\ell\ell'}\delta^K_{m-m'}C_\ell, \label{eq:psp}\\
        \langle a_{\ell_1m_1}a_{\ell_2m_2}a_{\ell_3m_3}\rangle =\gaunt{\ell_1}{\ell_2}{\ell_3}{m_1}{m_2}{m_3}b_{\ell_1\ell_2\ell_3},  \label{eq:bsp}
      \end{align}
      where angle brackets $\langle\cdots\rangle$ denote averaging over realisations of $a$, $\delta^K$ is the Kronecker delta, $\gaunt{\ell_1}{\ell_2}{\ell_3}{m_1}{m_2}{m_3}$ are the Gaunt coefficients,
      \begin{align}
        \gaunt{\ell_1}{\ell_2}{\ell_3}{m_1}{m_2}{m_3}
        &\equiv\int d\nv\,Y_{\ell_1m_1}(\nv)Y_{\ell_2m_2}(\nv)Y_{\ell_3m_3}(\nv) \nonumber \\ \nonumber
        &=\sqrt{\frac{(2\ell_1+1)(2\ell_2+1)(2\ell_3+1)}{4\pi}} \;\; \times\\
        &\hspace{12pt}\wtj{\ell_1}{\ell_2}{\ell_3}{0}{0}{0}\wtj{\ell_1}{\ell_2}{\ell_3}{m_1}{m_2}{m_3},
      \end{align}
      and the Wigner $3j$ symbols satisfy (among other things) the orthogonality relation
      \begin{equation} \label{eq:wtjortho}
        \sum_{m_1m_2}\wtj{\ell_1}{\ell_2}{\ell_3}{m_1}{m_2}{m_3}\wtj{\ell_1}{\ell_2}{\ell_3'}{m_1}{m_2}{m_3'}=\frac{\delta^K_{\ell_3\ell_3'}\delta^K_{m_3m_3'}}{2\ell_3+1} \, .
      \end{equation}

    \subsubsection{Power spectrum estimation}\label{sssec:meth.cl.pcl}
      A large body of literature exists regarding the estimation of angular power spectra from different astronomical datasets \citep{astro-ph/0105302,astro-ph/0303414,astro-ph/0410394,alonsoUnifiedPseudoC_2019}. In the simplest scenario, that of full-sky observations with homogeneous noise properties, a simple estimator can be built replacing the ensemble average in the equations above with an average over the symmetric azimuthal modes, e.g. for two fields $a$ and $b$:
      \begin{equation}\label{eq:pcl}
        \tilde{C}^{ab}_\ell\equiv\frac{1}{2\ell+1}\sum_{m=-\ell}^\ell a_{\ell m}b^*_{\ell m}.
      \end{equation}
      In realistic scenarios, however, in which observations have not been carried out over the whole celestial sphere, and in which the data have inhomogeneous and potentially correlated noise properties, this simple estimator is both biased and suboptimal (in terms of its recovered uncertainties). 
      
      A common solution to this problem is the use of the so-called ``pseudo-$C_\ell$'' estimator (PCL in what follows), also often called the ``MASTER'' algorithm \citep{astro-ph/0410394,alonsoUnifiedPseudoC_2019}. The main effects missed by the simplistic estimator above, and the ways in which the PCL estimator addresses them, are:
      \begin{enumerate}
        \item {\bf Mode loss.} Having observed a fraction of the sky, only an effective fraction of all modes will contribute to the average in Eq. \ref{eq:pcl}. This results in an overall decrement in the recovered power spectrum which is, to a good approximation, given by $1/\langle w_aw_b\rangle_{\rm pix}$, where $w_a$ and $w_b$ are the sky masks of both fields, and $\langle\cdots\rangle_{\rm pix}$ indicates averaging over all pixels on the sphere \citep{garcia-garciaDisconnectedPseudoC_2019,2010.09717}.
        \item {\bf Mode coupling.} The masked field $\tilde{a}\equiv w_a\,a$ is a product of two maps (the original map and its mask) and, therefore, its spherical harmonic coefficients are a convolution of the true harmonic coefficients and those of the mask. This result is also reproduced at the power spectrum level: the power spectrum of the masked maps is a convolution of the true power spectrum and that of the masks. Fortunately, the associated coupling kernel (often called the ``mode-coupling matrix'') can be calculated efficiently through analytical methods. By construction, this coupling matrix normally incorporates also the effects of mode loss.
        \item {\bf Suboptimal weighting.} Obtaining optimal uncertainties in the recovered power spectra ideally involves inverse-variance weighting the data (i.e. multiplying the maps by their inverse covariance matrix)\footnote{Note that this statement is only strictly true in the case of Gaussian random fields.}. While building and inverting the full pixel--pixel covariance matrix can be highly computationally demanding -- and is usually the bottleneck of optimal quadratic estimators \citep{astro-ph/9611174} -- particularly at high resolutions, the pseudo-$C_\ell$ estimator can be used without modification to take into account the local weighting of the map due to noise inhomogeneity. To do so, one need simply promote the mask $w_a$ from being a purely binary map, describing which regions of the sky have been observed ($w_a=1$) and which ones have not ($w_a=0$), to a weight map that tracks the inverse variance of the field in each pixel (with that inverse variance being zero where the field has not been observed). While this fails to recover optimal uncertainties on large scales, or for fields with large correlation lengths (or steep spectra), it is often a sufficiently good approximation for many large-scale structure observables \citep{1306.0005,alonsoUnifiedPseudoC_2019}.
      \end{enumerate}
            
      Besides addressing these issues, the ease with which the PCL estimator can be applied to multiple astrophysical datasets has motivated the development of additional extensions of the method, including the calculation of accurate covariance matrices (which we describe in the next section), the removal of sky systematics via linear deprojection methods and  $E/B$-mode purification for spin-$s$ fields. Extensive descriptions of these methods can be found in \cite{alonsoUnifiedPseudoC_2019}, \cite{garcia-garciaDisconnectedPseudoC_2019} and \cite{2010.09717}. This progress has made the precise estimation of angular power spectra from a large variety of astrophysical datasets, and their exploitation in accurate scientific analyses, a relatively simple and well understood procedure.

    \subsubsection{Power spectrum covariances}\label{sssec:meth.cl.cov}
      In the context of likelihood-based parameter estimation, an accurate estimate of the statistical uncertainties of power spectrum measurements is as important as the measurements themselves, since no reliable scientific constraints can be derived without the covariance matrix of the estimated data vector. For this reason, significant progress has been made in the last few years towards the accurate estimation of power spectrum covariances, particularly in the field of large-scale structure. Since power spectrum estimators are quadratic functions of the corresponding fields, their covariance matrix depends on the 4-point function or trispectrum of the fields involved. As such, the covariance matrix can be generally separated into three contributions \citep{1302.6994,1407.0060, mohammedPerturbativeApproachCovariance2017, liDisconnectedCovariance2point2019, wadekarGalaxyPowerSpectrum2020}:
      \begin{enumerate}
        \item {\bf Disconnected (``Gaussian'') covariance.} The ``disconnected'' contribution to the covariance matrix comes from all terms in the 4-point functions that involve disconnected products of the fields' two-point functions. This is the only non-zero contribution for purely Gaussian random fields (hence its second name), and in practice is usually the dominant contribution to the uncertainties on scales of cosmological interest for most large-scale structure tracers \citep{barreiraAccurateCosmicShear2018,liDisconnectedCovariance2point2019}.
        \item {\bf Connected non-Gaussian (cNG) terms,} corresponding to the remaining contribution from the connected trispectrum for sub-survey modes.
        \item {\bf Super-sample covariance (SSC).} In general, non-Gaussianities lead to non-zero statistical couplings between different Fourier modes \citep{1302.6994,1408.1081,1405.2666,2210.15647}. In particular, in the so-called ``squeezed limit'', the small-scale clustering of the field may be significantly affected by its local large-scale value\footnote{In the common ``separate Universe'' picture, a large-scale fluctuation of the density field is viewed by the small-scale fluctuations as locally residing in a Universe with an effectively different matter density and expansion rate.}. This effect leads to an additional contribution to the covariance matrix, sourced by the variance of modes larger than the surveyed volume and their correlation with the small-scale clustering.
      \end{enumerate}
      Note that the relative dominance of the disconnected covariance is not necessarily a statement about the Gaussianity of the tracers themselves: the galaxy overdensity, for instance, is highly non-Gaussian, even in projection. Nevertheless, its power spectrum uncertainties are normally dominated by the Gaussian component. Because of this, and due to the relative simplicity of the disconnected covariance, significant infrastructure has been built to estimate it accurately and efficiently, accounting for the impact of survey geometry on both signal and noise contributions. These techniques are particularly accurate in noise-dominated regimes.  

      Here we will employ the methodology introduced in \cite{garcia-garciaDisconnectedPseudoC_2019} (see also \cite{astro-ph/0307515,astro-ph/0410394}), the so-called ``improved Narrow-Kernel Approximation''. Under this approximation, as shown in \cite{2010.09717}, the disconnected covariance depends only on the power spectra of the fields under study. In particular, it can be estimated directly from the mode-coupled power spectra of the observed fields corrected for mode loss by multiplying the power spectra by $1/\langle w_aw_b\rangle_{\rm pix}$ as described under point 1 in the previous section\footnote{In fact, \cite{2010.09717} showed that, since this choice accounts for the impact of mode coupling, it leads to a more accurate estimate of the covariance than using the true underlying power spectrum.}, and thus can be calculated from the data in a model-independent way.

  \subsection{Estimator}\label{ssec:meth.fsb}
    \subsubsection{The Filter-Squared Power Spectrum estimator}\label{sssec:meth.fsb.cl}
      As a pedagogical warm-up, let us consider the filtered-squared approach for power spectrum estimation. We do not advocate this as a realistic method for power spectrum estimation, but it illustrates key aspects the proposed method. Consider a field $a(\nv)$ and a set of filters $W_\ell^{L_i}$ which, for a given integer label $i$, only have significant support over a compact range of angular scales $\ell \in L_i$. We assume both $a(\nv)$ and $W_\ell^{L_i}$ to be real-valued. We emphasise that, with our notation, $L_i$ labels a particular range of multipoles $\ell$, but does not correspond to a specific multipole itself. In what follows, we focus on one filter and drop the $i$ index.
      \begin{enumerate}
        \item We start by filtering $a$ in harmonic space, to preserve only a limited range of multipoles $L$. The filtered field is then:
        \begin{equation}
          a_{L}(\nv) \equiv {\mathcal S}^{-1}\left[W_\ell^{L}{\mathcal S}[a]_{\ell m}\right]_{\nv},
        \end{equation}
        \item We then square $a_L$ in real space: $a^2_L (\nv) \equiv(a_L(\nv))^2$.
        \item We calculate the mean of this field in real space, which can be shown to be given by
        \begin{equation}
          \left< a^2_L (\nv) \right> = \frac{1}{4\pi} \sum_\ell (2\ell+1) (W_\ell^L)^2 C_\ell. 
        \end{equation}
      \end{enumerate}
      We have, in effect, measured the power spectrum in the bin defined by $W^L_\ell$. If the variance of the field is equal to the power spectrum, the variance of the filtered field is given by the relevant portion of the power spectrum. The main advantage is that the method transparently integrates the power over the bin support and sums the number of modes that contribute -- the $(2\ell+1)$ factor. For the power spectrum, this is of course trivial, but for the bispectrum, looping over all possible triangles that belong to a certain bispectrum bin is significantly more complicated.

      Similarly, an estimator for any $n$-point correlator can be derived by multiplying $n$ filtered fields in real space and calculating the mean of the resulting field. However, we can multiply $n-1$ filtered fields and cross-correlate with the original field which short-cuts the operation of filtering and multiplying the last field many times as a function of scale. We elaborate on this approach for the $n=3$ case next.
    
    \subsubsection{The Filter-Squared Bispectrum Estimator}\label{sssec:meth.fsb.est}
      The FSB is given by the power spectrum between $(a_L^2)(\nv)$ and $a(\nv)$. In the simplest case (that of full-sky observations, with the power spectrum estimated at each integer $\ell$), this is simply
      \begin{equation}\label{eq:fsb1}
        \hat{\Phi}^L_\ell\equiv\frac{1}{2\ell+1}\sum_m(a_L^2)_{\ell m}a^*_{\ell m}.
      \end{equation}
      In the presence of a mask, and using bandpowers (i.e. bins of $\ell$) labelled by an index $b$, we replace this by the pseudo-$C_\ell$ estimator, as implemented in the \nmt code\footnote{\url{https://github.com/LSSTDESC/NaMaster}} \citep{alonsoUnifiedPseudoC_2019}.
      
      Let us expand the FSB estimator in the full sky case to find its relationship with the bispectrum of the fields involved. We start by writing the harmonic coefficients of the filtered-squared field $(a^2_{L})_{\ell m}$ in terms of the harmonic coefficients of the original field. Since $(a_L)_{\ell m}\equiv W_\ell^L\,a_{\ell m}$:
      \begin{align}\nonumber
        (a_{L}^2)_{\ell_3 m_3} &= \int d\nv\,(a_L^2)(\nv)\,Y^*_{\ell_3m_3}(\nv)\\\nonumber
        &= \sum_{\ell_1 m_1}\sum_{\ell_2m_2} a_{\ell_1m_1} a_{\ell_2m_2} W_{\ell_1}^{L} W_{\ell_2}^L(-1)^{m_3}\gaunt{\ell_1}{\ell_2}{\ell_3}{m_1}{m_2}{-m_3}.
      \end{align}
      Plugging this into Eq. \ref{eq:fsb1}, we obtain:
      \begin{equation}\label{eq:fsb2}
        \hat{\Phi}^L_{\ell} = \sum_{(\ell m)_{123}} {\cal G}_{\ell_1\ell_2\ell_3}^{m_1m_2 m_3} a_{\ell_1m_1} a_{\ell_2m_2} a_{\ell_3 m_3} \, K^{LL\ell}_{\ell_1\ell_2\ell_3},
      \end{equation}
      where we have defined the \emph{FSB kernel}:
      \begin{equation}
        K^{L_1L_2\ell}_{\ell_1\ell_2\ell_3} \equiv \frac{W_{\ell_1}^{L_1} W_{\ell_2}^{L_2}\delta_{\ell\ell_3}}{2\ell_3+1}.
      \end{equation}
      This is directly generalised to the case of bandpowers as
      \begin{equation}\label{eq:fsbkernel}
        K^{L_1L_2b}_{\ell_1\ell_2\ell_3} \equiv \frac{W_{\ell_1}^{L_1} W_{\ell_2}^{L_2}W^b_{\ell_3}}{N_b},
      \end{equation}
      where $W^b_{\ell_3}$ are the weights of the $b$-th bandpower, and $N_b\equiv\sum_{\ell\in b}(2\ell+1)W^b_\ell$ is a normalisation factor. The selection rules of Gaunt coefficients ensure that only combinations of $(\ell_1,\ell_2,\ell_3)$ that form a triangle can contribute to the estimator, while the FSB kernel ensures that the triangles in question are those that have two sides with scales $\ell_1\sim\ell_2\in L$ and one with scale $\ell_3\in b$.

      To derive the theoretical expectation for $\hat{\Phi}$, we take the expectation value of Eq. \ref{eq:fsb2}. Bearing in mind the definition of the bispectrum from Eq. \ref{eq:bsp}, and using the orthogonality of the Gaunt coefficients (Eq. \ref{eq:wtjortho}), we find the final relation between the FSB and the reduced bispectrum $b_{\ell_1\ell_2\ell_3}$:
      \begin{align}\label{eq:fsb_curved}
        \Phi^L_b = & \sum_{(\ell)_{123}} h^2_{\ell_1\ell_2\ell_3}K^{LLb}_{\ell_1\ell_2\ell_3}\,b_{\ell_1\ell_2\ell_3},
      \end{align}
      where
      \begin{equation}
        h^2_{\ell_1\ell_2\ell_3}\equiv\frac{(2\ell_1+1)(2\ell_2+1)(2\ell_3+1)}{4\pi}\wtj{\ell_1}{\ell_2}{\ell_3}{0}{0}{0}^2,
      \end{equation}
      can be interpreted as the number of possible triangles with sides $(\ell_1,\ell_2,\ell_3)$ on the celestial sphere \citep{1509.08107}.
    
    \subsubsection{Relation to other bispectrum estimators}\label{sssec:meth.fsb.other}
      Having a closed expression for the FSB estimator (Eq. \ref{eq:fsb2}) allows us to relate it to other bispectrum estimators previously used in the literature. First, consider the problem of estimating the amplitude $A$ of a bispectrum with a known scale dependence  (i.e. $b_{\ell_1\ell_2\ell_3}=A\,T_{\ell_1\ell_2\ell_3}$, with $T_{\ell_1\ell_2\ell_3}$ a known template). In the weakly non-Gaussian regime, the optimal estimator for $A$ takes the form \citep{astro-ph/0503375,astro-ph/0612571}:
      \begin{equation}\label{eq:bispec_optimal}
        \hat{A}=\frac{1}{S_{\rm norm}}\sum_{(\ell m)_{123}}\gaunt{\ell_1}{\ell_2}{\ell_3}{m_1}{m_2}{m_3}\left[\frac{T_{\ell_1\ell_2\ell_3}}{C_{\ell_1}C_{\ell_2}C_{\ell_3}}a_{\ell_1m_1}a_{\ell_2m_2}a_{\ell_3m_3}-{\cal L}\right],
      \end{equation}
      where $C_\ell$ is the power spectrum of the field under study, ${\cal L}$ is a linear term that we will discuss later, and $S_{\rm norm}$ is the normalisation factor
      \begin{equation}
        S_{\rm norm}\equiv\sum_{(\ell)_{123}}h^2_{\ell_1\ell_2\ell_3}\,\frac{T^2_{\ell_1\ell_2\ell_3}}{C_{\ell_1}C_{\ell_2}C_{\ell_3}}.
      \end{equation}
      The optimal estimator therefore has the same structure as the FSB (ignoring the linear term) with two caveats:
      \begin{itemize}
        \item The optimal kernel takes the form:
        \begin{equation}
          K^A_{\ell_1\ell_2\ell_3}=\frac{1}{S_{\rm norm}}\frac{T_{\ell_1\ell_2\ell_3}}{C_{\ell_1}C_{\ell_2}C_{\ell_3}}.
        \end{equation}
        \item The optimal estimator uses the inverse-variance weighted field $a_\ell/C_\ell$ (hence the presence of $C_\ell$ in the kernel).
      \end{itemize}

      Consider now a very specific choice of template, in which the bispectrum is zero everywhere except in a particular voxel in $(\ell_1,\ell_2,\ell_3)$, defined by bandpower/bin indices $L_1$, $L_2$, and $L_3$, i.e.:
      \begin{equation}
        T^{L_1L_2L_3}_{\ell_1\ell_2\ell_3}=\left\{
        \begin{array}{cc}
          1 & \ell_i\in L_i\,\forall\,i\\
          0 & {\rm otherwise}
        \end{array}
        \right..
      \end{equation}
      This is the basis for the binned bispectrum estimator of \cite{1509.08107} \citep[see also][]{2306.03915}, which aims to reconstruct the individual amplitudes of the bispectrum on different scales in a non-parametric way. In this case, assuming that the power spectrum does not vary significantly within each bandpower, the binned bispectrum kernel is:
      \begin{align}
        &K^{L_1L_2L_3}_{\ell_1\ell_2\ell_3}=\frac{\Theta^{L_1}_{\ell_1}\Theta^{L_2}_{\ell_2}\Theta^{L_3}_{\ell_3}}{S_{\rm norm}^{L_1L_2L_3}},\hspace{12pt}
        S_{\rm norm}^{L_1L_2L_3}=\sum_{\ell_i\in L_i}h^2_{\ell_1\ell_2\ell_3},
      \end{align}
      where $\Theta^L_\ell$ are top-hat filters. Comparing this with Eq. \ref{eq:fsbkernel}, we see that, in the full sky, for top-hat filters, and up to an irrelevant normalisation factor, the FSB is equivalent to the binned bispectrum estimator, in the case of triangular configurations that are close to isosceles (i.e. where $L_1=L_2$).
      
      Let us now briefly discuss the linear term ${\cal L}$ in Eq. \ref{eq:bispec_optimal}. In general, this term takes the form
      \begin{equation}
        {\cal L}=3T_{\ell_1\ell_2\ell_3}{\cal C}^{-1}_{\ell_1m_1,\ell_2m_2}\sum_{\ell_4m_4}{\cal C}^{-1}_{\ell_3m_3,\ell_4m_4}a_{\ell_4m_4},
      \end{equation}
      where ${\cal C}^{-1}_{\ell_1m_1,\ell_2m_2}$ is the general inverse covariance of map $a$ in harmonic space. For a full-sky, statistically isotropic field, this is simply ${\cal C}^{-1}_{\ell_1m_1,\ell_2m_2}=\delta_{\ell_1\ell_2}\delta_{m_1-m_2}/C_{\ell_1}$, in which case
      \begin{equation}
        {\cal L}=3T_{\ell_1\ell_2\ell_3}\frac{\delta_{\ell_1\ell_2}\delta_{m_1-m_2}}{C_{\ell_1}}\frac{a_{\ell_3m_3}}{C_{\ell_3}}.
      \end{equation}
      
      The properties of the Wigner 3$j$ symbols (specifically, their selection rules and particular values when $m_1=-m_2$) cancel out all contributions from the linear term except the monopole $\ell_3=m_3=0$ \citep{astro-ph/0503375}, making ${\cal L}$ irrelevant in the full-sky limit. In practical cases (e.g. in the presence of a sky mask or anisotropic noise components), the linear term can improve the uncertainties on specific triangle configurations (even if, being linear, its contribution to the mean of the estimator is zero). The FSB estimator presented here lacks this linear term, as do most estimators used for the analysis of the 3D bispectrum in galaxy surveys \citep{1407.5668,1606.00439}. We leave a more detailed study of the potential benefit of adding this term to the estimator for future work, and provide only a tentative recipe for how this could be implemented. As shown in \cite{1509.08107} \citep[see also][]{2107.06287}, the linear contribution can be estimated by simply substituting two of the fields in the estimator by Gaussian simulations uncorrelated with the data, and then averaging over simulations. In the case of the FSB, the contribution from the linear term can be written schematically as:
      \begin{equation}
        \label{eq:lintermfsb}
        \hat{\Phi}^L_{b,{\rm linear}}= \left\langle 2\hat{\Phi}^{L,(aGG)}_b+\hat{\Phi}^{L,(GGa)}_b\right\rangle_G,
      \end{equation}
       where $\hat{\Phi}^{L,(abc)}_\ell$ is the FSB for three fields $a$, $b$, and $c$, where $a$ and $b$ are filtered and multiplied, and then correlated with $c$. $G$ denotes a Gaussian simulation, and $\langle\cdots\rangle_G$ denotes averaging over several such simulations. We study the impact of this linear term on the statistical uncertainties of the FSB in Appendix \ref{app:linterm}. We find that it leads to a modest, but non-zero, reduction of the FSB errors on large scales (small multipoles $\ell$). Thus, although we will not use this linear term in the rest of this paper, including it in the estimator could be beneficial if aiming to constrain signals relevant on squeezed configurations.

      As we have seen, in the full-sky limit, for top-hat filters, and for close-to-isosceles configurations, the FSB estimator can mathematically coincide with the binned bispectrum estimator. However, the treatment of the FSB estimator as the power spectrum of two maps (one of them a filtered-squared map) allows us to make use of the significant infrastructure built by the community for power spectrum estimation, to address important challenges that arise in the practical application of the estimator to real data. We review two of these challenges in the next sections, namely the correction of mask-related effects in the estimator, and the calculation of accurate covariances.

    \subsubsection{Beyond isosceles configuration}
      \begin{figure}
          \centering
          \includegraphics[width=\linewidth]{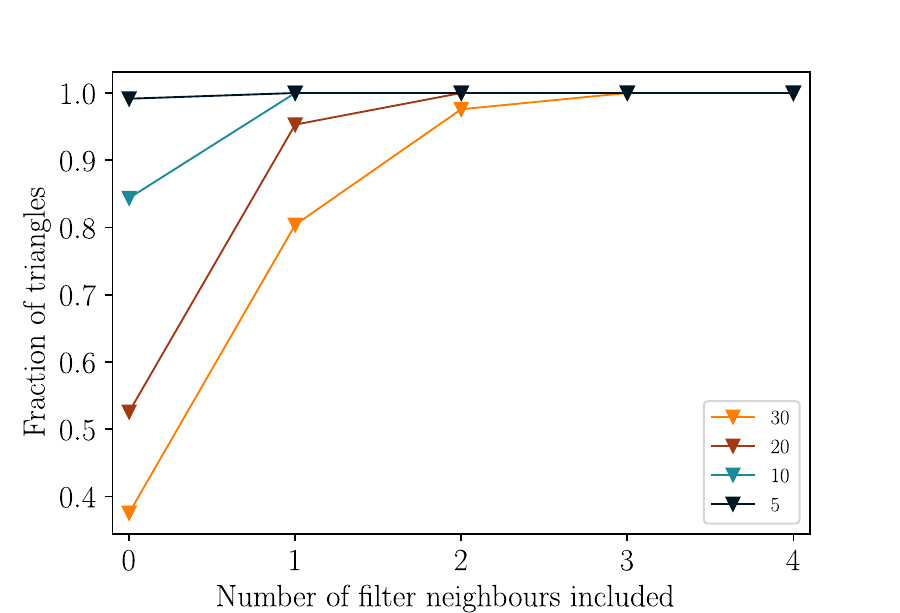}
          \caption{The number of of triangles captured by an FSB-like estimator as a function of the number of filter scale neighbours included for a different number of bins. Different line colours correspond to different number of primary $L$ bins between $\ell_{\rm min}=20$ and $\ell_{\rm max}=2000$ as denoted by the legend. We assume logarithmically-spaced filters, and use the same $\ell$ intervals to define filters and power spectrum bandpowers.}
          \label{fig:isocell}
      \end{figure}

    \begin{figure*} 
        \centering
        \includegraphics[width=\textwidth]{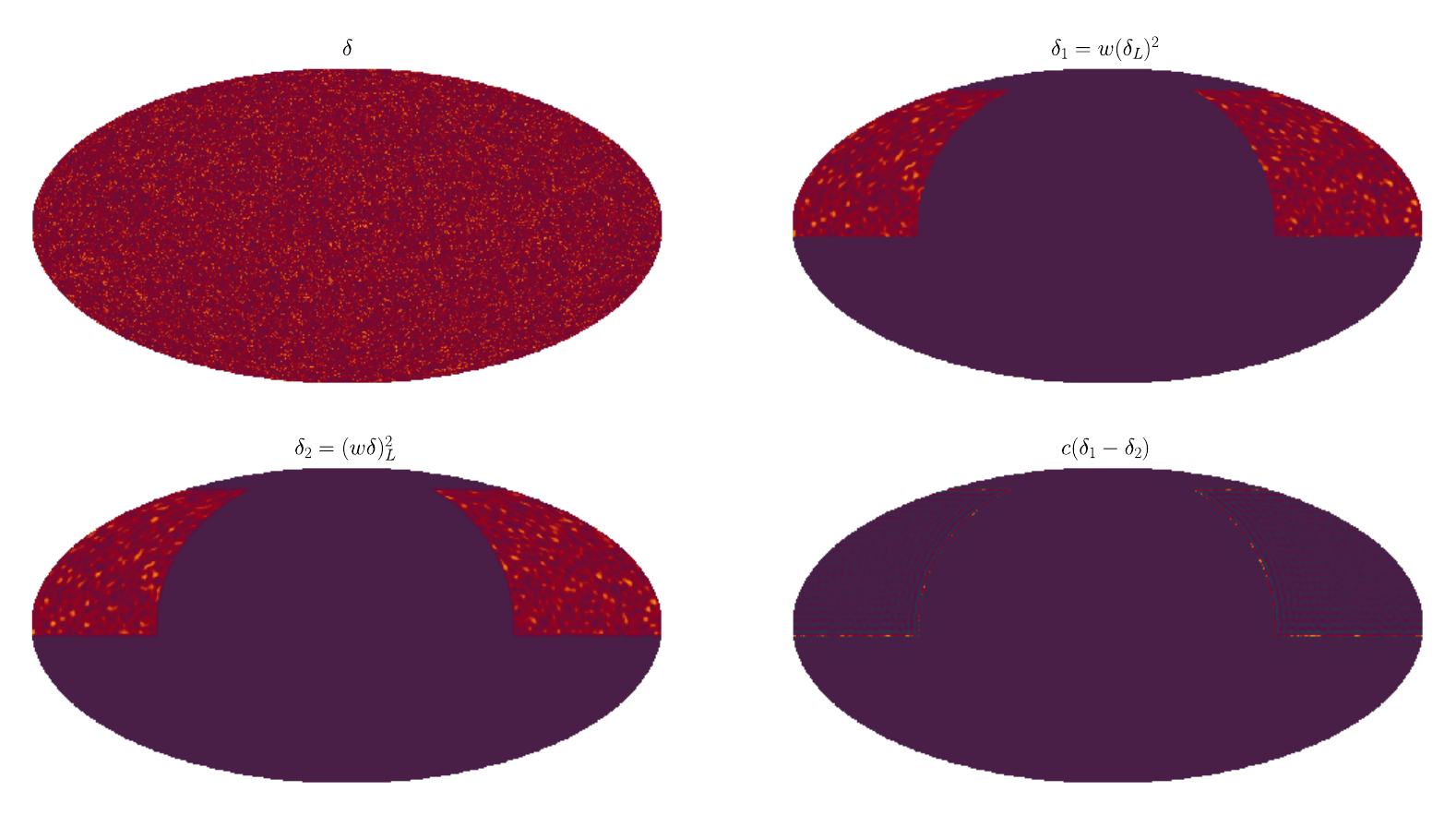}
        \caption{Mask effects: the back and forth transformation between harmonic and map space in the presence of a mask inevitably introduces some Gibbs-like effect -- the ringing pattern being most important on the edges of the mask (although the observed area is also affected to a smaller extent). The FSB is affected by this mask effect, as the filtering step implies computing the spherical harmonic expansion and its inverse.
        The mask effect can be seen by showing the difference (bottom right plot) between filtering the full-sky field \textit{then masking it} before squaring (top right plot), and filtering the \textit{already masked field} and putting it back into map space before squaring it (bottom left plot). All plots have identical colour bars, but the last map has been amplified by a factor of $c=2$ to better show the ringing effect.}\label{fig:maskeffects}
    \end{figure*}
      It is worth noting that the restriction in the type of triangular configurations the FSB can recover is not necessarily a strong limitation. Firstly, the estimator can be easily extended to cover all possible triangles by simply replacing the ``squaring'' step by a multiplication between maps with different filters, with all subsequent steps (including the interpretation in terms of the power spectrum of two maps) unchanged. Secondly, in some practical situations, the FSB is actually able to capture a majority of all available triangles. Specifically, in the case of projected large-scale structure tracers, both the power spectrum and the bispectrum are relatively featureless, and thus large $\ell$ bins can be used without losing much information. This is in fact desirable in the case of the bispectrum \citep{2306.03915}, since it leads to a significant reduction in the size of the resulting data vector. In this case, a given filter $L$ in the FSB actually contains a large number of non-isosceles triangles, since $L$ corresponds to an $\ell$ range and any triangle whose two sides $\ell_1$ and $\ell_2$ are both in $L$ would contribute to the measurement. The triangle condition $|\ell_1-\ell_2|\leq\ell_3\leq\ell_1+\ell_2$ places strong restrictions on the number of triangles that would not be covered by these bins.  To put this argument on a more concrete footing, consider a set of logarithmically-spaced bins between $\ell_{\rm min}=20$ and $\ell_{\rm max}=2000$.  In Fig. \ref{fig:isocell}, we show the fraction of all possible triangles whose sides are captured by an FSB-like estimator as a function of filter neighbours included. By filter neighbours we mean the combination of maps that are used: 0 corresponds to the FSB as presented here, 1 corresponds to adding products of neighbouring $L$-filtered maps, etc.  We see that as the number of bins decreases (and thus their width increases), the fraction of triangles captured by the FSB increases. For example, for the 10 bin case, the ``diagonal'' FSB discussed here covers 80\% of all triangles with 10 measured power spectra. Adding another 9 power spectra using adjacent filters would cover the remaining 20\% of triangles. If we start with 30 primary $L$ bins, we would need to go to $\sim90$ measurements to cover over 95\% of the triangles. We see that the FSB approach naturally affords for a flexible parametrisation of triangle shapes.
    
    \subsubsection{Mask effects}\label{sssec:meth.fsb.mask}
      As discussed in Section \ref{sssec:meth.cl.pcl}, the presence of a sky mask (an unavoidable fact in all astronomical observations) introduces a statistical coupling between different modes of the power spectrum. This is due to the convolution between harmonic coefficients of the mask and the original map when taking the spherical harmonic transform of a masked field. The FSB estimator presented here makes use of spherical harmonic transforms in two different steps: first, when filtering the field, before squaring it; secondly, when correlating the filtered-squared field with the unfiltered field in harmonic space. The latter step is also present in the power spectrum estimation of standard masked fields, and therefore the impact of masking in this step is taken into account exactly by the pseudo-$C_\ell$ estimator. The impact of the sky mask on the filtering step is illustrated in Fig. \ref{fig:maskeffects}. The figure shows a full-sky overdensity field, generated as described in Section \ref{sssec:res.sim.2dlpt} (top left), its filtered-squared version on a particular rectangular sky patch (top right), the result of filtering and squaring the same field after masking it on the same patch (bottom left), and the difference between both maps (bottom right). A top-hat filter covering the range of scales $\ell\in[0, 100]$ was used for this figure. As could be expected, the presence of a mask leads to differences that are mostly concentrated near the mask edges and then ``ring'' around zero, decaying sharply in amplitude further away from the edge.

      As we will show in Section \ref{sec:res}, the impact of these differences, even for more complex realistic sky masks, is fairly small, and does not significantly bias the estimator. For this reason, we have decided not to correct for the effect in the results presented here, although we take the following steps to mitigate it. Firstly, we work with zero-monopole fields, subtracting the mean of the field calculated within the mask. A large mean leads to significantly larger ringing, associated with what is effectively the harmonic transform of a sharp top-hat function. 
      We also re-mask the resulting filtered-squared field with the original mask, to cancel out any ringing outside of the unmasked area. Other strategies can be followed to further mitigate this effect if necessary. For instance, the mask applied to the filtered-squared field can be apodised, to further remove the impact of ringing within the observed area. We did not observe this to significantly improve what was already a negligible estimator bias for the cases studied here. Other strategies would involve making use of inpainting techniques to add simulated structure around the unmasked regions before filtering -- as has been done in the past to account for masking effects in standard bispectrum estimators \citep{1509.08107,Jung_2020}. We reiterate that these effects are subdominant for all cases studied here, and we therefore leave a more detailed quantitative study of these potential mitigation strategies for future work.

    \subsubsection{FSB covariances}\label{sssec:meth.fsb.cov}
      An accurate estimate of the covariance matrix of a given data vector is crucial for any reliable scientific analysis derived from it. This is a particularly challenging problem in the case of bispectra. First, the large size of the data vector for general bispectrum configurations makes the analytical calculation of its covariance a numerically intensive problem. Secondly, since the bispectrum is a third-order statistic, its covariance involves a particular configuration of the six-point correlator of the fields involved. A general theoretical calculation of these six-point functions, without a clear understanding of which contributions to them are relevant and which are negligible, is also highly computationally demanding. The alternative approach of using simulations to estimate the covariance matrix suffers from similar challenges as the large data vector size requires a large number of N-body simulations, making this approach computationally very demanding. Fortunately, significant progress has been made in theoretically estimating bispectrum covariance matrices in recent years \citep{biagettiCovarianceSqueezedBispectrum2022, salvalaggioBispectrumNonGaussianCovariance2024a}.

      In this sense, the interpretation of the bispectrum within the FSB formalism as the power spectrum of two fields, allows us to use much of the significant insight and infrastructure built by the community for the estimation of power spectrum covariances. A key insight in this area is the decomposition of the covariance for non-Gaussian fields into ``disconnected'' or ``Gaussian'' contributions -- those that would be present for purely Gaussian fields, dependent only on their two-point correlators -- and ``connected'' terms that involve higher-order correlators \citep{mohammedPerturbativeApproachCovariance2017}. It has generally been found that, even for highly non-Gaussian fields, such as the late-time matter overdensity, the power spectrum covariance is dominated by the disconnected contribution, and that the remaining connected terms are commonly dominated by the additional variance of a small number of modes (often just one, see e.g. \citealt{1407.0060} and \citealt{mohammedPerturbativeApproachCovariance2017}). This result has been exploited by the community to develop highly-accurate schemes to generate power spectrum covariances. In particular, accurate methods now exist to calculate the dominant disconnected power spectrum covariance fully accounting for survey geometry effects, particularly the mode-coupling caused by the presence of a sky mask \citep{liDisconnectedCovariance2point2019,garcia-garciaDisconnectedPseudoC_2019}. What is more, since this contribution depends only on the power spectra of the fields under study, it can be estimated directly from our own measurements in a model-independent way. The subdominant connected contributions, for which accuracy requirements are usually more lax, are commonly calculated by adopting a particular model (e.g. the halo model) to describe the higher-order correlators of the fields, and data-driven methods, such as jackknife resampling, can be used to validate the calculation.

      The question then arises: can a similar logic be applied to the calculation of bispectrum covariances? Within the FSB scheme, the bispectrum is treated as the power spectrum of a given field $a$ and its filtered-squared version $a_L^2$. Since often the fields involved are highly non-Gaussian to begin with, there is no immediate reason to believe that $a_L^2$ is significantly more non-Gaussian than $a$. Hence, the approximations described above for power spectrum covariances could be highly applicable to the FSB. As we will show in the rest of this paper, this is indeed the case: the FSB covariance is dominated by a disconnected component that can be calculated accurately in a model-independent way using power spectrum-based techniques. Most other connected components are subdominant, and the most relevant ones can be calculated accurately using the measured power spectra (i.e. also in a model-agnostic way).
      
      Note that the above is not a trivial statement: the disconnected FSB covariance is the covariance that arises from treating the two fields involved, $a$ and $a_L^2$, as Gaussian fields. This is different from the covariance that would arise from treating the original field, $a$, as a Gaussian field. For concreteness, we will distinguish both approximations by labelling them as ``disconnected'' and ``Gaussian'' respectively. To illustrate this point, consider the disconnected FSB covariance for full-sky observations. Since the FSB is simply the power spectrum of $a$ and $a_L^2$, we can use the so-called ``Knox formula'' \citep{knoxDeterminationInflationaryObservables1995}:
      \begin{align}\label{eq:knox_FSB}
        {\rm Cov}_{\rm dc} & \big(\hat{\Phi}^L_\ell, \hat{\Phi}^{L'}_{\ell'}\big) \nonumber \\ & =\frac{\delta_{\ell\ell'}}{2\ell+1}\left[C_\ell^{aa}C_\ell^{a_L^2a_{L'}^2}+C_\ell^{aa_L^2}C_\ell^{aa_{L'}^2}\right].
      \end{align}
      We can see that the term $C^{a_L^2a_{L'}^2}_\ell$ involves a 4-point function of $a$ and, once estimated from the data, it will include both a completely disconnected contribution (quadratic in $C_\ell^{aa}$), and a connected term involving the connected trispectrum of $a$. The latter would be zero for the Gaussian covariance. Likewise, the second term in the equation above involves the bispectrum of $a$, and would also be zero in the Gaussian approximation. Thus, we can see that the disconnected FSB covariance includes non-Gaussian contributions that would not be present in the purely Gaussian approximation. In fact, we can make a much stronger statement: \emph{the disconnected FSB covariance contains all Gaussian and non-Gaussian terms in the general bispectrum covariance that contribute exclusively to the diagonal FSB errors}. The mathematical proof of this statement is rather lengthy, and is fully fleshed out in Appendix \ref{app:gauscov}. The infrastructure built for the calculation of accurate power spectrum covariances (e.g. \citealt{garcia-garciaDisconnectedPseudoC_2019} and \citealt{2010.09717}) can therefore be used to estimate the bispectrum covariance (and its covariance with the power spectrum) within the FSB formalism in a straightforward and model-independent way\footnote{Model-independent since all ingredients of Eq. \ref{eq:knox_FSB} can be estimated directly from the data.}.

      As detailed in Appendices \ref{app:gauscov} and \ref{app:nongauscov}, the full FSB covariance contains additional terms that are not captured by the disconnected FSB covariance described above. These contributions are spread over all off-diagonal elements of the covariance matrix, and would correspond to the connected trispectrum components of the fields $a$ and $a_L^2$ in the usual power spectrum picture. Of these, the leading contribution to the FSB covariance, which we label $N_{222}$, takes the form
      \begin{align}\nonumber
        &{\rm Cov}_{N_{222}}\left(\hat{\Phi}^L_\ell,\hat{\Phi}^{L'}_{\ell'}\right)=\\\label{eq:covn222_1}
        &\hspace{20pt}\frac{C_\ell C_{\ell'}W_{\ell'}^LW^{L'}_\ell}{\pi}
        \sum_{\ell_1}(2\ell_1+1)W_{\ell_1}^LW_{\ell_1}^{L'}C_{\ell_1}\wtj{\ell}{\ell'}{\ell_1}{0}{0}{0}^2,
      \end{align}
      in the full sky. 
      
      Analogous results hold for the cross-covariance between the angular power spectrum and the FSB. In the full sky, the disconnected covariance takes the form:
      \begin{align}\label{eq:knox_FSB_cl}
        &{\rm Cov}_{\rm dc}\left(\hat{\Phi}^L_\ell,C_{\ell'}\right)=\frac{\delta_{\ell\ell'}}{2\ell+1}2C_\ell^{aa}C_\ell^{aa_L^2},
      \end{align}
      which would be exactly zero in the Gaussian approximation. The leading-order connected contribution to this cross-covariance, which we label $N_{32}$, takes the form
      \begin{equation}\label{eq:covn32_1}
        {\rm Cov}_{N_{32}}\left(\hat{\Phi}^L_\ell,C_{\ell'}\right)=\frac{4 W^L_{\ell'}C_{\ell'}\Phi^{L\ell}_{\ell'}}{(2\ell+1)},
      \end{equation}
      where $\Phi^{L\ell}_{\ell'}$ is a generalisation of the FSB obtained by multiplying the filtered field $a_L(\nv)$ with $a_\ell(\nv)$, the field filtered on a single multipole $\ell$ (i.e. for a delta-like kernel), and then correlated with the original field $a(\nv)$:
      \begin{equation}
        \Phi^{L\ell}_{\ell'}\equiv \langle (a_L\,a_\ell)_{\ell'm'}a^*_{\ell'm'}\rangle.
      \end{equation}

      From these two leading non-Gaussian contributions, we can already draw some conclusions as to the behaviour of the covariance off-diagonal elements: triangles sharing the same filtered modes are more strongly correlated ($N_{222}$), and triangles sharing the same filtered modes as the power spectrum also receive additional correlation ($N_{32}$). This is in agreement with the findings of \cite{biagettiCovarianceSqueezedBispectrum2022} when studying squeezed configurations in the 3D Euclidean case.

      Crucially, the leading connected contributions to the covariance depend only on the power spectrum and the generalised FSB of the field under study, which can be estimated directly from the data. This then allows us to define the following strategy to estimate the FSB covariance in the presence of realistic sky masks:
      \begin{enumerate}
        \item The disconnected contribution to the covariance is calculated using the incomplete-sky extensions of Knox's formula (Eqs. \ref{eq:knox_FSB} and \ref{eq:knox_FSB_cl}), through the so-called ``improved Narrow-Kernel Approximation'' (iNKA) described in detail in \cite{garcia-garciaDisconnectedPseudoC_2019} and \cite{2010.09717}.
        \item The leading-order connected contributions, ${\rm Cov}_{N_{222}}$ and ${\rm Cov}_{N_{32}}$ are calculated using Eqs. \ref{eq:covn222_1} and \ref{eq:covn32_1}, with all relevant ingredients ($C_\ell$ and $\Phi^{L\ell}_{\ell'}$) estimated from the data\footnote{Following the logic behind the iNKA, in practice we estimate these using the corresponding mode-coupled power spectra corrected for the mean mask product -- see \cite{2010.09717} for details.}. Since these equations are only exact in the full-sky limit, we account for the impact of survey geometry by scaling these contributions by the inverse observable sky fraction \citep{1004.4640}.
        \item The disconnected and connected contributions are then added together.
      \end{enumerate}
      The accuracy of this recipe at recovering the FSB covariance will be validated in Section \ref{sec:res}. 
  
  \subsection{Theory predictions}\label{ssec:meth.th}
    In the Limber approximation and for a flat universe, the reduced bispectrum for field $a$ can be computed as 
    \begin{equation}
      b_{\ell_1\ell_2\ell_3} = \int\mathrm{d}\chi\frac{(q^a(\chi))^3}{\chi^4}B_{aaa}\left(k_{\ell_1},k_{\ell_2},k_{\ell_3}, \chi\right),
    \end{equation}
    where $k_\ell\equiv(\ell+1/2)/\chi$, $B_{aaa}\left(k_1, k_2, k_3, \chi\right)$ is the 3D bispectrum for field $a$, $q^a$ denotes the field's window function and $\chi$ is the comoving distance \citep{Limber1953, loverdeExtendedLimberApproximation2008}.
    For galaxy tracers, which are the focus of this work, the window function is given by
    \begin{equation}
      q^{\delta_g, i}=\frac{H(z)}{c}p^i(z),
    \end{equation}
    where $p^i(z)$ denotes the normalised redshift distribution of galaxy sample $i$, $H(z)$ is the Hubble parameter and $c$ denotes the speed of light.

    In order to model the galaxy bispectrum, in this work, we make the simplifying assumption that galaxies are linearly biased with respect to the underlying dark matter distribution. This allows us to set
    \begin{equation}
      B_{ggg}(k_1, k_2, k_3) = b^3B_{mmm}(k_1, k_2, k_3),
    \end{equation}
    where $b$ denotes the linear, scale- and redshift-independent bias of the galaxy sample considered, and $B_{mmm}(k_1, k_2, k_3)$ is the 3D matter bispectrum. To model $B_{mmm}(k_1, k_2, k_3)$ we employ two different approaches.

    The first approach is based on perturbation theory: to lowest order in perturbation theory (tree-level or leading-order), the matter bispectrum is given by \citep[see e.g.][]{Desjacques:2018}:
    \begin{equation}
      B_{mmm}^{\mathrm{tree}}(k_1, k_2, k_3)=2F_2(\boldsymbol{k}_1, \boldsymbol{k}_2)P_L(k_1)P_L(k_2) + 2 \:\mathrm{perm.},
    \end{equation}
    where $P_L(k)$ is the linear matter power spectrum and $F_2(\boldsymbol{k}_1, \boldsymbol{k}_2)$ denotes the perturbation theory kernel defined as
    \begin{equation}
      F_2(\boldsymbol{k}_1, \boldsymbol{k}_2)=\frac{5}{7}+\frac{2}{7}\frac{(\boldsymbol{k}_1 \boldsymbol{k}_2)^2}{k_1^2k_2^2}+\frac{\boldsymbol{k}_1 \boldsymbol{k}_2}{2k_1k_2}\left(\frac{k_1}{k_2}+\frac{k_2}{k_1}\right).
    \end{equation}
    The tree-level matter bispectrum provides accurate fits to simulation results for scales up to $k_{\mathrm{max}}\lesssim 0.1 h$ Mpc$^{-1}$ \citep[see e.g.][]{Takahashi:2020}. In order to extend the validity of our theoretical model to smaller scales, we consider an alternative approach to model the matter bispectrum. Specifically, we employ the BiHalofit fitting function developed in \cite{Takahashi:2020}. We model the remaining cosmological quantities using the Core Cosmology Library (\texttt{CCL}\footnote{\url{https://github.com/LSSTDESC/CCL}}) \citep{Chisari:2019}.

    Galaxies are discrete tracers of the underlying density field and the observed galaxy FSB thus receives stochastic contributions. As shown in Appendix \ref{app:stochasticity}, assuming Poisson statistics for galaxies, these two terms are given by
    \begin{align}\label{eq:fsb_stoch}
      \Phi^L_{b, \mathrm{stoch}} = & \sum_{(\ell)_{123}} h^2_{\ell_1\ell_2\ell_3}K^{LLb}_{\ell_1\ell_2\ell_3}\,\left[\frac{1}{\bar{n}}\left(C_{\ell_1}+C_{\ell_2}+C_{\ell_3}\right)+\frac{1}{\bar{n}^2}\right],
    \end{align}
    where $\bar{n}$ denotes the angular number density of galaxies per steradian in the survey. It is worth noting that the first term in Eq.~\ref{eq:fsb_stoch} scales as the galaxy bias squared and could thus potentially help break parameter degeneracies.

\section{Results and validation}\label{sec:res}
  \subsection{Validation suite}\label{ssec:res.sims}
    In order to validate the FSB estimator and the methods described in Section \ref{sssec:meth.fsb.cov} to calculate its covariance, we must make use of simulations with a bispectrum that is non-zero, known and, in as much as possible, realistic (i.e. representative of the type of bispectrum signal present in realistic LSS probes). The precision with which the estimator can be validated is ultimately limited by the number of simulations used. For these reasons, our validation is carried out in three stages using different types of simulations that balance realism and computational cost. These three stages are described in the next three subsections. The outcome of each simulation type is a two-dimensional map of the galaxy overdensity. The analysis of these maps to validate the FSB estimator is described in Section \ref{sssec:res.sims.val}.

    \subsubsection{Stage 1: 3D LPT simulations}\label{sssec:res.sims.3dlpt}
      We first apply the FSB estimator to fast, full-sky simulations generated using first-order Lagrangian Perturbation Theory as a basis for the structure formation model. These simulations were generated using \texttt{CoLoRe} \citep{ramirez-perezCoLoReFastCosmological2022}, a fast simulation code that can be used to generate full-sky realisations of cosmological surveys. Large-scale structures are reproduced using a perturbative scheme rather than a full N-body approach; it is therefore significantly less computationally expensive to run than standard N-body simulations. \texttt{CoLoRe} implements first- and second-order LPT to generate a three-dimensional matter overdensity field in the lightcone that is then biased and Poisson-sampled to generate simulated galaxy catalogues (as well as a large variety of other cosmological probes).
      
      We generate 100 fast simulations of a galaxy sample with clustering properties (number density and linear galaxy bias) similar to those of the DECaLS DR8 sample, specifically the sample selected in \cite{2010.00466}, assuming a subsample with photometric redshifts in the range $0.3\leq z_p\leq0.8$. Simulations were generated on a box encompassing the comoving volume out to $z=0.9$, with a grid size $N_{\rm grid}=1024$, corresponding to a moderate spatial resolution of $\Delta x=4.2\,h^{-1}{\rm Mpc}$. Maps of the projected galaxy overdensity were then constructed from these simulated catalogues.

      The main advantage of using {\tt CoLoRe} is the ability to generate full-sky realisations at relatively cheap cost, which are vital to quantify any bias in the FSB by comparing the result of applying it to full- and incomplete-sky observations with masks of arbitrary complexity. The main drawbacks are the limitations of LPT to reproduce the small-scale clustering of matter and galaxies, and their relatively high computational cost compared with purely two-dimensional approaches. These limitations are addressed in stages 2 and 3.

\begin{figure}[b]
    \centering
    \includegraphics[width=0.5\textwidth]{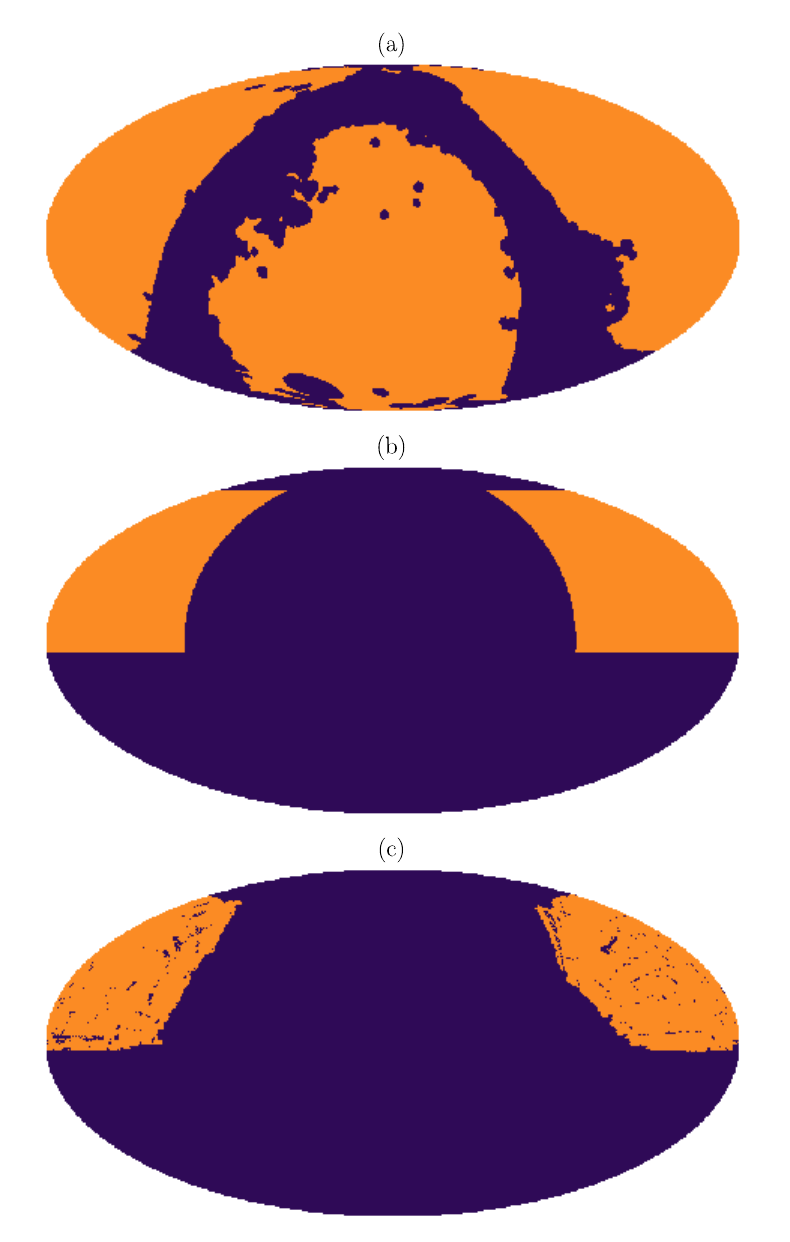}
    \caption{Binary masks used throughout this study: Galactic mask (a), square mask (b), and the BOSS survey mask (c).}
    \label{fig:masks}
\end{figure}
    
    \subsubsection{Stage 2: N-body simulations}\label{sssec:res.sims.glam}

      Our second validation suite consists of 600 of the 2000 publicly available GLAM simulated galaxy catalogues, often referred to as lightcones \citep{2024MNRAS.tmp.1526E}. 
      These lightcones were constructed from the GLAM suite of N-body simulations \citep{GLAMsimu}, which follow the evolution of $2000^3$ dark matter particles in a $1h^{-1}\mathrm{Gpc}$ box, corresponding to a particle mass resolution of $1.06\times10^{10}\,h^{-1}\mathrm{M}_\odot$. The cosmological parameters adopted are $\Omega_\mathrm{m,0}=0.309$, $\Omega_\mathrm{b,0}=0.0486$, $\Omega_\mathrm{\Lambda,0}=0.691$, $h=0.677$, $n_\mathrm{s}=0.9667$, and $\sigma_8=0.816$, representing the best-fit $\Lambda$CDM parameters corresponding to the \textit{Planck} 2015 cosmology \citep{planck15}.

      GLAM lightcones are built by assigning galaxies to distinct halos using the halo occupation distribution method \citep[HOD;][]{PeacockSmith,BerlindWeinberg,Kravtsov}, projecting the positions and radial velocities to observable coordinates (angles and redshifts). The GLAM lightcones used here reproduce the observed clustering statistics of the CMASS sample from the Baryon Oscillation Spectroscopic Survey (BOSS), achieving an excellent fidelity on second, third, and fourth-order statistics \citep{2024arXiv240109523P}. Furthermore, they replicate the redshift and stellar mass-dependent evolution of observed BOSS galaxy clustering. For a detailed description of the construction of these lightcones, we refer readers to \cite{2024MNRAS.tmp.1526E}. The CMASS sample has an effective redshift $z_{\rm eff}=0.57$, and a linear galaxy bias  $b_g=2$ \citep{1202.6057}. 

      The GLAM simulations allow us to validate the FSB estimator for data with a completely realistic bispectrum signal, free from any of the approximations present in the simulations used in stages 1 and 3. Their main drawbacks, in turn, are:
      \begin{itemize}
        \item The number of realisations, while large and sufficient to validate the FSB estimator, limits the accuracy with which we can validate our covariance matrix estimation methods. This limitation is addressed in Stage 3.
        \item The presence of shot noise increases the statistical uncertainties of the FSB, especially on small scales, and reduces the accuracy to which the estimator is tested. This limitation is addressed in Stage 3.
        \item The simulated catalogues do not cover the complete celestial sphere, but rather a rectangular region large enough to encompass the BOSS footprint. When using the GLAM simulations, we will therefore compare the result of the FSB estimator on maps defined on this rectangular region against maps with the more complex BOSS mask (see details in Section \ref{ssec:res.glam}). This limitation (our inability to test our estimator against full-sky simulations) is addressed in stages 1 and 3.
      \end{itemize}
      We note that, while the GLAM suite contains simulated catalogues that include the impact of survey incompleteness and fibre assignment weights on the BOSS sample, we use simpler versions of these catalogues that populate the full rectangular region described above, before any of these effects are included. We build galaxy overdensity maps from these catalogues and then impose the BOSS mask to validate the estimator. The reason for this choice is that the presence of incompleteness and fibre assignment weights modifies the effective shot noise of the sample, which changes the shape of the bispectrum in a non-trivial way as described in Section \ref{ssec:meth.th}. This would thus prevent an accurate comparison between the rectangular-mask measurements (in which these effects are not present) and measurements using the more complex BOSS mask, as well as any comparison against perturbative theoretical expectations (which we attempt in Section \ref{ssec:res.glam}).

    \subsubsection{Stage 3: 2D LPT simulations}\label{sssec:res.sim.2dlpt}
      Our final suite of simulations aims to address the two main shortcomings of the previous two stages: the presence of shot noise in the maps (and its contribution to the statistical uncertainties), and the possibility of generating large numbers of low-cost simulations with a roughly realistic 2, 3, and 4-point correlation function, in order to test our covariance matrix estimator.

      We achieve this following a method similar to that used in \cite{alonsoRecoveringTidalField2016} to reconstruct the projected tidal field from 2D maps of the galaxy overdensity in photometric redshift surveys. In essence, the method is a direct application of first-order Lagrangian perturbation theory to 2-dimensional spaces. The procedure is described in more detail in Appendix \ref{app:2dlpt}. In short, a Gaussian realisation of the projected overdensity is generated. This is interpreted as the covariant divergence of a vector field on the sphere, describing the displacement of matter elements from their initial Lagrangian positions. The overdensity field resulting from this displacement can then be constructed from the original Gaussian realisation by differentiation in harmonic space followed by a simple non-linear transformation. This procedure is thus very fast and memory-efficient, and allows us to generate large numbers of simulations with non-zero (and, to some extent, realistic) higher-order statistics. We do not Poisson-sample the resulting overdensity fields, and thus these realisations are free from shot noise.

      We generate 6000 such simulations, with the initial Gaussian overdensity field generated from a power spectrum compatible with the clustering of the BOSS CMASS galaxies. These are generated on the full sky, and the estimator is then applied to the same maps with different sky masks.
    
    \subsubsection{Validation procedure}\label{sssec:res.sims.val}
      In each of the stages we just described, the outcome is a set of simulated maps of the galaxy overdensity defined over either the full sky (in stages 1 and 3) or over a large rectangular sky region (in stage 2). We apply the FSB estimator to these simulations, as well as to versions of them in which a more restrictive sky mask has been applied. Specifically, we will consider three different sky masks:
      \begin{itemize}
        \item The {\bf rectangular mask} (in RA-Dec coordinates) enclosing the BOSS footprint, which we described in Section \ref{sssec:res.sims.glam}.
        \item A large {\bf Galactic mask}, similar to that used in the analysis of full-sky photometric surveys (see e.g. \citealt{1805.11525} and \citealt{1909.09102}). The mask removes regions of high dust extinction and star contamination, as well as a number of bright extragalactic sources. 
        \item The {\bf BOSS mask}, constructed from the positions of random catalogues in the North Galactic Cap, made publicly available with the tenth BOSS data release \citep{1312.4877}. This is the most complex of the masks studied here, with significant small-scale features.
      \end{itemize}
      These three masks are shown in Fig. \ref{fig:masks}. We note that all masks explored here are purely binary for simplicity. This is not in itself a limiting factor in the estimator: employing masks that track the local noise variance of the fields involved leads to an estimator with more optimal statistical uncertainties, without introducing additional biases \citep{alonsoUnifiedPseudoC_2019}. The design and usage of optimal masks in the context of the FSB will be the subject of future work.
   
    \begin{figure*}
      \centering
      \includegraphics[width=\linewidth]{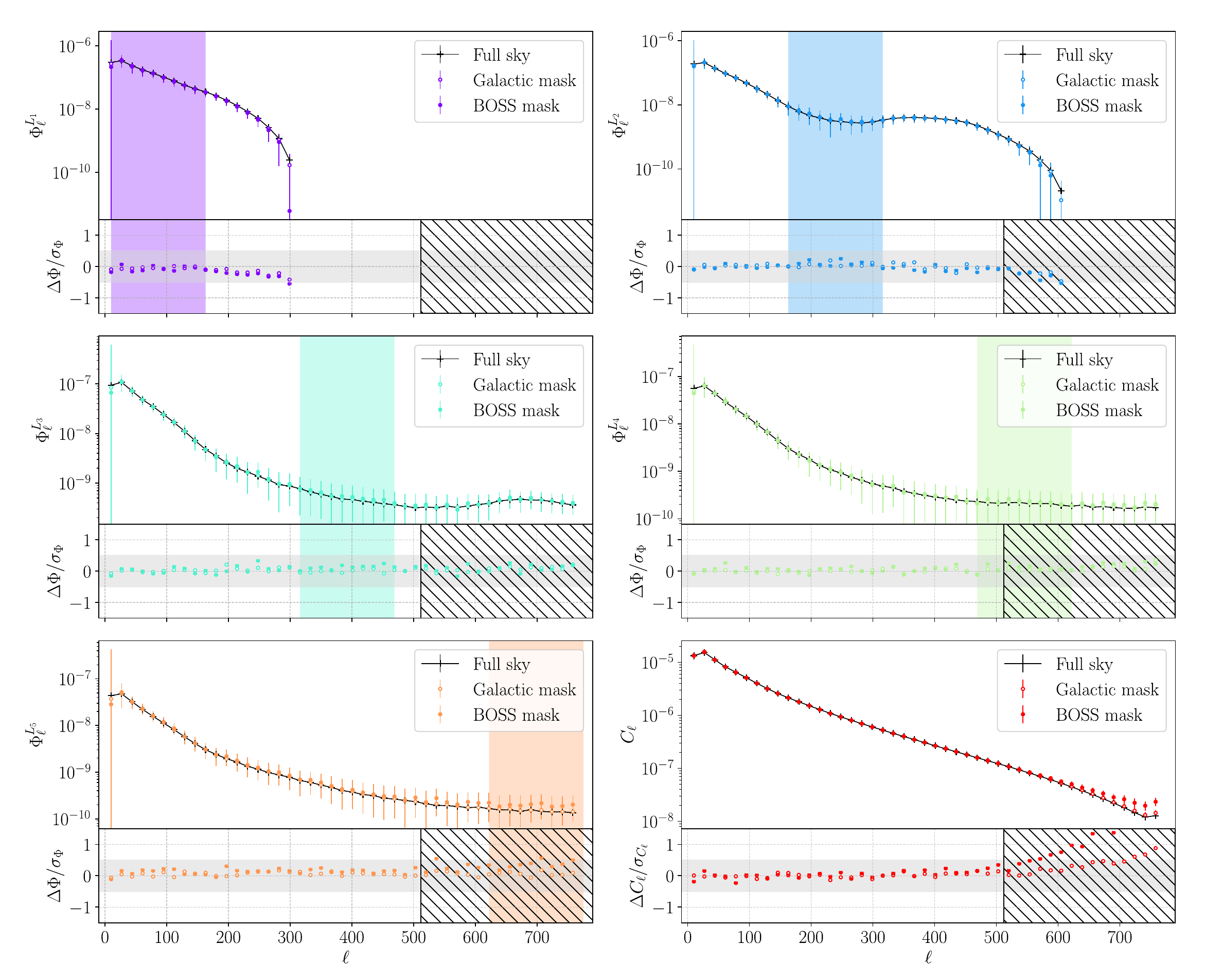}
      \caption{Mean FSB signal in the 3D LPT simulations for 5 different filters, in the presence of two different masks (Galactic mask and BOSS footprint -- empty and full circles, respectively) and in the full-sky scenario. The highlighted $\ell$ range in every plot shows the multipole range of the top-hat filter used for the corresponding FSB. The residuals plots show the bias of the estimator, which remains within the $[-0.5\sigma; 0.5\sigma]$ range (indicated by the grey band) for all filters explored. We ignore the multipole range  $\ell > 2 N_\text{side}$ (hatched in the residual plots), where the numerical instability of the HEALPix SHTs limits the accuracy of the estimator.} \label{fig:lpt3d_fsb_bias}
    \end{figure*}

      Having computed the FSB and angular power spectrum of all simulations, we estimate the mean and covariance matrix of these measurements, and compare the recovered mean FSB in masked and full-sky simulations to quantify the estimator bias. The measured covariance is then compared with the analytical covariance, estimated as described in Section \ref{sssec:meth.fsb.cov}. A few technical points must be made regarding how this comparison is carried out in practice. In the presence of a sky mask and power spectra binned into bandpowers, residual mode-coupling effects must be taken into account when comparing the result of our estimator to its theoretical expectation. This is further complicated by the fact that no exact analytical prediction exists for the expected FSB. Since, ultimately, we wish to ensure that the estimator is able to account for the impact of a sky mask, we proceed by measuring the FSB and power spectra on the full-sky version of all simulations and at all integer multipoles, and use the average of these measurements as a theoretical prediction to compare against. This theoretical prediction is then convolved with the bandpower window functions, accounting for residual mode-coupling effects as described in Eq. 19 of \cite{alonsoUnifiedPseudoC_2019}. Since spherical harmonic transforms of HEALPix maps \citep[the pixelisation scheme we use in this work;][]{gorskiHEALPixFrameworkHigh2005} lose accuracy on multipoles $\ell>2N_{\rm side}$ \citep{1306.0005}, this procedure means that our ``theoretical'' expectations are also inaccurate in this regime. We therefore restrict our validation to scales $\ell\leq2N_{\rm side}$. Finally, as discussed in \cite{2010.09717}, the accuracy of the analytical power spectrum covariance can be significantly improved by using, as input, the mode-coupled power spectra scaled by the inverse mean of the squared mask. Thus, for each type of simulation and mask, we use the mode-coupled pseudo-$C_\ell$, appropriately scaled, and averaged over simulations, as input to calculate the analytical FSB covariance.

      In what follows, we use HEALPix maps with a resolution parameter $N_{\rm side}=256$ (corresponding to $\sim13$-arcminute pixels). We estimate the FSB using five top-hat filters with equal width spanning the range $2\leq\ell\leq3N_{\rm side}-1$. For this proof of concept, this choice of filter shape and numbers was, to a certain extent, arbitrary; the simple top-hat filter was our first choice as it allows for easy tracking of the triangle configurations contributing to each FSB. However, it is not the only valid filter choice; we will leave a comparison of filter shapes and their impact on the estimator to future work.
          All power spectra (including both $C_\ell$ and $\Phi^L_\ell$) are calculated in 45 equi-spaced bandpowers covering the same range of scales. As mentioned above, we restrict our validation to scales $\ell<2N_{\rm side}=512$. All power spectra between fields, including the original field $a$ and its filtered-squared counterpart $a_L^2$, were calculated using the pseudo-$C_\ell$ algorithm as implemented in \nmt \citep{alonsoUnifiedPseudoC_2019}. Disconnected covariance matrices were calculated using \nmt through the Narrow-Kernel Approximation described in \cite{garcia-garciaDisconnectedPseudoC_2019} and \cite{2010.09717}.
  
  \subsection{Validation on 3D LPT simulations}\label{ssec:res.3dlpt}
  
    \begin{figure*}
      \centering
      \includegraphics[width=\textwidth]{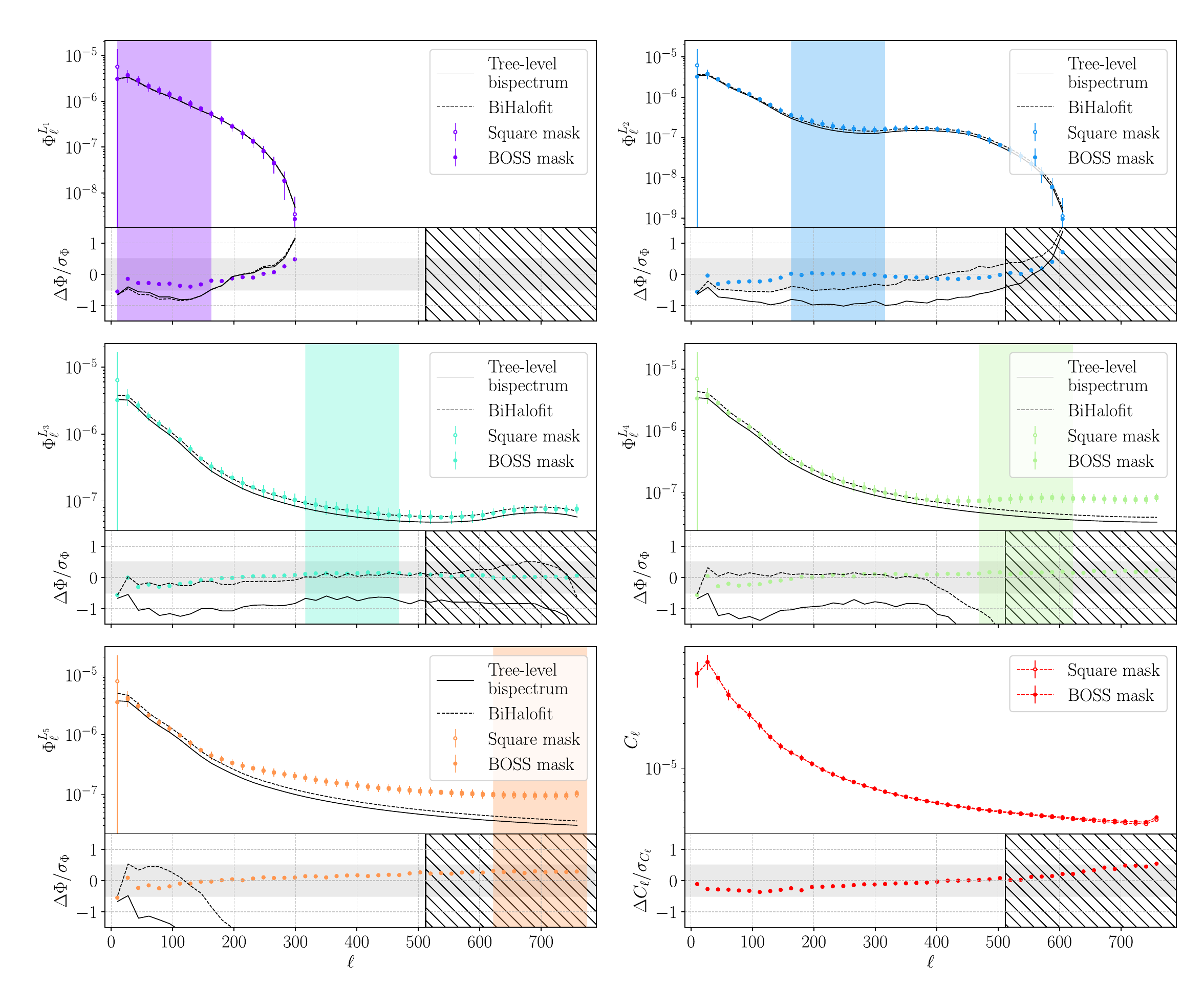}
      \caption{As Fig. \ref{fig:lpt3d_fsb_bias} for the GLAM simulations. Theoretical predictions for the bispectrum are also plotted in solid (Tree-level approach) and dashed (BiHalofit approach) lines.} \label{fig:glam_fsb_bias}
    \end{figure*}

    \begin{figure*}
      \centering
      \includegraphics[width=0.9\textwidth]{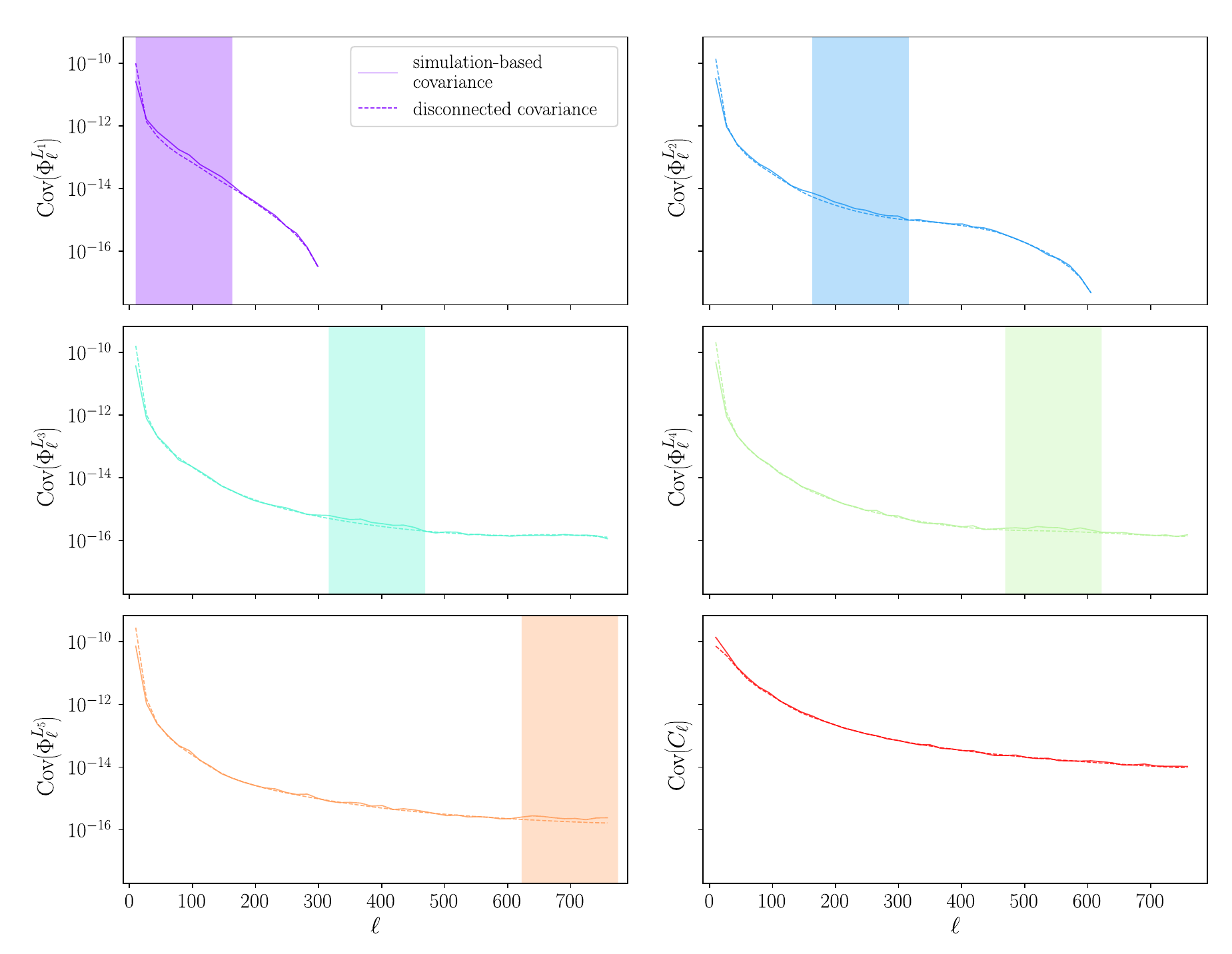}
      \caption{Diagonal of the covariance matrix from GLAM simulations. The solid lines show the main diagonal of the simulation-based covariance, while the dashed lines represent the purely disconnected analytical covariance. The slight discrepancy between the disconnected and simulation-based diagonals within the filtered range are the consequence of an additional non-Gaussian term described in Section \ref{ssec:res.sims}.} \label{fig:glam_covdiags}
    \end{figure*}

    The FSB and power spectrum averaged over the 3D LPT simulations are shown in Fig. \ref{fig:lpt3d_fsb_bias}. Residual plots show the bias of the estimator with respect to its theoretical value which, as described above, is obtained from convolving the full-sky, unbinned FSB with the bandpower window functions. The estimator bias is then obtained by computing the difference between the theoretical and measured values, in units of measurement variance $\sigma$:
    \begin{equation}
      b = \frac{\Phi_{\ell, \text{full sky}}^L - \Phi_{\ell, \text{masked}}^L}{\sigma_{\ell, \text{masked}}} \; .
    \end{equation}
    In our validation range ($\ell < 2N_\text{side}$), the FSB estimator remains within a $\pm 0.5 \sigma$ range (shown as a greyed out horizontal band in Fig. \ref{fig:lpt3d_fsb_bias}) from the theoretical prediction -- a value similar to the bias observed for the power spectrum. At higher $\ell$ (hatched area), the accuracy of our analysis becomes limited by the map resolution: this is particularly striking for the power spectrum (lower right plot), which becomes biased at high $\ell$ in the presence of complex masks. As we mentioned before, this is caused by the instability of the spherical harmonic transforms for HEALPix on scales $\ell>2N_{\rm side}$, which has been documented in the literature \citep{1306.0005,alonsoUnifiedPseudoC_2019}. This also affects the theoretical expectation we compare against, since it is calculated from the SHT of the full-sky realisations. We therefore also discard FSB values above that $2N_\text{side}$ threshold, and will do so for the rest of this analysis. Note that, in this comparison, we have subtracted the shot-noise contribution to the power spectrum, in order to further reduce edge effects at the end of the band. The shot noise pseudo-$C_\ell$ can be calculated analytically as 
    \begin{equation} \label{eq:shotnoise}
        \tilde N_\ell = \frac{\langle w \rangle_\text{pix}}{\bar{n}},
    \end{equation}
    where $\bar{n}$ is the mean number density of the sample in ${\rm srad}^{-1}$ and $\langle w\rangle$ is the mean value of the mask across the full sky (see \cite{Nicola2020} for details). As discussed in Section \ref{sssec:meth.fsb.other}, we do not account for shot noise in the case of the FSB, as its contribution is not as straightforward to estimate and depends on the signal power spectrum -- see Appendix \ref{app:stochasticity}. Despite the remaining shot noise contribution, the relative FSB bias remains low, at levels comparable to that of the power spectrum in the absence of noise. 

    It is interesting to note that in the first two filters (upper two plots in Fig. \ref{fig:lpt3d_fsb_bias}), the FSB signal goes to zero when it becomes impossible to form closed triangles with the scales selected during the filtering step: the maximum value allowed for the third side $\ell_3$ in Eq. \ref{eq:fsb_curved} is twice that of the largest of $\ell_1$ and $\ell_2$ (constrained by the filter), forming a linear triangle in this configuration. Were we to use higher resolution maps, we would also expect the signal to die off in the other filters when that limit is reached. 

    This first comparison thus reassures us that, for the type of bispectrum signal present in these LPT simulations, the FSB estimator is sufficiently accurate, and residual contamination from the filtering stage in the presence of a mask does not lead to significant biases, even in the case of complex masks.

\subsection{Validation on GLAM simulations}\label{ssec:res.glam}

    Fig. \ref{fig:glam_fsb_bias} shows the mean FSB signal and the corresponding estimator bias for the N-body GLAM simulations. Where it is defined, and in the range $\ell<2N_{\rm side}$, the FSB estimator and power spectrum applied to the BOSS mask remain within $0.5 \sigma$ of the measurements made on the larger rectangular mask.

    There is one caveat to this particular validation test; ideally, the correct approach would be to compare the FSB measurements of masked realisations against the FSB obtained in the case of full-sky realisations. As we described earlier, the GLAM lightcones used here cover only a portion of the sky, and we instead compare the FSB estimated on a simple rectangular mask with the FSB measured on the significantly more complex BOSS mask. However, the rectangular mask being relatively simple, we do not expect to see complicated mask effects that would have escaped the validation against the full-sky LPT simulations in Sections \ref{ssec:res.3dlpt} and \ref{ssec:res.2dlpt} (see Fig. \ref{fig:maskeffects}).
    
    In addition to comparing results for the BOSS and rectangular masks, having measured the FSB in the realistic GLAM simulations allows us to attempt a comparison of our measurements against simple theoretical models. Specifically, we compute theoretical predictions using tree-level perturbation theory and the BiHalofit fitting function \citep{Takahashi:2020}, as described in Sec.~\ref{ssec:meth.th}. As our aim is not yet to obtain cosmological constraints from these measurements, we do not attempt to fit our results. Instead, we employ the cosmological parameters and redshift distributions from the GLAM simulations, assume a constant galaxy bias $b_g=2$ for the CMASS sample, and use these to obtain rough theoretical predictions. The results are shown in Fig.~\ref{fig:glam_fsb_bias} alongside the measurements for both sky masks (tree-level and BiHalofit predictions in solid and dashed lines, respectively). As can be seen, the theoretical predictions fit the measurements relatively well, at least on the largest scales, in particular those obtained using BiHalofit. As we consider larger bandpowers, $\ell$, and smaller filter scales, the predictions become increasingly less accurate. This is probably due to the fact that these measurements probe increasingly small scales, where some of the approximations in our theoretical calculations break down. The most important one in this context is the assumption of a linear bias \citep{Desjacques:2018}, as well as having fixed it to a specific value rather than fitting it to the data. The other approximation worth bearing in mind is the Limber approximation, which impact on the power spectrum was thoroughly discussed in \cite{simonHowAccurateLimber2007} and \cite{lemosEffectLimberFlatsky2017}. 
    This is a significantly smaller effect, given the relatively large width of the CMASS redshift bin. For example, the Limber approximation under-estimates the power spectrum at $\ell\sim19$ (the edge of our first bin) by only 2\%, which is much smaller than the statistical uncertainties due to cosmic variance. Furthermore, the impact of this approximation is limited to large scales, instead of the small scales in which we see the main differences with respect to our theoretical prediction.  We leave a thorough comparison of the FSB to theoretical predictions for future work.

    \begin{figure*}
      \centering
      \includegraphics[width=0.9\textwidth]{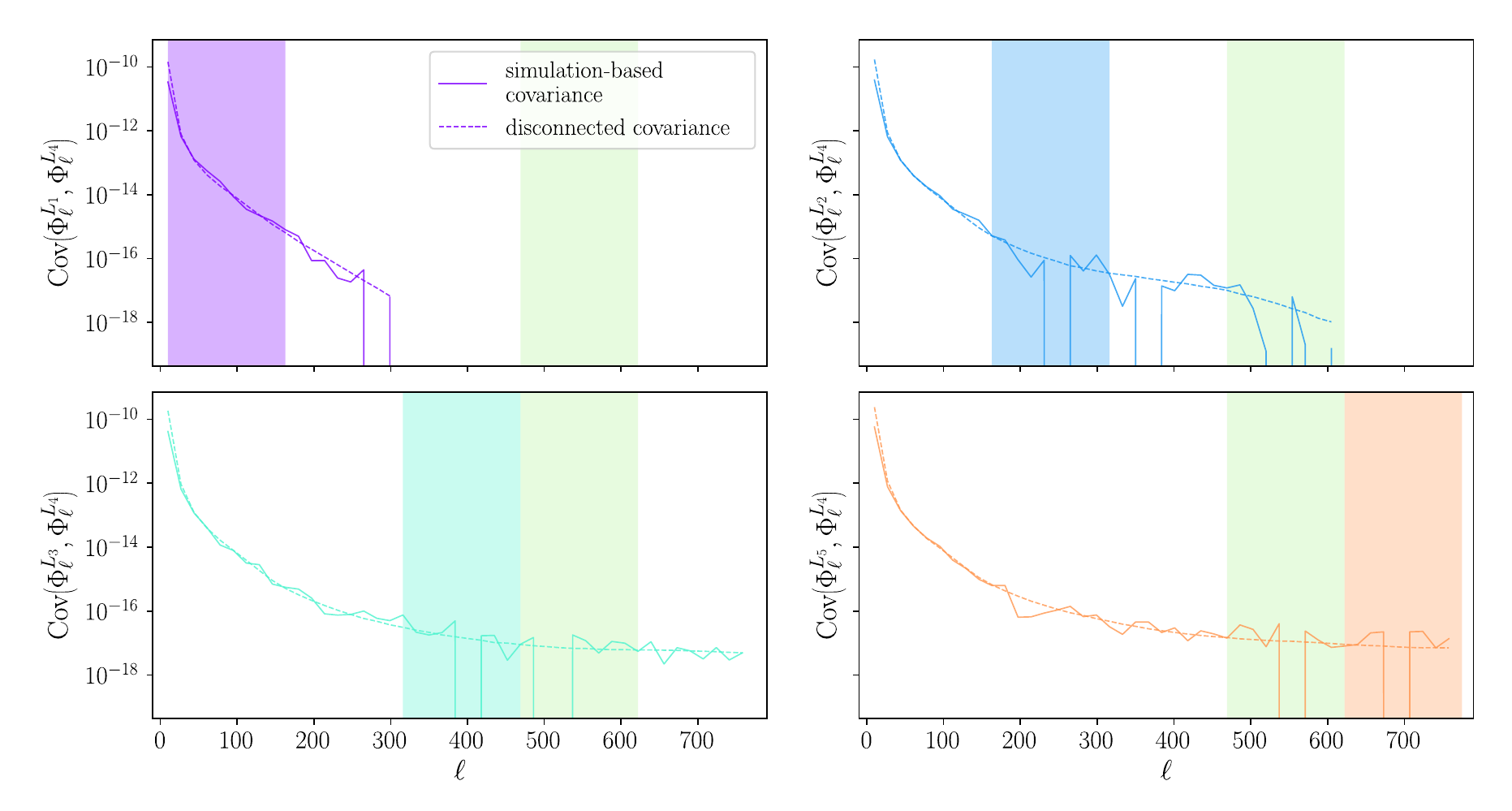}
      \caption{Diagonals of the covariance matrix blocks corresponding to the cross-covariance of the GLAM $L_4$ FSB with the remaining four FSBs.}\label{fig:glam_f3diags}
    \end{figure*}

    \begin{figure*}
      \centering
      \includegraphics[width=0.9\textwidth]{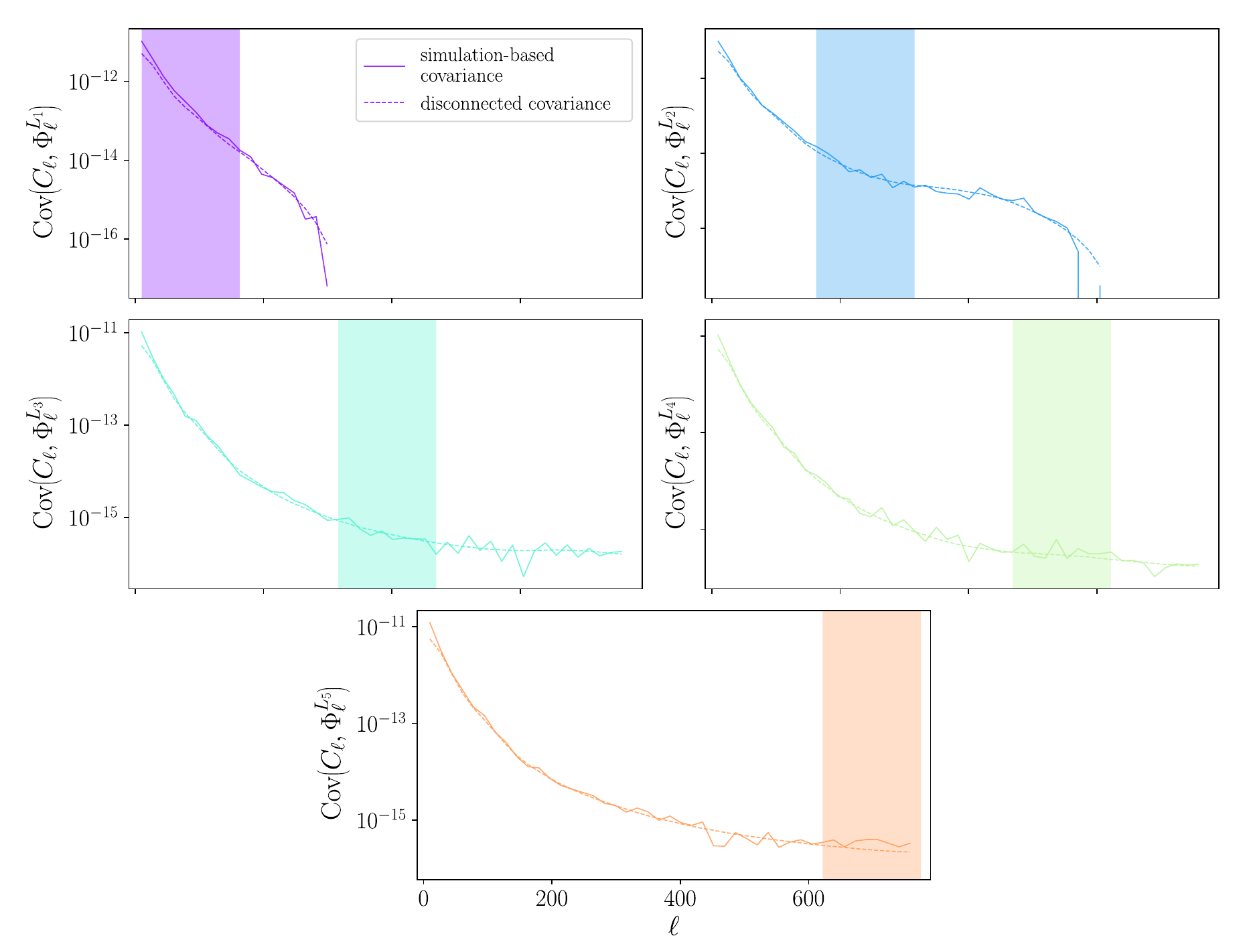}
      \caption{Diagonals of the covariance matrix blocks corresponding to the cross-covariance of the GLAM power spectrum with the five FSBs. \vspace{6mm}}\label{fig:glam_cldiags}
    \end{figure*}

    Having access to 600 simulations allows us to attempt a first assessment of the FSB covariance matrix. We do so for the full data vector containing 5 FSBs and the power spectrum of the original map, and thus the covariance will contain (FSB, FSB), (FSB, $C_\ell$), and ($C_\ell$, $C_\ell$) components. We compute the covariance using two different approaches: the first one is the simulation-based covariance, estimated explicitly from the 600 FSBs and power spectra obtained for each lightcone. The second one is the analytical disconnected covariance as an initial approximation for the full covariance. As described in Section \ref{sssec:meth.fsb.cov}, the analytical disconnected covariance requires, as input, the power spectrum of all fields involved. In our case, this includes the power spectrum of the original field, $C_\ell^{aa}$, its cross-correlation with the filtered-squared field (i.e. the FSB $C^{aa_L^2}_\ell\equiv\Phi^L_\ell$), and the auto-correlation of the filtered-squared field $C_\ell^{a_L^2a_L^2}$. These are estimated from the data, following the iNKA as described in Section \ref{sssec:meth.cl.cov}. This approach, exhaustively discussed in Appendix \ref{app:gauscov}, should capture all the purely-diagonal contributions to the covariance -- regardless of whether those contributions are Gaussian or non-Gaussian in nature. We show the diagonal of a few blocks of the covariance matrix: the auto-covariance of the FSBs and power spectrum in Fig. \ref{fig:glam_covdiags}, the cross-covariance of the fourth FSB and all other FSBs in Fig. \ref{fig:glam_f3diags} and the cross-covariance of the power spectrum and all other FSBs in Fig. \ref{fig:glam_cldiags}. These show reasonable agreement; however, we can identify some mild discrepancies between the two covariances. Most evidently, we can appreciate that the simulated (FSB, FSB) auto-covariance is slightly larger than the analytical disconnected covariance (by about $\sim20\%$) on the scales corresponding to the filter window function. As we will demonstrate in the next section, this additional contribution is strongly dominated by the $N_{222}$ term (and by the $N_{32}$ term in the case of the (FSB, $C_\ell$) covariance -- although this is hardly visible in the last subset of Fig. \ref{fig:glam_f3diags}). In fact, it was the presence of these differences that prompted us to investigate the leading connected contributions to the FSB covariance. The analytical expressions for these terms are given in Section \ref{sssec:meth.fsb.cov}, and their derivation is detailed in Appendix \ref{app:nongauscov}.

    \begin{figure*}
      \centering
      \includegraphics[width=\textwidth]{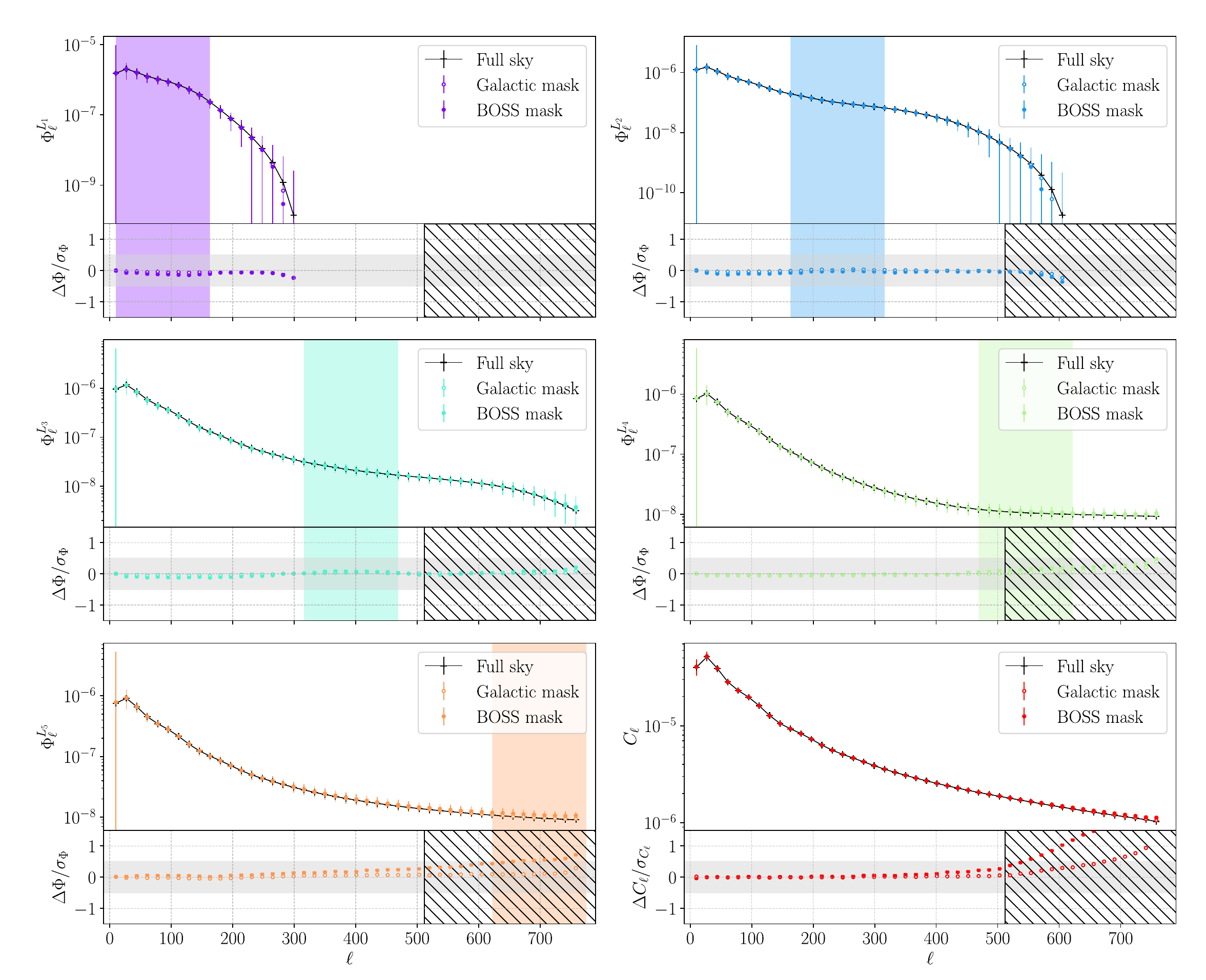}
      \caption{As Fig. \ref{fig:lpt3d_fsb_bias} for the 2D LPT simulations, showing results for the Galactic and BOSS masks. The estimator bias remains within the $[-0.5\sigma; 0.5\sigma]$ range in all filters for $\ell \leq 2N_\text{side}$.}\label{fig:lpt_fsb_bias}
    \end{figure*}

    This initial exploration of the FSB covariance shows that the analytical disconnected covariance, calculated using the methodologies developed in the context of power spectra, is able to capture the most relevant contributions to the true FSB covariance. In order to quantify this more exhaustively, we require a larger number of simulations, which motivates the validation against 2D LPT realisations described in the next section.

  \subsection{Validation on 2D LPT simulations}\label{ssec:res.2dlpt}

    \begin{figure*}
      \centering
      \includegraphics[width=\textwidth]{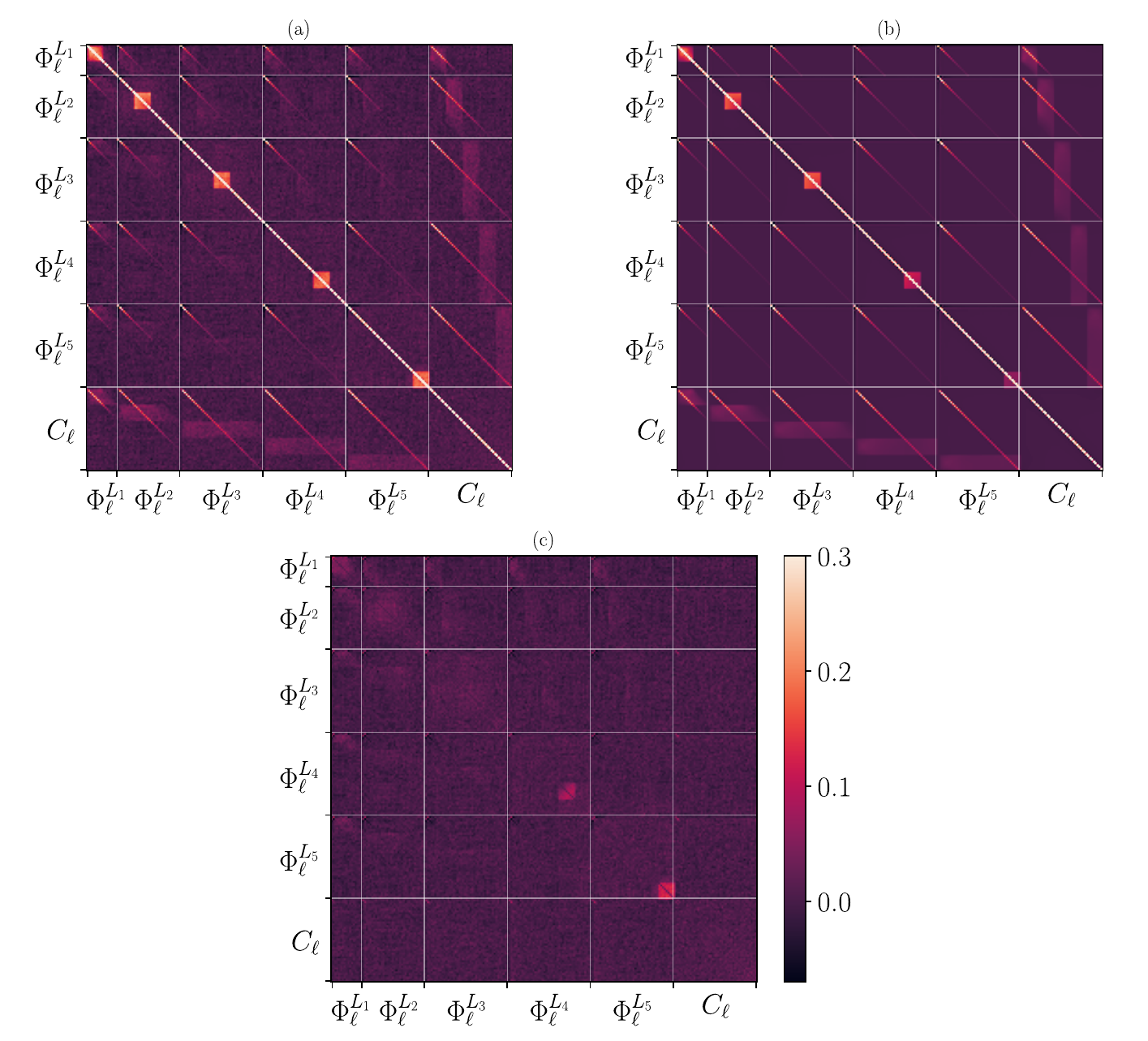}
      \caption{Exploring the non-Gaussian contributions to the covariance: (a) 2D LPT simulation-based correlation matrix with BOSS mask; (b) disconnected correlation matrix computed from the mean FSB (computed over all simulations and using BOSS mask) with additional non-Gaussian terms $N_{222}$ and $N_{32}$; (c) difference between (a) and (b). 
      The colour bar is the same for all three subplots. Each matrix shown is made of $(5+1)\times(5+1)$ blocks; these correspond to the FSB in the five different filters described previously, and to the $C_\ell$s of the original map. The blocks in the first two rows and columns are not the same size as the remaining blocks: we decided to remove data points from the data vector where the FSB signal is identically zero (as explained in Section \ref{sec:meth}). } \label{fig:lpt_cors}
    \end{figure*}

    \begin{figure}
      \centering
      \includegraphics[width=\linewidth]{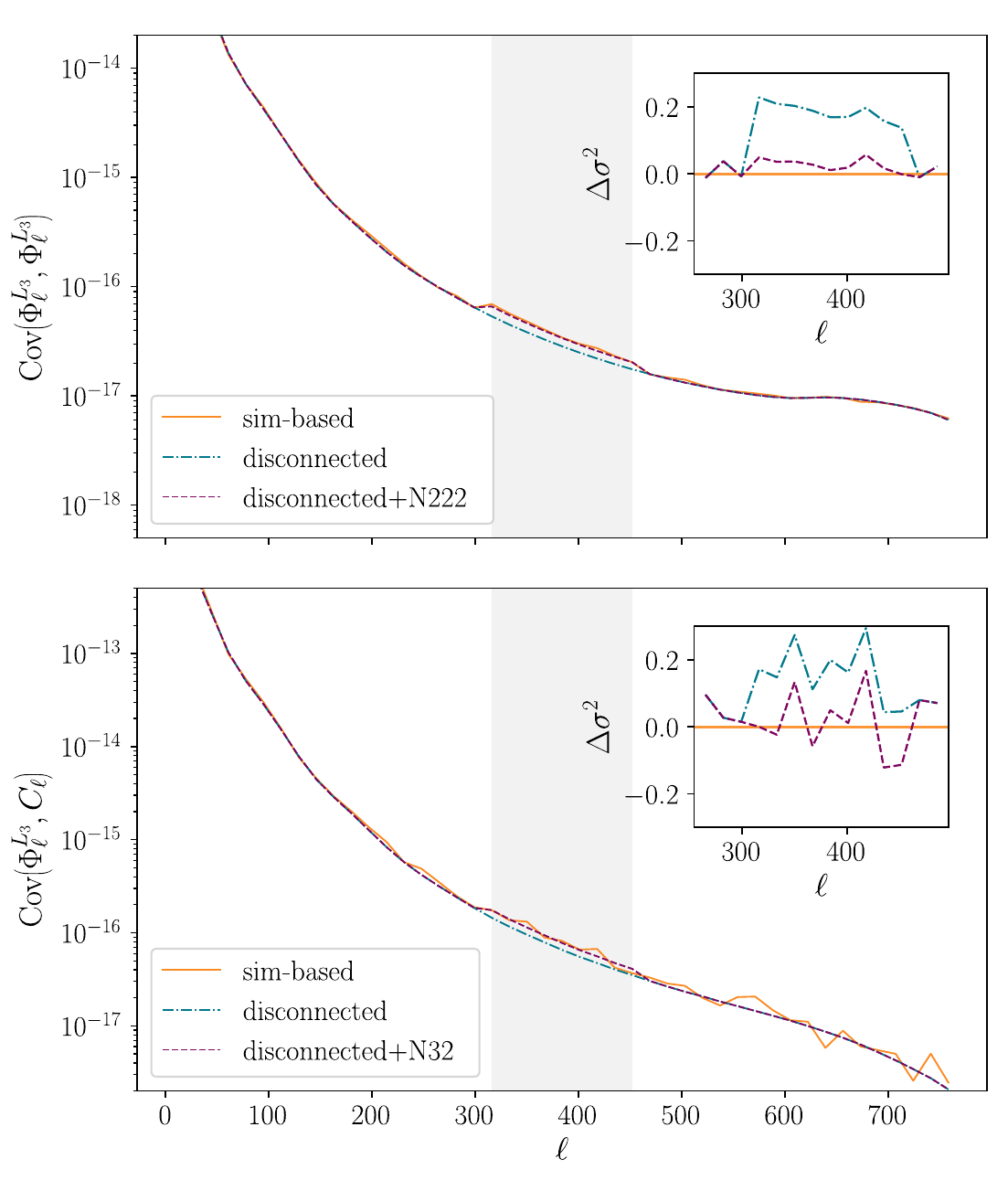} 
      \caption{Corrections for the $N_{222}$ (top) and $N_{32}$ (bottom) terms in the diagonal of two covariance blocks from Fig. \ref{fig:lpt_cors}, computed using the BOSS footprint: auto-covariance $(\Phi_\ell^{L_3}, \Phi_\ell^{L_3})$ and cross-covariance $(\Phi_\ell^{L_3}, C_\ell)$. The greyed out vertical band shows the range of the $L_3$ filter. Insets show the relative change $\Delta \sigma^2 = (\sigma^2_\text{sim-based} - \sigma^2_\text{analytic}) / \sigma^2_\text{sim-based}$ ; the addition of the $N_{222}$ and $N_{32}$ analytical corrections reproduce the full simulation-based covariance more closely in that range.}\label{fig:correction}
    \end{figure}

    \begin{figure}
      \centering
      \includegraphics[width=\linewidth]{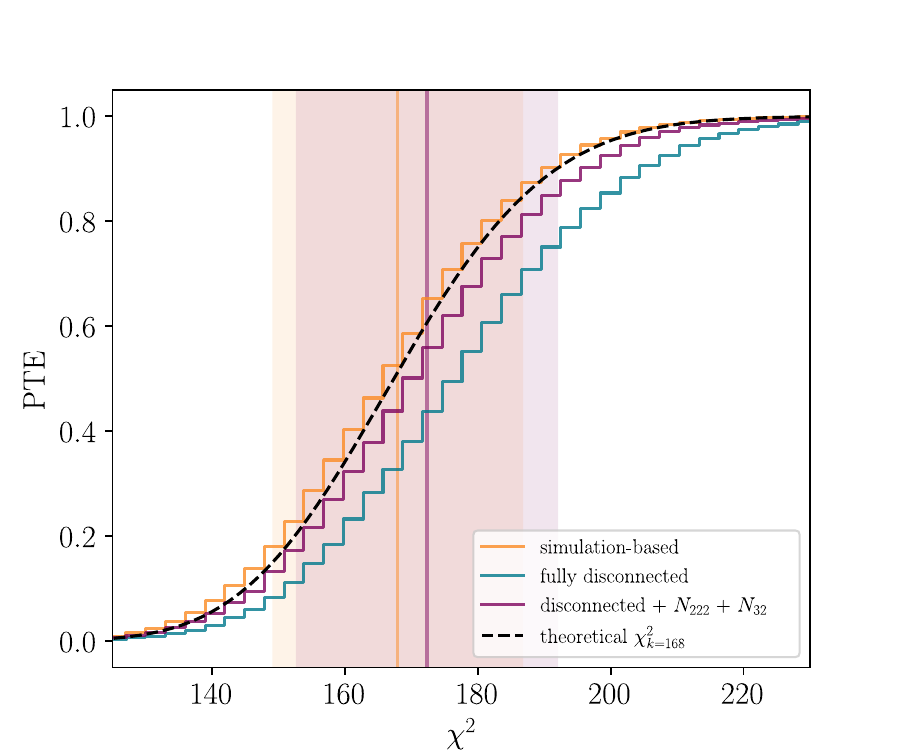} 
      \caption{Probability-to-exceed plot showing the residuals' cumulative $\chi^2$ distribution for each covariance matrix: the simulation-based covariance matrix (orange histogram), the disconnected covariance (turquoise) and the disconnected covariance with $N_{222}$ and $N_{32}$ corrections (burgundy). We also plot the mean and $1\sigma$ interval for each distribution, using the same colour scheme. Note that here we use the full covariance and data vectors, comprising of multipoles below $2 N_\text{side}$ and satisfying the triangle inequality. The black dashed line shows the expected $\chi^2$ distribution for a Gaussian data vector with the same number of degrees of freedom ($N_{\rm dof}=168$).} \label{fig:pte_master}
    \end{figure}

    \begin{figure*}
      \centering
      \includegraphics[width=\textwidth]{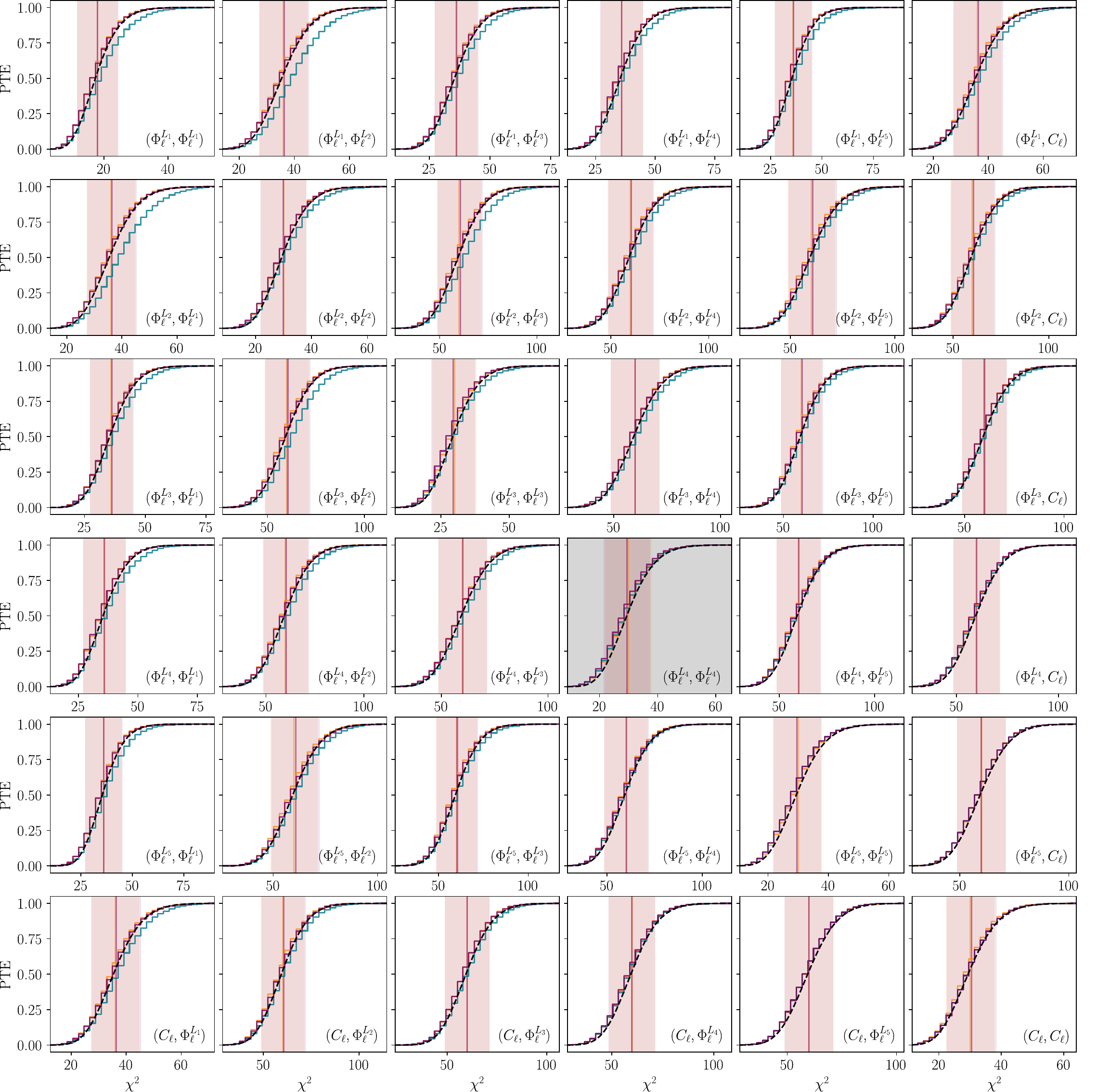} 
      \caption{Probability-to-exceed plots in blocks corresponding to the covariance matrix blocks in Fig. \ref{fig:lpt_cors}. In each block we compare the residuals' cumulative $\chi^2$ distribution for different covariance matrices: the simulation-based covariance matrix (orange histogram), the disconnected covariance with $N_{222}$ and $N_{32}$ corrections (burgundy histogram) and the fully disconnected covariance (turquoise histogram). We also plot the mean and $1\sigma$ interval for the first two distributions, using the same colour scheme. The black dashed line shows the expected chi-square distribution for the relevant number of degree of freedoms which, due to the uneven size of blocks in the covariance matrix, ranges from $N_{\rm dof}=18$ to $N_{\rm dof}=30$ in the diagonal blocks, and from $N_{\rm dof}=36$ to $N_{\rm dof}=60$ in the off-diagonal blocks. Taking account of non-Gaussian contributions narrows the gap between the residuals' distribution obtained with the simulation-based and with the approximated covariance matrix. The colour of the subplot background indicates whether the distributions  of the simulation-based and and analytical covariances pass the KS test -- white blocks for a positive outcome, greyed out blocks for negative.} \label{fig:pte}
    \end{figure*}

    In order to validate the FSB estimator and its analytical covariance to a higher level of accuracy, we repeat our measurements on the 6000 2D LPT simulations using the Galactic and BOSS masks shown in Fig. \ref{fig:masks}, comparing the results against the full-sky measurements in the same simulations. Confirming our previous results, we find that the FSB estimator is unbiased, with all deviations from the full-sky results remaining significantly smaller than $0.5\sigma$ on scales $\ell<2N_{\rm side}$. This is shown in Fig. \ref{fig:lpt_fsb_bias}.
    
    The main advantage of the significantly larger number of simulations available in this case, is our ability to explore the full FSB covariance matrix (without restricting ourselves to individual block diagonal elements).  Fig. \ref{fig:lpt_cors} shows the correlation matrix $r_{ij}$ of our full data vector. The correlation coefficient is defined as
    \begin{equation} \label{eq:cormatrix}
      r_{ij} = \frac{\mathsf C_{ij}}{\sqrt{\mathsf C_{ii} \mathsf C_{jj}}} \; ,
    \end{equation}
    where ${\sf C}_{ij}$ is the covariance between elements $i$ and $j$ of the data vector. Results are shown for the covariance matrix estimated from all 6000 simulations (subpanel (a)), for the theoretical covariance described in Section \ref{sssec:meth.fsb.cov}, containing the disconnected terms and the leading-order $N_{222}$ and $N_{32}$ terms (subpanel (b)), and the difference between both (subpanel (c)). In this figure, the data vector contains the FSB measurements in the 5 different filters first, followed by the angular power spectrum. We only show the combinations where the FSB is not zero by construction (i.e. those satisfying the triangular inequality, as described above), and hence the sectors corresponding to the first two filters are noticeably shorter than the last three.

    Focusing first on subpanel (a) of this figure, we see that the correlation matrix displays an interesting structure:
    \begin{itemize}
      \item Each covariance block, of any type (i.e. $(\Phi^{L_i}_\ell,\Phi^{L_j}_{\ell'})$, $(\Phi^{L_i}_\ell,C_{\ell'})$, or $(C_\ell,C_{\ell'})$), is dominated by a diagonal component $(\ell=\ell')$. As we have seen in the previous section, this component is well captured by the analytical disconnected covariance.
      \item In sub-blocks involving two FSBs with the same filter $L_i$, we see an additional contribution, relevant at the level of $r_{ij}\sim0.3$, that couples all pairs $(\ell,\ell')$ corresponding to the same filter $L_i$. This contribution is well described by the $N_{222}$ term, as can be clearly seen in the theoretical prediction shown in subpanel (b).
      \item In blocks of the form $(\Phi^{L_i}_\ell,C_{\ell'})$, we observe an additional contribution in the form of a band covering power spectrum scales $\ell'\in L_i$. This contribution is significantly smaller than the disconnected and $N_{222}$ terms, but still clearly noticeable in the figure. It is well described by the $N_{32}$ term, which is also included in the theoretical prediction of subpanel (b).
    \end{itemize}
    The two connected terms ($N_{222}$ and $N_{32}$) were computed analytically, using as inputs the mean power spectrum and mean generalised FSB $\Phi_{\ell'}^{L\ell}$ measured in the simulations. Note that the generalized FSB $\Phi^{L\ell}_{\ell'}$ arising in the $N_{32}$ term (Eq. \ref{eq:covn32_1}) is defined at each multipole $\ell'$ for the 5 filters $L$ and for single-multipole filters $\ell$. In practice, we bin FSBs and power spectra into bandpowers, and so we also substitute these narrow filters $\ell$ by the same bandpowers used in our binning scheme for $\ell'$, giving us a total of 225 $\Phi_{\ell'}^{L_i\ell_j}$ per simulation. Because of this larger number of configurations, estimating the generalised FSB is a significantly costlier operation (although by no means unmanageable). It is also worth noting that the analytical expression for the $N_{222}$ term (Eq. \ref{eq:covn222_1}) is very similar to the expression for the mode coupling matrix \citep[MCM; Eq. 14 in ][]{alonsoUnifiedPseudoC_2019}, so we take advantage of the fast implementation of this type of operations in \nmt to compute this term.
    
    When the $N_{222}$ and $N_{32}$ terms are added to the disconnected covariance, estimated from the measured power spectra following the methodology described in \cite{garcia-garciaDisconnectedPseudoC_2019} and \cite{Nicola2020}, and Section \ref{sssec:meth.cl.cov}, we obtain the analytical covariance shown in subpanel (b) of the figure. As subpanel (c) shows, our analytical prediction does an excellent job at recovering the true FSB-$C_\ell$ covariance. The only visible residuals correspond to scales $\ell \geq 2N_{\rm side}$, where the inaccuracy of the HEALPix SHTs prevents us from calculating the inputs to the different terms in the analytical covariance with sufficient precision.
    
    To better visualise the impact of the connected terms, Fig. \ref{fig:correction} shows the diagonal of two covariance matrix blocks taken from Fig. \ref{fig:lpt_cors}: the $(\Phi^{L_3}_\ell,\Phi^{L_3}_{\ell})$ and  $(\Phi^{L_3}_\ell,C_{\ell})$ blocks. Results are shown for the true, simulation-based covariance (orange), the disconnected-only analytical covariance (turquoise), and for the full analytical covariance, containing the $N_{222}$ and $N_{32}$ terms (burgundy). The inset in the figure shows the relative difference between the two analytical predictions and the simulation-based estimate. We see that the disconnected-only estimate under-predicts the variance by $\sim20\%$ on the filtered scales, and that this difference is made up for by the $N_{222}$ and $N_{32}$ terms at high accuracy.

    As a final quantitative test of the accuracy with which we are able to estimate the FSB-$C_\ell$ covariance, we examine the distribution of $\chi^2$ values obtained from different covariance matrices. For a given choice of covariance matrix ${\sf C}$, we compute, for each simulation, the quantity
    \begin{equation}
      \chi^2 \equiv (\mathbf d - \bar{\mathbf d})^\mathsf{T} {\sf C}^{-1} (\mathbf d - \bar{\mathbf d}),
    \end{equation}
    where $\mathbf d$ is the measured data vector, and $\bar{\mathbf d}$ is its average over all simulations. We do this for three different choices of ${\sf C}$: the simulation-based covariance, the disconnected covariance, and the disconnected covariance corrected for the $N_{222}$ and $N_{32}$ terms. If the data vector was normally distributed, and given the correct covariance, the $\chi^2$ quantity above should follow a $\chi^2$ distribution with a number of degrees of freedom equal to the size of ${\bf d}$. Ultimately, in order to avoid relying on the assumption that ${\bf d}$ be Gaussianly distributed, we will instead compare the distribution of the $\chi^2$ values for the two analytical covariances with that obtained from the simulation-based one (which is the manifestly correct covariance). To remain consistent with the rest of the analysis, we do not use the full covariance matrix in the computation, but instead remove $\ell > 2 N_\text{side}$ (this corresponds to the $30^\text{th}$ bandpower) and the range in the first filter where the signal is zero by the triangular inequality (corresponding to the $18^\text{th}$ bandpower). 

    The logical choice for ${\bf d}$ is to use the full data vector to assess how well the different covariance approximations replicate the simulation-based one overall. The resulting cumulative $\chi^2$ distributions are shown in Fig. \ref{fig:pte_master}. As shown in the figure, the distribution of $\chi^2$ values estimated from the simulation-based covariance (orange steps) does follow closely the expected $\chi^2$ distribution (dashed black line) for a purely Gaussian data vector. The turquoise steps, showing the distribution of $\chi^2$ estimated using a covariance matrix containing only the purely disconnected contributions, follow a similar trend, albeit displaying significant differences with respect to the true distribution. As shown by the burgundy steps, adding the $N_{222}$ and $N_{32}$ contributions narrows this gap significantly. The difference in the mean of the orange and burgundy $\chi^2$ distributions is less than $2.5\%$, corresponding to less than $\sim20\%$ of their standard deviation. This level of accuracy is enough to provide a fair assessment of the goodness of fit of any given model, and would lead to negligible biases on any inferred model parameters.
    
    It is also instructive to examine the validity of the analytical covariance matrix estimator used here across different parts of the data vector. Specifically, in order to explore different blocks of the full covariance matrix individually, the data vector ${\bf d}$ will now only contain a single FSB (i.e. a single filter) or $C_\ell$ measurement when exploring the diagonal blocks of the covariance, or an FSB--FSB (with distinct filters) or FSB--$C_\ell$ pair when exploring the off-diagonal blocks.
    
    Fig. \ref{fig:pte} shows the cumulative distributions for the simulation-based covariance (orange), the analytical covariance including disconnected, $N_{222}$ and $N_{32}$ terms (burgundy), and the purely disconnected prediction (turquoise). The vertical bands, following the same colour pattern, show the 68\% interval around the mean of the first two distributions. Overall, we see a significant qualitative improvement when including the $N_{222}$ and $N_{32}$ corrections. We evidently do not see an improvement over the purely disconnected approach in the block corresponding to the last filter, where the $2N_\text{side}$ cut-off removed the extra non-Gaussian contribution, effectively making both covariance approximations identical. To quantify this qualitative agreement, we also compute the Kolmogorov-Smirnov (KS) statistic, which estimates whether two PDFs can belong to the same underlying distribution \citep{KStestDarling1957}. We do so using a threshold which correspond to a $p$-value of $p = 0.001$. The blocks for which this test fails have their backgrounds shaded in gray. We see that the analytical approximation passes this test in almost all cases, with a single failure for the $(\Phi_\ell^{L_4}, \Phi_\ell^{L_4})$ auto-covariance -- where the filter includes angular scales in the range $\ell \geq 2N_\text{side}$ and numerical inaccuracies in the spherical harmonic transforms can be significant. Note that, in spite of this failure, the mean and standard deviation of the distribution recovered by the analytical covariance are very close to the simulation-based one, and both distributions are very similar. Thus, any $\chi^2$ value estimated with the analytical covariance would be very unlikely to lead to an incorrect assessment regarding the goodness of fit of a given theoretical prediction for the FSB. It is also worth pointing out that the small differences that can be identified visually between the analytical and simulation-based distributions in the blocks involving the FSB are similar to those we can observe in the power spectrum-only block (bottom right panel), for which the analytical approximations used here are routinely used in the scientific analysis of current large-scale structure data \citep{mohammedPerturbativeApproachCovariance2017, barreiraAccurateCosmicShear2018, garcia-garciaDisconnectedPseudoC_2019, liDisconnectedCovariance2point2019}.
    
    Thus, we conclude that the analytical approximations presented here to estimate the bispectrum covariance within the FSB picture, enabled by the infrastructure developed previously for the computation of power spectrum covariances, are accurate enough to be employed in the analysis bispectra from projected large-scale structure data, removing one of the main roadblocks in the efficient analysis of third-order statistics for cosmology.

\section{Conclusion}\label{sec:conclusion}
  This paper presents the ``filtered-squared bispectrum'' (FSB), an estimator of the binned bispectrum that consists of computing the power spectrum between the target field and the square of the same field filtered over a range of scales. Since it is phrased in the form of a power spectrum, we are able to make use of the significant infrastructure built by the community for the estimation of power spectra and their covariance matrices. The main results presented here can be summarised as follows:
  \begin{itemize}
    \item The use of pseudo-$C_\ell$ techniques allows us to take into account the most relevant effects caused by a sky mask, and the resulting estimator is largely unbiased. We have suggested techniques to mitigate residual survey geometry biases should they become relevant in specific applications not considered here.
    \item The disconnected power spectrum covariance, when applied to the FSB, includes both purely Gaussian as well as non-Gaussian contributions. What is more, it contains {\sl all} Gaussian and non-Gaussian components that contribute only to the diagonal uncertainties. Using the methods developed in the past to estimate disconnected power spectrum covariances, we are thus able to compute highly-accurate FSB-FSB and FSB-$C_\ell$ covariances in a completely data-driven way, without any need for large simulation suites.
    \item We have derived expressions for the leading-order connected contributions to the covariance, and shown that they can also be estimated accurately in a data-driven manner. The resulting covariances achieve the same level of accuracy as those routinely used in state-of-the-art power spectrum-based cosmological analyses.
  \end{itemize}
  We expect the FSB estimator to be useful in the cosmological analysis of existing and future projected LSS probes. A {\tt python} package implementing the FSB estimator (including its covariance matrix), making use of the \nmt module, is made publicly available\footnote{\url{https://github.com/ashaeres/pyfsb.git}}.

  As presented here, the FSB is limited in several ways. First, by construction the FSB is able to recover bispectrum amplitudes for triangular configurations that are close to isosceles. Although, as we have argued, this can include a large fraction of all available triangles, the FSB could be easily generalised to a completely lossless ``filtered-multiply'' approach, in which the squaring step is simply replaced by the multiplication of two filtered fields with different filters. Secondly, to simplify the notation we have considered only power spectra and bispectra of a single field. The multi-field version of the FSB estimator would be a straightforward generalisation, however\footnote{The only potential difference might be in the ``multiplication'' step, where the two fields multiplied may have to first be restricted to the same spatial footprint. This should not incur a significant loss of information.}. Thirdly, the paper has only explored the case of scalar fields defined on relatively compact sky footprints. An important probe that strongly deviates from these two assumptions is cosmic shear. The spin-2 nature of cosmic shear should be relatively straight-forward to accommodate. However, the shear is measured only at the positions of observed sources that are subject to stringent selection criteria that aggressively mask the footprint in the presence of diffraction spikes and other observational vagaries. These effects lead to a mask that is considerably more inhomogeneous, structured and irregular on small scales compared to the masks considered in this work, leading to an increased level of mode coupling across scales. The validity of the FSB estimator as presented here is far from evident. Given the importance of cosmic shear as a powerful probe to access the matter bispectrum largely without bias, this application should be studied in detail. Recent progress in the area of power spectrum estimation for discretely sampled fields may prove useful in this direction \citep{2312.12285,2407.21013}. In the context of cosmic shear, it will also be important to quantify and potentially consider additional contributions to the covariance matrix from super-sample modes \citep{1302.6994}. Finally, as mentioned in Section \ref{sssec:meth.fsb.other}, in its current incarnation the FSB estimator does not contain the linear term included by optimal bispectrum estimators. Although the estimator is able to achieve close to optimal uncertainties on most configurations for LSS-like signals, the potential of this linear term to reduce the variance of specific configurations (e.g. squeezed-limit) should be explored in detail, as it would be important, for example, in searches for primordial non-Gaussianity.

  In summary, although this paper has focused on the validation of the FSB, there are multiple avenues for further research that can be explored in future work. We hope the FSB will facilitate the exploitation of third-order statistics in current and upcoming experiments in a practical and efficient manner.

\section*{Acknowledgements}
  We thank Matteo Biagetti, Lina Castiblanco, Jacopo Salvalaggio, Gerrit Farren, Robert Reischke, Emiliano Sefusatti, and Blake Sherwin for useful discussions. LH is supported by a Hintze studentship, which is funded through the Hintze Family Charitable Foundation.  JAC is funded by a Kavli/IPMU PhD Studentship. JE acknowledges financial support from the Spanish MICINN funding grant PGC2018-101931-B-I00. The GLAM catalogues are publicly available at \href{https://skun.iaa.csic.es/SUsimulations/UchuuDR2/Uchuu_GLAM_BOSS-eBOSS/GLAM-Uchuu/}{Skies and Universe}. DA acknowledges support from STFC and the Beecroft Trust. We made extensive use of computational resources at the University of Oxford Department of Physics, funded by the John Fell Oxford University Press Research Fund.

\bibliography{main}

\appendix

\section{Linear term}

\label{app:linterm}

\begin{figure}
  \centering
  \includegraphics[width=0.8\linewidth]{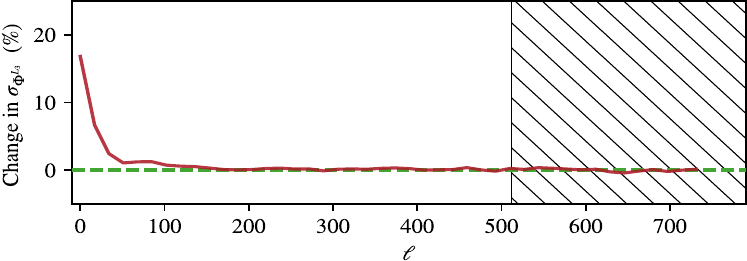}
  \caption{Difference between the original estimator, and the version in which we subtract the linear term. The plot shows the relative change $(\sigma_\Phi - \sigma_\Phi^\text{linear}) / \sigma_\Phi \times 100$ for $\Phi_\ell^{L_3}$.}
  \label{fig:linterm}
\end{figure}

As discussed in Section \ref{sssec:meth.fsb.other}, optimal cubic estimators for the bispectrum include a linear term which effectively lowers the covariance of the estimator in the presence of a mask. In the case of the FSB, the linear term introduced in \cite{1509.08107} is given by Eq. \ref{eq:lintermfsb}: 
\begin{equation} \nonumber
    \hat{\Phi}^L_{b,{\rm linear}}= \left\langle 2\hat{\Phi}^{L,(aGG)}_b+\hat{\Phi}^{L,(GGa)}_b\right\rangle_G ,
\end{equation}
where the filtering step applies to the first two fields in brackets. The average is taken over Gaussian simulations, which can be generated in two ways: 
\begin{enumerate}
    \item by generating a Gaussian map at the pixel-level, where each pixel follows a normal distribution with variance equal to that of field $a$; 
    \item by generating a Gaussian realisation following the power spectrum of field $a$.
\end{enumerate}
We selected $400$ 2D-LPT simulations on which we applied the BOSS mask, and for each we generated $200$ Gaussian (masked) simulations. For each simulation we then computed $\hat \Phi_{b, \text{linear} }^L$ for each FSB, and subtracted them from the data vector. From this we then computed the corresponding covariance matrix over all $400$ simulations. We only show results for the second approach to generating Gaussian fields (i.e. following the same power spectrum as the data), as the pixel-based approach yielded unreliable results: while it reduced the variance at the largest scales, it increased it on the smaller ones. In Fig. \ref{fig:linterm}, we show the change introduced by subtracting the linear term on the variance of $\Phi^{L_3}_\ell$; the impact of the linear term is of similar or lower significance for the remaining FSBs, and is also observed in the diagonal entries of the off-diagonal (cross-FSB) blocks of the covariance. The linear term noticeably reduces the variance by up to $16\%$ around the first multipoles. For $\ell > 50$ the linear term does not provide significant improvement to the FSB variance. The linear term does not change the correlation structure significantly outside of the diagonal $(b=b')$ entries in each block.

We thus see that the linear term leads to a modest, but non-zero improvement in the statistical uncertainties of the FSB, particularly on the largest multipoles. Including it could therefore be relevant when employing the FSB to search for primordial non-Gaussianity, or any signal peaking on squeezed triangle configurations.

\section{Gaussian power spectrum covariances and bispectra}\label{app:gauscov}

  Let us now consider the different contributions to the FSB covariance matrix. Because working with fully connected correlation functions in harmonic space can become quite cumbersome for orders higher than 3, we will first compute these contributions using the flat-sky approximation. More precisely, we will assume a Euclidean space in $N$ dimensions.

  \subsection{Preliminaries}\label{ssec:gauscov.prelim}
    We start by introducing some notation that describes the tools used in the upcoming derivations:
  
    \begin{itemize}
      \item Fourier transforms and their inverses are:
      \begin{equation}
        f_{\bf k}\equiv\int d^N x\,e^{-i{\bf k}\cdot{\bf x}}\,f({\bf x}) \, \hspace{12pt}
        f({\bf x})=\int \dfk{k}\,e^{i{\bf k}\cdot{\bf x}}\,f_{\bf k} \,
      \end{equation}
      where we use the shorthand:
      \begin{equation}
        \dfk{k}\equiv\frac{d^N k}{(2\pi)^N}.
      \end{equation}
      \item The power spectrum, bispectrum, and trispectrum are defined via:
      \begin{align}
        &\langle \delta_{\bf k}\delta_{\bf q}\rangle = (2\pi)^N\delta^D({\bf k}+{\bf q})\,P_k,\\
        &\langle \delta_{\bf k}\delta_{\bf q}\delta_{\bf p}\rangle = (2\pi)^N\delta^D({\bf k}+{\bf q}+{\bf p})\,B({\bf k}, {\bf q}, -{\bf k}-{\bf q}),\\
        &\langle \delta_{\bf k}\delta_{\bf q}\delta_{{\bf k}'}\delta_{{\bf q}'}\rangle = (2\pi)^N\delta^D({\bf k}+{\bf q}+{\bf k}'+{\bf q}')\,T({\bf k}, {\bf q}, {\bf k}', {\bf q}'=-{\bf k}-{\bf q}-{\bf k}'),
      \end{align}
      where $\delta^D({\bf x})$ is the Dirac delta function.
      \item The ``filtered-squared'' map $\delta_L^2\left({\bf x} \right)$ and its Fourier transform $\left(\delta_L^2\right)_{\bf k}$ are related to the overdensity field itself via:
      \begin{align}
        &\delta_L^2({\bf x})=\int {\cal D}^N k\,{\cal D}^N k' W_k^L\,W_{k'}^L\,\delta_{\bf k}\delta_{{\bf k}'}e^{i({\bf k}+{\bf k}')\cdot{\bf x}},\\
        &\left(\delta_L^2\right)_{\bf k}=\int {\cal D}^N q\,W_q^LW_{|{\bf k}-{\bf q}|}^L\delta_{\bf q}\delta_{{\bf k}-{\bf q}},
      \end{align}
      where $W_k^L$ is a filter function (which typically will pick up only values of $k\in L$). To obtain the last equation we have used the identity
      \begin{equation}
        \int d^N xe^{i{\bf k}\cdot{\bf x}}=(2\pi)^N \delta^D({\bf k})
      \end{equation}
      (which will be used repeatedly in what follows).
      \item The discrete version of the Dirac delta function in Fourier space is:
      \begin{equation}
        (2\pi)^N\delta^D({\bf k}-{\bf k}') \leftrightarrow V\,\delta^K_{{\bf k},{\bf k}'}, \label{eq:discretedelta}
      \end{equation}
      where $\delta^K$ is the Kronecker delta, and $V$ is the volume of the region covered.
   
      With this notation, our estimator for the FSB is:
      \begin{equation}\label{eq:fsbest}
        \hat{\Phi}^L_k \equiv V^{-1}V_k^{-1}\int_{V_k} d^N k\,\delta_{\bf k}^* \left(\delta_L^2\right)_{\bf k}=V^{-1}V_k^{-1}\int_{V_k} d^N k\,\int {\cal D}^N q\,W_q^LW_{|{\bf k}-{\bf q}|}^L\delta_{-{\bf k}}\delta_{\bf q}\delta_{{\bf k}-{\bf q}},
      \end{equation}
      where $V_k$ is the Fourier-space volume of the ``ring'' of equivalent $k$s we are averaging over:
      \begin{equation}
        V_k\equiv \int^{V_k} d^N k.
      \end{equation}
      For example, in 2D, for a ring of radius $k$ and infinitesimal width $\Delta k$, $V_k=2\pi k\,\Delta k$. Taking the expectation value of the estimator above, the relation between the FSB and the bispectrum is
      \begin{equation}
        \Phi^L_k \equiv\langle \hat{\Phi}^L_k \rangle=\int \dfk{q}\,W^L_q\,W^L_{|{\bf k}-{\bf q}|}\,B(-{\bf k},{\bf q},{\bf k}-{\bf q}).
      \end{equation}
    \end{itemize}

  \subsection{General covariance expression}\label{ssec:gauscov.gen}
    The FSB covariance is:
    \begin{align}
      {\rm Cov}\left(\hat{\Phi}^L_k,\hat{\Phi}^{L'}_{k'}\right)
      & \equiv \left\langle\hat{\Phi}^{L}_{k} \hat{\Phi}^{L'}_{k'} \right\rangle-\left\langle\hat{\Phi}^{L}_{k} \right\rangle \left\langle\hat{\Phi}^{L'}_{k'} \right\rangle\\
      &=V^{-2}V_k^{-2}\int_{V_k} d^N k\int_{V_k}d^N k' \int\dfk{q} \int\dfk{q'}W^L_q W^L_{|{\bf k}-{\bf q}|}W^{L'}_{q'}W^{L'}_{|{\bf k}'-{\bf q}'|}\\
      &\hspace{40pt} \left(\left\langle \delta_{-{\bf k}} \delta_{\bf q} \delta_{{\bf k}-{\bf q}} \delta_{-{\bf k}'} \delta_{{\bf q}'} \delta_{{\bf k}'-{\bf q}'}\right\rangle-\left\langle\delta_{-{\bf k}}\delta_{\bf q}\delta_{{\bf k}-{\bf q}}\right\rangle \left\langle\delta_{-{\bf k}'}\delta_{{\bf q}'}\delta_{{\bf k}'-{\bf q}'}\right\rangle\right)\\\label{eq:cov_fsb_grl}
      &={\cal I}\left[\left\langle\delta_{-{\bf k}}\delta_{\bf q}\delta_{{\bf k}-{\bf q}}\delta_{-{\bf k}'}\delta_{{\bf q}'}\delta_{{\bf k}'-{\bf q}'}\right\rangle\right]-\Phi^{L}_{k}\Phi^{L'}_{k'},
    \end{align}
    where, for short, we have defined the integral operator
    \begin{equation}
      {\cal I}[f]\equiv V^{-2}V_k^{-2}\int_{V_k} d^Nk\int_{V_k}d^Nk' \int\dfk{q} \int\dfk{q'}\,W^L_qW^L_{|{\bf k}-{\bf q}|}W^{L'}_{q'}W^{L'}_{|{\bf k}'-{\bf q}'|}\,f({\bf k},{\bf q},{\bf k}',{\bf q}').
    \end{equation}
    The rest of the calculation is then based on, first, splitting the six-point function above into its different disconnected and fully connected terms, and then simplifying the action of ${\cal I}$ on each of them.

    A 6-point function $\langle abcdef \rangle$ can be split into 4 different contributions:
    \begin{itemize}
      \item $``2+2+2"$ terms of the form $\langle ab\rangle\langle cd\rangle\langle ef\rangle$. In our case, these are the only terms that contribute if the field $\delta_{\bf k}$ is Gaussian. There are 15 different possible terms:
      \begin{align}\nonumber
        \avg{abcdef}_{222}
        =&
        \avg{ab}\avg{cd}\avg{ef}+\avg{ab}\avg{ce}\avg{df}+\avg{ab}\avg{cf}\avg{de}+\\\nonumber
        &\avg{ac}\avg{bd}\avg{ef}+\avg{ac}\avg{be}\avg{df}+\avg{ac}\avg{bf}\avg{de}+\\\nonumber
        &\avg{ad}\avg{bc}\avg{ef}+\avg{ad}\avg{be}\avg{cf}+\avg{ad}\avg{bf}\avg{ce}+\\\nonumber
        &\avg{ae}\avg{cd}\avg{bf}+\avg{ae}\avg{bc}\avg{df}+\avg{ae}\avg{cf}\avg{bd}+\\\label{eq:222all}
        &\avg{af}\avg{cd}\avg{eb}+\avg{af}\avg{ce}\avg{db}+\avg{af}\avg{bc}\avg{de},
      \end{align}
      where terms 4--15 are just copies of terms 1--3 swapping field $b$ by any of $\{c,d,e,f\}$.
      \item $``3+3"$ terms of the form $\langle abc\rangle\langle def\rangle$. There are 10 different terms:
      \begin{align}\nonumber
        \avg{abcdef}_{33}=&
        \avg{abc}\avg{def}+\avg{abd}\avg{cef}+\avg{abe}\avg{cdf}+\avg{abf}\avg{cde}+\\\nonumber
        &\avg{acd}\avg{bef}+\avg{ace}\avg{bdf}+\avg{acf}\avg{bde}+\\\nonumber
        &\avg{ade}\avg{bcf}+\avg{adf}\avg{bce}+\\\label{eq:33all}
        &\avg{aef}\avg{bcd}.
      \end{align}
      \item $``4+2"$ terms of the form $\avg{ab}\avg{cdef}_c$, where $\avg{cdef}_c$ is the connected 4-point function (i.e. trispectrum if all fields are in Fourier space). Again, there are 15 different terms, corresponding to the 15 different pairs of fields that we can put in the ``2'' term:
      \begin{align}\nonumber
        \avg{abcdef}_{42}=&
        \avg{ab}\avg{cdef}+\avg{ac}\avg{bdef}+\avg{ad}\avg{bcef}+\avg{ae}\avg{bcdf}+\avg{af}\avg{bcde}+\\\nonumber
        &\avg{bc}\avg{adef}+\avg{bd}\avg{acef}+\avg{be}\avg{acdf}+\avg{bf}\avg{acde}+\\\nonumber
        &\avg{cd}\avg{abef}+\avg{ce}\avg{abdf}+\avg{cf}\avg{abde}+\\\nonumber
        &\avg{de}\avg{abcf}+\avg{df}\avg{abce}+\\\label{eq:42all}
        &\avg{ef}\avg{abcd}.
      \end{align}
      \item The fully connected 6-point function $\avg{abcdef}_c$. We will use the following notation:
      \begin{equation}
        \langle \delta_{{\bf k}_1}\delta_{{\bf k}_2}\delta_{{\bf k}_3}\delta_{{\bf k}_4}\delta_{{\bf k}_5}\delta_{{\bf k}_6}\rangle=(2\pi)^N\,\delta^D\left(\sum_i{\bf k}_i\right)\,{\cal P}_{(6)}({\bf k}_i).
      \end{equation}
    \end{itemize}

    In the literature, these terms are usually designated by the initials of the corresponding correlation functions (Power spectrum, Bispectrum and Trispectrum): the $``2+2+2"$ term corresponds to the PPP term, $``3+3"$ to the BB term, and $``4+2"$ to the TP term \citep{barreiraSqueezedMatterBispectrum2019, biagettiCovarianceSqueezedBispectrum2022, salvalaggioBispectrumNonGaussianCovariance2024a}. The following subsections work out each of these terms, and simplify them by making use of symmetries.

    \subsubsection{The ``$2+2+2$'' terms}\label{sssec:gauscov.gen.222}
      We start with the purely Gaussian terms, those which only contain 2-point correlation functions -- the only terms that survive if we assume $\delta_{\bf k}$ to be a Gaussian field. Using the notation above, we must remember that $(a,b,c)$ correspond to fields that contribute to the first FSB (i.e. $(a=\delta_{-{\bf k}},\,b=\delta_{{\bf q}},\,c=\delta_{{\bf k}-{\bf q}})$). As such, their wavenumbers add up to zero. This means that any term that involves two of these fields in the same ensemble average cancels out automatically, since it involves evaluating the other field in the trio at a zero wavenumber. For example, the first term in Eq. \ref{eq:222all} involves $\avg{ab}\equiv\avg{\delta_{-{\bf k}}\delta_{\bf q}}\propto\delta^D({\bf k}-{\bf q})$, and therefore will only receive contributions for which field $c$ is $\delta_{{\bf k}-{\bf q}\rightarrow0}$; correlating this term with any other remaining field will give a vanishing term. With this consideration, we can thus cancel out terms 1-6, 7, 11, and 15 in Eq. \ref{eq:222all}, leaving us with 6 remaining terms (8, 9, 10, 12, 13, 14). We can in fact further simplify these expressions, by considering symmetries in the indices we use for each field. Our FSB estimator (Eq. \ref{eq:fsbest}) contains a dummy wavenumber ${\bf q}$ which we integrate over. Therefore, a change of variables on that integral of the form ${\bf q}\rightarrow{\bf k}-{\bf q}$ will leave the result unchanged. Consequently, all terms in the covariance should be symmetric under the exchange of fields $b\leftrightarrow c$ and $e\leftrightarrow f$. It is then straightforward to split these 6 remaining terms into two groups: terms 8 and 9 are equivalent, and terms 10, 12, 13 14 are also identical. We can write the $``2+2+2"$ contribution as: 
      \begin{align}\nonumber
        \avg{abcdef}_{222} = & \quad 2 \avg{ad}\avg{be}\avg{cf} \quad + \quad 4 \avg{ae}\avg{cd}\avg{bf}.
      \end{align}

      Let us now focus on the first term of this new expression:
      \begin{align}
        {\cal I}[\avg{ad}\avg{be}\avg{cf}]={\cal I}\left[(2\pi)^{3N}\delta^D({\bf k}+{\bf k}')\delta^D({\bf q}+{\bf q}')\delta^D({\bf k}+{\bf k}'-{\bf q}-{\bf q}')P_kP_qP_{|{\bf k}-{\bf q}|}\right].
      \end{align}
      Because of the first two delta functions, the third delta function is just $\delta^D(0)$, which reduces to $V/(2\pi)^N$ in Eq. \ref{eq:discretedelta}. A term like this will always appear in all cases, and can be used to cancel out one of the volume factors implicit in ${\cal I}$. Then, we will use one of the remaining delta functions to cancel out one of the integrals over either ${\bf q}$ or ${\bf q}'$. After doing this, and expanding ${\cal I}$, the result is:
      \begin{align}
        {\cal I}[\avg{ad}\avg{be}\avg{cf}]
        &=(2\pi)^N V^{-1}V_k^{-2} \int_{V_k}d^N k\, d^N k'\,\delta^D({\bf k}+{\bf k}') \, P_k \int \dfk{q}\,W_q^LW_q^{L'}W^L_{|{\bf k}-{\bf q}|}W^{L'}_{|{\bf k}-{\bf q}|}P_qP_{|{\bf k}-{\bf q}|}.
      \end{align}
      The remaining Dirac delta function can be used to cancel out the integral over ${\bf k}'$, although only when the moduli of ${\bf k}$ and ${\bf k}'$ are in the same ring, otherwise the whole expression vanishes -- so we will introduce a Kronecker delta $\delta^{K}$ to account for that. The remaining integral over the angles of ${\bf k}$ then simply cancels one of the $V_k^{-1}$ factors, since the integrand should only depend on $k\equiv|{\bf k}|$ by statistical isotropy. The result is thus:
      \begin{equation}
        {\cal I}[\avg{ad}\avg{be}\avg{cf}]
        =(2\pi)^N V^{-1}V_k^{-1} \delta^K_{k,k'}\,P_k\int\dfk{q}\,W_q^LW_q^{L'}W^L_{|{\bf k}-{\bf q}|}W^{L'}_{|{\bf k}-{\bf q}|}P_qP_{|{\bf k}-{\bf q}|}.
      \end{equation}

      We repeat the same process with the second term:
      \begin{align}
        {\cal I}[\avg{ae}\avg{cd}\avg{bf}]={\cal I}\left[(2\pi)^{3N}\delta^D(-{\bf k}+{\bf q}')\delta^D({\bf k}-{\bf q}-{\bf k}')\delta^D({\bf q}+{\bf k}'-{\bf q}')P_kP_qP_{|{\bf k}-{\bf q}|}\right].
      \end{align}
      As before, the last delta function is just $\delta^D(0)$, and we use it to cancel one volume factor. Then, we use the first delta function to cancel the integral over ${\bf q}'$, and the second delta function to cancel the integral over ${\bf q}$. We could not do this for the other terms because one of the delta functions only involved external legs (i.e. ${\bf k}$ and ${\bf k}'$). The result is:
      \begin{align}\nonumber
        {\cal I}[\avg{ae}\avg{cd}\avg{bf}]
        &=V^{-1}V_k^{-2} \int_{V_k} d^N k \int_{V_k} d^N k'W_k^{L'}W_{k'}^LP_kP_{k'}\,W_{|{\bf k}-{\bf k}'|}^LW_{|{\bf k}-{\bf k}'|}^{L'}P_{|{\bf k}-{\bf k}'|}\\
        &=V^{-1}W_k^{L'}W_{k'}^LP_kP_{k'}\,\left\langle W_{|{\bf k}-{\bf k}'|}^LW_{|{\bf k}-{\bf k}'|}^{L'}P_{|{\bf k}-{\bf k}'|}\right\rangle_{\varphi},
      \end{align}
      where $\langle\cdots\rangle_\varphi$ implies averaging over the angle between ${\bf k}$ and ${\bf k}'$,
      \begin{equation}
        \langle f({\bf k},{\bf k}')\rangle_\varphi\equiv \frac{1}{V_k}\int_{V_k}d^N k'\,f({\bf k},{\bf k}'),
      \end{equation}
      and we implicitly assume that $f$ is a statistically isotropic function (i.e. it only depends on scalars such as $k\equiv|{\bf k}|$, $k'\equiv|{\bf k}'|$, and ${\bf k}\cdot{\bf k}'$).

      With this, the contribution to the FSB covariance from all terms of the form $2+2+2$ is:
      \begin{align}\nonumber
        {\rm Cov}_{222}\left( \hat{\Phi}^{L}_{k} , \hat{\Phi}^{L'}_{k'} \right) =
        & \quad 2 (2\pi)^N \delta_{k,k'}^K\,V^{-1}V_k^{-1} \,P_k\int\dfk{q}\,W_q^LW_q^{L'}W^L_{|{\bf k}-{\bf q}|}W^{L'}_{|{\bf k}-{\bf q}|}P_qP_{|{\bf k}-{\bf q}|}\quad +\\
        &\quad 4V^{-1}W_k^{L'}W_{k'}^LP_kP_{k'}\,\left\langle W_{|{\bf k}-{\bf k}'|}^LW_{|{\bf k}-{\bf k}'|}^{L'}P_{|{\bf k}-{\bf k}'|}\right\rangle_{\varphi}.
      \end{align}
      The two terms are qualitatively different. The first term contributes only to the diagonal of the covariance, and drops down like the inverse of the number of modes $V_k^{-1}\propto k^{N-1}$. The second term, on the other hand, contributes to all diagonal and off-diagonal elements, and does not scale as strongly with $k$. This situation is familiar in the context of the covariance matrix of power spectra of non-Gaussian fields. The first term mirrors the properties of the purely Gaussian terms in the power spectrum covariance, while the second term is similar to the ``connected'' non-Gaussian contributions. This makes sense: the FSB is nothing but the power spectrum between two fields, $\delta$ and $\delta_L^2$. Even if $\delta$ was exactly Gaussian, $\delta_L^2$ would not be, and this gives rise to the second term in the expression above.

      It is worth keeping this structure in mind, since it will be reproduced by all the other contributions below.
  
    \subsubsection{The ``$3+3$'' terms}\label{sssec:gauscov.gen.33}
      Let us first determine which terms in Eq. \ref{eq:33all} we need to consider. The first term is cancelled out by the counter-term in the definition of the covariance (Eq. \ref{eq:cov_fsb_grl}). Unfortunately, since none of these terms involve a two-point correlator, we can't cancel out any terms by virtue of them involving evaluating one of the fields at ${\bf k}=0$, as we did in the previous section. We can, however, use the symmetries we discussed in the previous section (i.e. the exchanges $(b\rightarrow c)$ and $(e\rightarrow f)$) to determine how many of the remaining 9 terms are distinct:
      \begin{itemize}
        \item Term 5 is equivalent to term 2.
        \item Terms 3, 4, 6, and 7 are all equivalent.
        \item Terms 8 and 9 are equivalent. In fact, they are the same as terms 2 and 5 if we swap ${\bf k}$ and ${\bf k}'$ (i.e. they're not equivalent, but we can compute them automatically from term 2).
        \item Term 10 is distinct from the rest, since all internal legs (i.e. ${\bf q}$ and ${\bf q}'$) fall neatly in separate ensemble averages.
      \end{itemize}
      Thus, we only need to compute terms 2, 3, and 10:
      \begin{equation} 
        \avg{abcdef}_{33} \quad = \quad 4 \avg{abd}\avg{cef} \quad + \quad 2 \big( \avg{abe}\avg{cdf} + ((b, c) \leftrightarrow (e, f)) \big) \quad + \quad \avg{aef}\avg{bcd} \; . \nonumber 
      \end{equation}

      Let us start with the last term:
      \begin{align}
        {\cal I}[\avg{aef}\avg{bcd}]
        &={\cal I}[\avg{\delta_{-{\bf k}}\delta_{{\bf q}'}\delta_{{\bf k}'-{\bf q}'}}\avg{\delta_{-{\bf k}'}\delta_{{\bf q}}\delta_{{\bf k}-{\bf q}}}]\\
        &={\cal I}\left[(2\pi)^{2N} \delta^D({\bf k}-{\bf k}')\delta^D({\bf k}-{\bf k}')B(-{\bf k},{\bf q}',{\bf k}-{\bf q}')B(-{\bf k},{\bf q},{\bf k}-{\bf q})\right].
      \end{align}
      As before, one of the delta functions is just $\delta^D(0)$, and will cancel out a volume factor. The remaining delta function in this case only involves ${\bf k}$ and ${\bf k}'$, and can be used to cancel out one of the ring integrals (e.g. over $k'$), swapping it for a Kronecker delta. The remaining integral over the angles of ${\bf k}$ will then cancel out one of the $V_k^{-1}$ factors. The result is then:
      \begin{align}
        {\cal I}[\avg{aef}\avg{bcd}]
        & = (2\pi)^N \delta^K_{k,k'} V^{-1}V_k^{-1} \int\dfk{q}W^L_qW^L_{|{\bf k}-{\bf q}|}B(-{\bf k},{\bf q},{\bf k}-{\bf q})\int\dfk{q'}W^L_{q'}W^L_{|{\bf k}-{\bf q}'|}B(-{\bf k},{\bf q}',{\bf k}-{\bf q}'),
      \end{align}
      which can be written succinctly as
      \begin{align}
        {\cal I}[\avg{aef}\avg{bcd}]
        &=(2\pi)^N \delta^K_{k,k'} V^{-1}V_k^{-1} \Phi^{L}_{k} \Phi^{L'}_{k'}.
      \end{align}

      Now let us look at the first term:
      \begin{align}
        {\cal I}[\avg{abd}\avg{cef}]={\cal I}\left[(2\pi)^{2N}\delta^D(-{\bf k}-{\bf k}'+{\bf q})\delta^D(0)B(-{\bf k},{\bf q},-{\bf k}')B(-{\bf k}',{\bf q}',{\bf k}'-{\bf q}')\right].
      \end{align}
      As before: we use $\delta^D(0)$ to cancel out a volume factor, and use the remaining delta function to cancel out the integral over ${\bf q}$. The result is:
      \begin{align}\nonumber
        {\cal I}[\avg{abd}\avg{cef}]
        & = V^{-1} V_k^{-2} \int_{V_k}d^N k\int_{V_k}d^N k' \, W^L_{k'}W^L_{|{\bf k}+{\bf k}'|}B(-{\bf k},-{\bf k}',{\bf k}+{\bf k'})\int\dfk{q'}\,W^{L'}_{q'}W^L_{|{\bf k}'-{\bf q}'|}B(-{\bf k}',{\bf q}',{\bf k}'-{\bf q}').\\
        &=V^{-1}\left\langle W^L_{k'}W^L_{|{\bf k}+{\bf k}'|}B(-{\bf k},-{\bf k}',{\bf k}+{\bf k'})\,\Phi^{L'}_{k'} \right\rangle_\varphi.
      \end{align}

      And finally, the middle term is:
      \begin{equation}
        {\cal I}[\avg{abe}\avg{cdf}]={\cal I}\left[(2\pi)^{2N} \delta^D(-{\bf k}+{\bf q}+{\bf q}')\delta^D(0)B(-{\bf k},{\bf q},{\bf k}-{\bf q})B({\bf q}',-{\bf k}', {\bf k}'-{\bf q}' ) \right].
      \end{equation}
      Using the same logic as above, this results in:
      \begin{equation}
        {\cal I}[\avg{abe}\avg{cdf}]=V^{-1}\left\langle \int\dfk{q}\,W^L_qW^L_{|{\bf k}-{\bf q}|}W^{L'}_qW^{L'}_{|{\bf k}'-{\bf q}|}\,B(-{\bf k},{\bf q},{\bf k}-{\bf q})\,B(-{\bf k}',{\bf q},{\bf k}'-{\bf q})\right\rangle_{\varphi}.
      \end{equation}

      Thus, the contribution to the FSB covariance from all terms of the form $3+3$ is:
      \begin{align}\nonumber
        {\rm Cov}_{33}\left(\hat{\Phi}^{L}_{k} ,\hat{\Phi}^{L'}_{k'}\right) \quad =
        &(2\pi)^N \delta^K_{k,k'}V^{-1}V_k^{-1}\Phi^{L}_{k} \Phi^{L'}_{k'} \quad + \\\nonumber
        & 2\,V^{-1}\left[\left\langle W^L_{k'}W^L_{|{\bf k}+{\bf k}'|}B(-{\bf k},-{\bf k}',{\bf k}+{\bf k'})\,\Phi^{L'}_{k'} \right\rangle_\varphi+((k,L)\leftrightarrow(k',L'))\right] \quad + \\
        &4\,V^{-1}\left\langle \int\dfk{q}\,W^L_qW^L_{|{\bf k}-{\bf q}|}W^{L'}_qW^{L'}_{|{\bf k}'-{\bf q}|}\,B(-{\bf k},{\bf q},{\bf k}-{\bf q})\,B(-{\bf k}',{\bf q},{\bf k}'-{\bf q})\right\rangle_{\varphi}.
      \end{align}
  
    \subsubsection{The ``$4+2$'' terms}\label{sssec:gauscov.gen.42}
      As before, we start by identifying equivalent and vanishing terms in Eq. \ref{eq:42all}. Terms 1, 2, 6, 13, 14, and 15 are automatically zero, since they involve two-point correlators of the form $\langle ab\rangle$, which imply evaluating a third field of the same triad at zero wavenumber. Of the 9 surviving terms:
      \begin{itemize}
        \item Term 3 is distinct from all others, since it includes a power spectrum involving the two outgoing modes (${\bf k}$ and ${\bf k}'$).
        \item Terms 4 and 5 are equivalent.
        \item Terms 7 and 10 are equivalent, and can be computed from terms 4 and 5 by swapping $(k,L)\leftrightarrow(k',L')$.
        \item Terms 8, 9, 11, and 12 are equivalent.
      \end{itemize}
      Thus, as in the $3+3$ case, we must just calculate 3 different terms. The calculation involves the same techniques used in the previous sections, and we will not repeat them here. The result is as follows:

      \begin{itemize}
        \item Term 3 is
        \begin{align}\nonumber
          {\cal I}[\avg{ad}\avg{bcef}]
          &={\cal I}\left[(2\pi)^{2N} \delta^D({\bf k}+{\bf k}')\delta^D(0) \, P_k \, T({\bf q},{\bf k}-{\bf q},{\bf q}',-{\bf k}-{\bf q}'\right]\\
          & = (2\pi)^N \delta^K_{k,k'} V^{-1} V_k^{-1} P_k\int\dfk{q}\dfk{q'}W^L_qW^L_{|{\bf k}-{\bf q}|}W^{L'}_{q'}W^{L'}_{|{\bf k}+{\bf q}'|}T({\bf q},{\bf k}-{\bf q},{\bf q}',-{\bf k}-{\bf q}').
        \end{align}
        \item Terms 4 and 5 are:
        \begin{align}\nonumber
          {\cal I}[\avg{ae}\avg{bcdf}]
          &={\cal I}\left[(2\pi)^{2N} \delta^D(-{\bf k}+{\bf k}'-{\bf q}')\delta^D(0) P_k\,T({\bf q},{\bf k}-{\bf q},-{\bf k}',{\bf k}'-{\bf k})\right]\\
          &=V^{-1}P_k\left\langle\int \dfk{q}\,W^L_qW^L_{|{\bf k}-{\bf q}|}W^{L'}_kW^{L'}_{|{\bf k}-{\bf k}'|}T({\bf q},{\bf k}-{\bf k},-{\bf k}',{\bf k}'-{\bf k})\right\rangle_\varphi,
        \end{align}
        from which terms 7 and 10 can also be computed.
        \item Terms 8, 9, 11, and 12 are:
        \begin{align}\nonumber
          {\cal I}[\avg{be}\avg{acdf}]
          &={\cal I}\left[(2\pi)^{2N} \delta^D({\bf q}+{\bf q}')\,\delta^D(0) P_q\,T(-{\bf k},{\bf k}-{\bf q},-{\bf k}',{\bf k}'+{\bf q})\right]\\
          &=V^{-1}\left\langle\int\dfk{q}W^L_qW^L_{|{\bf k}-{\bf q}|}W^{L'}_qW^{L'}_{|{\bf k}'+{\bf q}|}P_q\,T(-{\bf k},{\bf k}-{\bf q},-{\bf k}',{\bf k}'+{\bf q})\right\rangle_\varphi.
        \end{align}
      \end{itemize}

      Thus, the contribution to the FSB covariance from all terms of the form $4+2$ is:
      \begin{align}\nonumber
        {\rm Cov}_{42} \left( \hat{\Phi}^{L}_{k} ,\hat{\Phi}^{L'}_{k'} \right) \quad = 
        & (2\pi)^N \delta^K_{k,k'} V^{-1}V_k^{-1} P_k \int\dfk{q}\dfk{q'}W^L_qW^L_{|{\bf k}-{\bf q}|}W^{L'}_{q'}W^{L'}_{|{\bf k} + {\bf q}'|}T({\bf q},{\bf k}-{\bf q},{\bf q}',-{\bf k}-{\bf q}') \quad + \\ \nonumber
        & 2\,V^{-1}\left[P_k\left\langle\int \dfk{q}\,W^L_qW^L_{|{\bf k}-{\bf q}|}W^{L'}_kW^{L'}_{|{\bf k}-{\bf k}'|}T({\bf q},{\bf k}-{\bf k},-{\bf k}',{\bf k}'-{\bf k})\right\rangle_\varphi \; + \; ((k,L)\leftrightarrow(k',L'))\right] \quad +\\
        &4\,V^{-1}\left\langle\int\dfk{q}W^L_qW^L_{|{\bf k}-{\bf q}|}W^{L'}_qW^{L'}_{|{\bf k}'+{\bf q}|}P_q\,T(-{\bf k},{\bf k}-{\bf q},-{\bf k}',{\bf k}'+{\bf q})\right\rangle_\varphi.
      \end{align}
  
    \subsubsection{The ``$6$'' term}\label{sssec:gauscov.gen.6}
      The last term, corresponding to the fully connected six-point function, is simply:
      \begin{equation}
        {\rm Cov}_6\left(\hat{\Phi}^{L}_{k} ,\hat{\Phi}^{L'}_{k'}\right)=V^{-1}\left\langle\int\dfk{q}\dfk{q'}W^L_qW^L_{|{\bf k}-{\bf q}|}W^{L'}_{q'}W^{L'}_{|{\bf k}'-{\bf q}'|}{\cal P}_{(6)}(-{\bf k},{\bf q},{\bf k}-{\bf q},-{\bf k}',{\bf q}',{\bf k}'-{\bf q}')\right\rangle_\varphi.
      \end{equation}
    
    \subsubsection{Summary}\label{sssec:gauscov.gen.sum}
      Putting the results from the previous sections together, we find that the FSB covariance has the following structure:
      \begin{align}
        {\rm Cov}\left(\hat{\Phi}^{L}_{k} ,\hat{\Phi}^{L'}_{k'}\right)=(2\pi)^D\delta^K_{k,k'}V^{-1}V_k^{-1}\left[D_{222}+D_{33}+D_{42}\right]+V^{-1}[N_{222}+N_{33}+\tilde{N}_{33}+N_{42}+\tilde{N}_{42}+N_6],
      \end{align}
      where the three diagonal terms are:
      \begin{align}\label{eq:d222}
        &D_{222}(k)\equiv2P_k\int\dfk{q}\,W_q^LW_q^{L'}W^L_{|{\bf k}-{\bf q}|}W^{L'}_{|{\bf k}-{\bf q}|}P_qP_{|{\bf k}-{\bf q}|},\\\label{eq:d33}
        &D_{33}(k)\equiv {\Phi}^{L}_{k} ,{\Phi}^{L'}_{k},\\\label{eq:d42}
        &D_{42}(k)\equiv P_k\int\dfk{q}\,\dfk{q'}\,W^L_qW^L_{|{\bf k}-{\bf q}|}W^{L'}_{q'}W^{L'}_{|{\bf k}+{\bf q}'|}T({\bf q},{\bf k}-{\bf q},{\bf q}',-{\bf k}-{\bf q}'),
      \end{align}
      and the different non-diagonal terms are:
      \begin{align}
        &N_{222}(k,k')\equiv4W_k^{L'}W_{k'}^LP_kP_{k'}\,\left\langle W_{|{\bf k}-{\bf k}'|}^LW_{|{\bf k}-{\bf k}'|}^{L'}P_{|{\bf k}-{\bf k}'|}\right\rangle_{\varphi},\\
        &N_{33}(k,k')\equiv4\left\langle \int\dfk{q}\,W^L_qW^L_{|{\bf k}-{\bf q}|}W^{L'}_qW^{L'}_{|{\bf k}'-{\bf q}|}\,B(-{\bf k},{\bf q}{\bf k}-{\bf q})\,B(-{\bf k}',{\bf q}{\bf k}'-{\bf q})\right\rangle_{\varphi},\\
        &\tilde{N}_{33}(k,k')\equiv2\left[\left\langle W^L_{k'}W^L_{|{\bf k}+{\bf k}'|}B(-{\bf k},-{\bf k}',{\bf k}+{\bf k'})\,\Phi^{L'}_{k'})\right\rangle_\varphi+((k,L)\leftrightarrow(k',L'))\right],\\
        &N_{42}(k,k')\equiv4\left\langle\int\dfk{q}W^L_qW^L_{|{\bf k}-{\bf q}|}W^{L'}_qW^{L'}_{|{\bf k}'+{\bf q}|}P_q\,T(-{\bf k},{\bf k}-{\bf q},-{\bf k}',{\bf k}'+{\bf q})\right\rangle_\varphi,\\
        &\tilde{N}_{42}(k,k')\equiv2\left[P_k\left\langle\int \dfk{q}\,W^L_qW^L_{|{\bf k}-{\bf q}|}W^{L'}_kW^{L'}_{|{\bf k}-{\bf k}'|}T({\bf q},{\bf k}-{\bf k},-{\bf k}',{\bf k}'-{\bf k})\right\rangle_\varphi+((k,L)\leftrightarrow(k',L'))\right],\\
        &N_6(k,k')\equiv\left\langle\int\dfk{q}\dfk{q'}W^L_qW^L_{|{\bf k}-{\bf q}|}W^{L'}_{q'}W^{L'}_{|{\bf k}'-{\bf q}'|}{\cal P}_{(6)}(-{\bf k},{\bf q},{\bf k}-{\bf q},-{\bf k}',{\bf q}',{\bf k}'-{\bf q}')\right\rangle_\varphi
      \end{align}
  
  \subsection{Gaussian covariances}\label{ssec:gauscov.gaus}
    Our philosophy so far has been to understand the FSB as the cross-correlation between two maps: $\delta_{\bf k}$ and $(\delta_L^2)_{\bf k}$. Here we explore whether we can reuse one of the key results we have from the treatment of power spectrum covariances to obtain a relatively accurate prediction for the FSB covariance. It is well known that the so-called ``Gaussian'' power spectrum covariance, in which one treats all fields involved as if they were Gaussian random fields, captures the dominant contributions to the power spectrum uncertainties of realistic (i.e. non-Gaussian) tracers of the large-scale structure, particularly on sufficiently large scales (although the result is usually valid even on scales where the fields are already markedly non-Gaussian). We can thus explore whether treating {\sl both} $\delta_{\bf k}$ and $(\delta^2_L)_{\bf k}$ as Gaussian fields provides a good approximation to the FSB covariance.

    We will start by reviewing the main results regarding Gaussian power spectrum covariances, and then we will apply them to the FSB and compare the result to the full calculation described in the previous section.

    \subsubsection{Gaussian power spectrum covariances}\label{sssec:gauscov.gaus.cl}
      Consider the following general unbiased estimator of the power spectrum between two fields $a$ and $b$:
      \begin{equation}
        \hat{P}^{ab}_k\equiv V^{-1}\,V_k^{-1}\int_{V_k}d^Dk\,a_{\bf k}^*{b}_{\bf k}.
      \end{equation}
      The covariance between two such power spectra is:
      \begin{align}\nonumber
        {\rm Cov}\left(\hat{P}^{ab}_k,\hat{P}^{cd}_{k'}\right)
        &\equiv\left\langle\hat{P}^{ab}_k\,\hat{P}^{cd}_{k'}\right\rangle-\left\langle\hat{P}^{ab}_k\right\rangle\left\langle\hat{P}^{cd}_{k'}\right\rangle\\
        &=V^{-2}V_k^{-2}\int_{V_k}d^Dk\,\int_{V_k}d^Dk'\left(\langle a_{-{\bf k}}b_{\bf k}c_{-{\bf k}'}d_{{\bf k}'}\rangle-\langle a_{-{\bf k}}b_{\bf k}\rangle\langle c_{-{\bf k}'}d_{{\bf k}'}\rangle\right)
      \end{align}
      Assuming all fields to be Gaussian, we can expand the 4-point correlator into its three different pairings to obtain
      \begin{align}\nonumber
        {\rm Cov}\left(\hat{P}^{ab}_k,\hat{P}^{cd}_{k'}\right)
        &=V^{-2}V_k^{-2}\int_{V_k}d^Dk\,\int_{V_k}d^Dk'\left(\langle a_{-{\bf k}}c_{-{\bf k}'}\rangle\langle b_{\bf k}d_{{\bf k}'}\rangle+\langle a_{-{\bf k}}d_{{\bf k}'}\rangle\langle b_{\bf k}c_{-{\bf k}'}\rangle\right)\\\nonumber
        &=V^{-2}V_k^{-2}(2\pi)^{2D}\int_{V_k}d^Dk\,\int_{V_k}d^Dk'\delta^D({\bf k}+{\bf k}')\delta^D(0)\,P^{ac}_kP^{bd}_k+(c\leftrightarrow d)\\
        &=(2\pi)^D\delta^K_{k,k'}V^{-1}V_k^{-1}\left[P^{ac}_kP^{bd}_k+P^{ad}_kP^{bc}_k\right].
      \end{align}

      Thus, the Gaussian covariance is purely diagonal in $k$, and scales with number of modes as $\propto (2\pi)^DV^{-1}V_k^{-1}$, exactly the same way as all the purely diagonal terms in the full covariance calculation of the previous section.
  
    \subsubsection{Gaussian FSB covariance}\label{sssec:gauscov.gaus.fsb}
      Applying the result we just derived to the FSB, assuming both $\delta$ and $\delta_L^2$ to be Gaussian, we obtain:
      \begin{align}\label{eq:covG0}
        {\rm Cov}_G\left(\hat{\Phi}^{L}_{k} ,\hat{\Phi}^{L'}_{k'}\right)=(2\pi)^D\delta^K_{k,k'}V^{-1}V_k^{-1}\left[P^{\delta\delta}_kP^{\delta_L^2\delta_{L'}^2}_k+P^{\delta\delta_L^2}_kP^{\delta\delta_{L'}^2}_k \right].
      \end{align}
      Above $P^{\delta\delta}_k$ is simply $P_k$, and $P^{\delta\delta_L^2}_k=\Phi^L_k$. We only need to calculate $P^{\delta_L^2\delta_{L'}^2}_k$:
      \begin{align}\nonumber
        P^{\delta_L^2\delta_{L'}^2}_k
        &=V^{-1}\langle (\delta_L^2)_{-{\bf k}}(\delta_{L'}^2)_{\bf k}\rangle\\
        &=V^{-1}\int \dfk{q}\,\dfk{q'}\,W_q^L\,W_{|{\bf k}-{\bf q}|}^L\,W_{q'}^{L'}\,W_{|{\bf k}+{\bf q}'|}^{L'}\langle \delta_{\bf q}\delta_{-{\bf k}-{\bf q}}\delta_{{\bf q}'}\delta_{{\bf k}-{\bf q}'}\rangle.
      \end{align}
      The expression above will receive contributions from the disconnected and connected trispectrum. Of the three disconnected terms, the first one ($\langle\delta_{\bf q}\delta_{-{\bf k}-{\bf q}}\rangle\langle\delta_{{\bf q}'}\delta_{{\bf k}-{\bf q}'}\rangle$) is zero for all ${\bf k}\neq0$, and can be discarded. The two other terms are equivalent (under the change of variables ${\bf q}\rightarrow{\bf k}+{\bf q}$, and yield:
      \begin{align}\nonumber
        \left[P^{\delta_L^2\delta_{L'}^2}_k\right]_{22}
        &=2V^{-1}\int \dfk{q}\,\dfk{q'}\,W_q^L\,W_{|{\bf k}+{\bf q}|}^L\,W_{q'}^{L'}\,W_{|{\bf k}-{\bf q}'|}^{L'}(2\pi)^{2N}\delta^D({\bf q}+{\bf q}')\delta^D(0)P_qP_{|{\bf k}+{\bf q}|}.\\\label{eq:pd2_22}
        &=2\int \dfk{q}\,W_q^L\,W_{|{\bf k}-{\bf q}|}^L\,W_q^{L'}\,W_{|{\bf k}-{\bf q}|}^{L'}P_qP_{|{\bf k}-{\bf q}|}.
      \end{align}
      In turn, the fully-connected term is:
      \begin{align}\label{eq:pd2_4}
        \left[P^{\delta_L^2\delta_{L'}^2}_k\right]_{4}
        &=\int \dfk{q}\,\dfk{q'}\,W_q^L\,W_{|{\bf k}+{\bf q}|}^L\,W_{q'}^{L'}\,W_{|{\bf k}-{\bf q}'|}^{L'}T({\bf q},{\bf k}-{\bf q}, {\bf q}',-{\bf k}-{\bf q}').
      \end{align}

      Comparing Eqs. \ref{eq:covG0}, \ref{eq:pd2_22}, and \ref{eq:pd2_4} with Eqs. \ref{eq:d222}, \ref{eq:d33}, and \ref{eq:d42}, we thus find:
      \begin{equation}
        {\rm Cov}_G\left(\hat{\Phi}^{L}_{k} ,\hat{\Phi}^{L'}_{k'}\right)=(2\pi)^N \delta^K_{k,k'}V^{-1}V_k^{-1}\left[D_{222}(k)+D_{42}(k)+D_{33}(k)\right].
      \end{equation}

      Thus, the Gaussian power spectrum covariance applied to the FSB (treating both $\delta$ and $\delta_L^2$ as Gaussian fields) recovers all the diagonal terms obtained in the fully non-Gaussian calculation. Note that this is distinct from the covariance one would obtain treating $\delta$ (but not $\delta_L^2$) as a Gaussian field, which would correspond to the combination of all $2+2+2$ contributions, missing all the $3+3$ and $4+2$ terms contributing to the diagonal errors. The Gaussian power spectrum covariance applied to the FSB is thus able to recover all the relevant non-Gaussian purely-diagonal contributions, and neglects all the non-diagonal terms. Given the scaling of the diagonal terms with $V_k^{-1}\propto 1/k^{N-1}$, we can expect them, in general, to be the most relevant contributions on large scales, but to be superseded by the non-diagonal terms, which do not scale with $V_k$, on sufficiently small scales.

\section{Non-Gaussian covariances}\label{app:nongauscov}
  Here we give equivalent derivations to terms $D_{222}$ and $N_{222}$ from Appendix \ref{app:gauscov} in the curved-sky picture. We also derive the $D_{32}$ and $N_{32}$ terms (BP term in the literature) contributing to the bispectrum-power spectrum cross-covariance. In what follows, we will ignore the counter term in the definition of the covariance, as it cancels out with one of the $3+2$ contributions for the power spectrum-bispectrum cross-covariance, and one of the $3+3$ contributions for the bispectrum covariance, as shown in Appendix \ref{app:gauscov}.

  \subsection{Bispectrum covariance}\label{ssec:nongauscov:fsb}
    The full covariance of the FSB is given by
    \begin{equation}
      \text{Cov} \left(\hat{\Phi}^{L}_{k} ,\hat{\Phi}^{L'}_{k'}\right) \equiv \left\langle \hat \Phi^{L}_{k} \; \hat \Phi ^{L'}_{k'} \right\rangle - \left\langle \hat \Phi^{L}_{k} \right\rangle \; \left\langle \hat \Phi ^{L'}_{k'} \right\rangle \; .
    \end{equation}
    As described in the flat-sky approach (Appendix \ref{app:gauscov}), the first term will yield a 6-point correlation function. To see this in the spherical case, we start by writing the definition of the FSB estimator, given in Section \ref{ssec:meth.fsb}: 
    \begin{align}
      \hat{\Phi}^{L}_{\ell} & \equiv \frac{1}{2\ell +1}\sum_{ m} \,(\delta_L^2)_{\ell m}\,\delta^*_{\ell m} = \frac{1}{2\ell +1}\sum_{ \ell_{1, 2}}  W_{\ell_1}^L W_{\ell_2}^L \,\sum_{m, m_1, m_2} \mathcal G^{\ell, \ell_1, \ell_2}_{-m, m_1, m_2} \; \delta_{\ell_1 m_1} \delta_{\ell_2 m_2}\,\delta_{\ell -m} \, .
    \end{align}
    The ensemble average for a product of two FSBs with filters $L$ and $L'$ is then 
    \begin{align}\label{eq:cov6}
      \langle \hat{\Phi}^{L}_{\ell} \hat{\Phi}^{L'}_{\ell'} \rangle = \frac{1}{(2\ell +1)(2\ell' +1)} \sum_{ \ell_{1, 2} , \ell'_{1, 2}} &  W_{\ell_1}^L W_{\ell_2}^L W_{\ell_1'}^{L'} W_{\ell_2'}^{L'} \, \nonumber \\
      & \sum_{m, m', m_{1, 2}, m'_{1, 2}} \mathcal G^{\ell, \ell_1, \ell_2}_{-m, m_1, m_2} \mathcal G^{\ell', \ell_1', \ell_2'}_{-m', m_1', m_2'} \; \langle \delta_{\ell_1 m_1} \delta_{\ell_2 m_2}\,\delta_{\ell -m} \delta_{\ell_1' m_1'} \delta_{\ell_2' m_2'}\,\delta_{\ell' -m'} \rangle  \, ,
    \end{align}
    thus recovering the same 6-point function as in the flat-sky approach. Using the same reasoning, we decompose the 6-point correlation function into its connected components; in this section, we will only repeat the derivation for the $``2+2+2"$ decomposition, as it is empirically the most important non-Gaussian contribution to the FSB covariance. If we write the 6-point correlation function as $\langle \delta_{\ell_1 m_1} \delta_{\ell_2 m_2}\,\delta_{\ell -m} \delta_{\ell_1' m_1'} \delta_{\ell_2' m_2'}\,\delta_{\ell' -m'} \rangle \equiv \langle abcdef \rangle$ for simplicity, there are 15 unique permutations of $\langle ab \rangle \langle cd \rangle \langle ef \rangle$; only 6 terms remain after we make the observation that computing $\langle a_{\ell_1 m_1}b_{\ell_2 m_2} \rangle$ implies $\ell_1 = \ell_2$. Indeed, as $\ell_1$, $\ell_2$ and $\ell_3$ make up a triangle contributing to the bispectrum $b_{\ell_1 \ell_2 \ell_3}$, setting $\ell_1 = \ell_2$ only leaves one possible value for the last leg: $\ell_3 = 0$. Now the next pair evaluated would be $$ \langle c_{\ell_3 m_3} d_{\ell_4 m_4} \rangle = \langle c_{00} d_{\ell_4 m_4} \rangle =  c_{00} \langle d_{\ell_4 m_4} \rangle = 0 \times 0 = 0 $$as we assume we are dealing with fields with zero mean. Therefore all terms with at least one 2-point correlation function, evaluated for a pair of fields among $\{a, b, c\}$ or among $\{ d, e, f \}$, will vanish. Evaluating a 2-point correlation function with any other pairing will decrease the number of indices summed over, and ultimately the expression reduces to a much simpler form: we obtain 2 identical terms -- equivalent to the $D_{222}$ diagonal term -- and 4 terms which make up the $N_{222}$ non-Gaussian term. 
    
    One of the two terms in $D_{222}$ is
    \begin{equation}
      \langle ad \rangle \langle be \rangle \langle cf \rangle = \langle \delta_{\ell_1 m_1} \delta_{\ell_1' m_1'} \rangle \langle \delta_{\ell_2 m_2} \delta_{\ell_2' m_2'} \rangle \langle \,\delta_{\ell -m} \delta_{\ell' -m'} \rangle = \delta^K_{\ell_1 \ell'_1} C_{\ell_1} \, \times \, \delta^K_{\ell_2 \ell'_2} C_{\ell_2} \, \times \, \delta^K_{\ell \ell'} C_{\ell} \, ,
    \end{equation}
    while one of the four terms of $N_{222}$ would be 
    \begin{equation}
      \langle ae \rangle \langle bf \rangle \langle cd \rangle = \langle \delta_{\ell_1 m_1} \delta_{\ell_2' m_2'} \rangle \langle \delta_{\ell_2 m_2}  \delta_{\ell' -m'} \rangle \langle \,\delta_{\ell -m}   \delta_{\ell_1' m_1'} \rangle = \delta^K_{\ell_1 \ell'_2} C_{\ell_1} \, \times \, \delta^K_{\ell_2 \ell'} C_{\ell'} \, \times \,  \delta^K_{\ell \ell'_1} C_{\ell} \, .
    \end{equation}
    Plugging this back into Eq. \ref{eq:cov6} gives us the Gaussian and non-Gaussian $``2+2+2"$ contributions for the FSB covariance: 
    \begin{align} 
      D_{222} (\ell, \ell') & = \frac{2}{(2\ell +1)(2\ell' +1)} \sum_{ \ell_{1, 2} , \ell'_{1, 2}}  W_{\ell_1}^L W_{\ell_2}^L W_{\ell_1'}^{L'} W_{\ell_2'}^{L'} \,\sum_{m, m', m_{1, 2}, m'_{1, 2}} \mathcal G^{\ell, \ell_1, \ell_2}_{-m, m_1, m_2} \mathcal G^{\ell', \ell_1', \ell_2'}_{-m', m_1', m_2'} \; \delta^K_{\ell_1 \ell'_1} C_{\ell_1}\delta^K_{\ell_2 \ell'_2} C_{\ell_2}\delta^K_{\ell \ell'} C_{\ell} \nonumber \\
      & = \frac{2  C_{\ell} \delta^K_{\ell \ell'}}{(2\ell +1)^2} \sum_{ \ell_{1, 2}}  W_{\ell_1}^L W_{\ell_2}^L W_{\ell_1}^{L'} W_{\ell_2}^{L'} C_{\ell_1} C_{\ell_2}  \,\sum_{m, m_{1, 2}} \mathcal G^{\ell, \ell_1, \ell_2}_{-m, m_1, m_2} \mathcal G^{\ell, \ell_1, \ell_2}_{-m, m_1, m_2} \nonumber \\
      & = \frac{2  C_{\ell} \delta^K_{\ell \ell'}}{4\pi (2\ell +1)} \sum_{ \ell_{1, 2}}  \left[ (2\ell_1+1) W_{\ell_1}^L W_{\ell_1}^{L'} C_{\ell_1} \right] \left[ (2\ell_2+1) W_{\ell_2}^{L} W_{\ell_2}^{L'} C_{\ell_2} \right] \wtj{\ell}{\ell_1}{\ell_2}{0}{0}{0}^2 \, ,
    \end{align}

    \begin{align} 
      N_{222} (\ell, \ell') & = \frac{4}{(2\ell +1)(2\ell' +1)} \sum_{ \ell_{1, 2} , \ell'_{1, 2}}  W_{\ell_1}^L W_{\ell_2}^L W_{\ell_1'}^{L'} W_{\ell_2'}^{L'} \,\sum_{m, m', m_{1, 2}, m'_{1, 2}} \mathcal G^{\ell, \ell_1, \ell_2}_{-m, m_1, m_2} \mathcal G^{\ell', \ell_1', \ell_2'}_{-m', m_1', m_2'} \; \delta^K_{\ell_1 \ell'_2} C_{\ell_1} \delta^K_{\ell_2 \ell'} C_{\ell'} \delta^K_{\ell \ell'_1} C_{\ell} \nonumber \\
      & = \frac{4  C_{\ell} C_{\ell'} }{(2\ell +1)(2\ell' +1)} \sum_{ \ell_{1}}  W_{\ell_1}^L W_{\ell'}^L W_{\ell}^{L'} W_{\ell_1}^{L'} C_{\ell_1}  \,\sum_{m, m', m_{1}} \mathcal G^{\ell, \ell_1, \ell'}_{-m, m_1, m'} \mathcal G^{\ell, \ell_1, \ell'}_{-m, m_1, m'} \nonumber \\
      & = \frac{C_{\ell} C_{\ell'} W_{\ell'}^L W_{\ell}^{L'}}{\pi} \sum_{ \ell_{1}} (2\ell_1+1) W_{\ell_1}^L W_{\ell_1}^{L'} C_{\ell_1} \wtj{\ell}{\ell_1}{\ell'}{0}{0}{0}^2 \, .
    \end{align}

    As highlighted in Appendix \ref{app:gauscov}, the counter term in the expression of the full covariance cancels out with a term from the $``3+3"$ contribution.

    The $D_{222}$ is a purely diagonal term, due to the Kronecker delta in front of the expression. $N_{222}$ is however much more interesting: it contributes only when the filters are identical ($W_\ell^L W_{\ell'}^{L'}$ factor) and can therefore account, in theory, for the heightened correlations we observe in Fig. \ref{fig:lpt_cors} along the main diagonal where the two filters match. 
  
  \subsection{Power spectrum $\times$ bispectrum covariance}\label{ssec:nongauscov:fsbcl}
    We repeat the same process for the cross-correlation between the FSB and power spectra; this time the first term in the covariance will give a 5-point correlation function:
    \begin{align}
      \left\langle \hat{\Phi}^{L}_{\ell} \; \hat C_{\ell'} \right\rangle & = \left\langle \frac{1}{2\ell +1}\sum_{ m} \,(\delta_L^2)_{\ell m}\,\delta^*_{\ell m} \;\; \frac{1}{2\ell' +1}\sum_{ m'} \,\delta_{\ell' m'}\,\delta^*_{\ell' m'} \right\rangle \nonumber \\ 
      & = \frac{1}{(2\ell +1)(2\ell' +1)} \; \sum_{\ell_{1} \ell_{2}} \; \sum_{ m, m', m_1, m_2} \mathcal G_{m, m_1, m_2}^{\ell, \ell_1, \ell_2} \, \langle \delta^L_{\ell_1 m_1} \delta^L_{\ell_2 m_2} \,\delta^*_{\ell m} \,\delta^{}_{\ell' m'}\,\delta^*_{\ell' m'} \rangle \, .
    \end{align}
    We write $\langle \delta^L_{\ell_1 m_1} \, \delta^L_{\ell_2 m_2} \delta^{*}_{\ell_3 m_3} \; \delta^{}_{\ell_4 m_4}\,\delta^*_{\ell_5 m_5} \rangle \equiv \langle abcde \rangle$; using the same reasoning as in the previous section, we can similarly simplify the number of terms to evaluate. 

    Once again, we assume we are dealing with zero-mean fields: all decompositions that contain the mean of a field will vanish. This means the $1 + 4$, $1 + 2 + 2$ and $1+1+1+ 2$ contributions vanish; we are left with just the fully connected ($5$) and the $2 + 3$ contributions to the full covariance. We will look into the $2+ 3$ term first, as we suspect it is the main contribution to the cross covariance -- from perturbation theory, we expect the connected tetraspectrum to have a lower signal than the product of a power spectrum and bispectrum.
    As in Section \ref{ssec:nongauscov:fsb}, we can discard a certain number of terms by recalling the closed triangle rule; the remaining terms are
    \begin{align} 
      \langle abcde \rangle_{32} = &  \quad \; \cancel{\langle cde \rangle \langle ab \rangle} + \cancel{\langle bde \rangle \langle ac \rangle} + \langle bce \rangle \langle ad \rangle + \langle bcd \rangle \langle ae \rangle \nonumber \\ 
      & + \cancel{\langle ade \rangle \langle bc \rangle} + \langle ace \rangle \langle bd \rangle + \langle acd \rangle \langle be \rangle \nonumber \\ 
      & + \langle abd \rangle \langle ce \rangle + \langle abe \rangle \langle cd \rangle \nonumber \\ & + \langle abc \rangle \langle de \rangle \nonumber
    \end{align}
    Plugging in $a = b \equiv \delta_{\ell m}^L$ and $c = d = e \equiv \delta_{\ell m}$ , we notice two things: swapping fields $c, d, e$ keeps quantities unchanged, as well as swapping $a$ and $b$. Thus we can group terms as such: terms 3,4,6,7 together, and terms 8,9,10 in a second contribution. 

    The second group is in fact a term which we recover in the Gaussian-limit estimate of the cross-covariance: treating the FSB as a cross power spectrum of $\delta$ and $\delta_L^2$, we have that 
    \begin{align} \nonumber 
      {\rm Cov} \left( C_\ell^{\delta \delta_L^2}, C_{\ell'}^{\delta \delta} \right) \; = \;  & \left\langle C_\ell^{\delta \delta_L^2} C_{\ell'}^{\delta \delta} \right\rangle - \left\langle C_\ell^{\delta \delta_L^2} \right\rangle \left\langle C_{\ell'}^{\delta \delta} \right\rangle \\ \nonumber
      \; = \;  & \delta^K_{\ell \ell_1} \delta^K_{\ell' \ell_2} \left\langle \delta_{\ell m} \left(\delta_L^2\right)_{\ell_1 m_1} \delta_{\ell' m'} \delta_{\ell_2 m_2} \right\rangle - \delta^K_{\ell \ell_1} \delta^K_{\ell' \ell_2} \left\langle \delta_{\ell m} \left(\delta_L^2\right)_{\ell_1 m_1} \right\rangle \left\langle \delta_{\ell' m'} \delta_{\ell_2 m_2} \right\rangle \\ \nonumber
      \; = \;  & \bigg[ \delta^K_{\ell \ell_1} \delta^K_{\ell' \ell_2} \left\langle \delta_{\ell m} \left(\delta_L^2\right)_{\ell_1 m_1} \right\rangle \left\langle \delta_{\ell' m'} \delta_{\ell_2 m_2} \right\rangle + \delta^K_{\ell \ell'} \delta^K_{\ell_1 \ell_2} \left\langle \delta_{\ell m} \delta_{\ell' m'} \right\rangle \left\langle \left(\delta_L^2\right)_{\ell_1 m_1} \delta_{\ell_2 m_2} \right\rangle \\ \nonumber
      & \; + \; \delta^K_{\ell \ell_2} \delta^K_{\ell' \ell_1} \left\langle \delta_{\ell m} \delta_{\ell_2 m_2} \right\rangle \left\langle \delta_{\ell' m'} \left(\delta_L^2\right)_{\ell_1 m_1} \right\rangle \bigg] - \delta^K_{\ell \ell_1} \delta^K_{\ell' \ell_2} \left\langle \delta_{\ell m} \left(\delta_L^2\right)_{\ell_1 m_1} \right\rangle \left\langle \delta_{\ell' m'} \delta_{\ell_2 m_2} \right\rangle \\
      \; = \;  & 2 \, \delta^K_{\ell \ell'} \, \left\langle C_\ell^{\delta \delta_L^2} \right\rangle \left\langle C_{\ell'}^{\delta \delta} \right\rangle \; ,
    \end{align}
    which is exactly what this second group is, when you notice that the first term in the group also cancels out with the counter-term from the definition for the covariance, as it is simply the product of the original FSB and power spectrum. This diagonal term in the FSB-power spectrum cross-covariance is thus also recovered by the Gaussian approach described in \ref{sssec:gauscov.gaus.fsb} for the FSB covariance; using the same notation, we label it $D_{32}$. 

    The first group however is a non-Gaussian contribution to the covariance: 
    \begin{align} \nonumber
      N_{32} (\ell, \ell') & \equiv \frac{1}{(2\ell +1)(2\ell' +1)}\sum_{ m, m'} \sum_{(\ell m)_{1, 2}} \mathcal G_{m, m_1, m_2}^{\ell, \ell_1, \ell_2} \, \left[ \langle \delta^L_{\ell_2 m_2} \,\delta^*_{\ell m} \delta_{\ell' m'} \rangle \, \langle \delta^L_{\ell_1 m_1}\,\delta^*_{\ell' m'} \rangle + \text{3 perm.} \right]  \\ \nonumber
      & = \frac{4}{(2\ell +1)(2\ell' +1)}\sum_{ m, m'} \sum_{(\ell m)_{1, 2}} \mathcal G_{m, m_1, m_2}^{\ell, \ell_1, \ell_2} \, \left[ W_{\ell_2}^L \mathcal G_{m_2, m, m'}^{\ell_2, \ell, \ell'} b_{\ell_2, \ell, \ell'} \, W_{\ell'}^L C_{\ell'} \delta_{\ell_1 \ell'} \right]  \\ \nonumber
      & = \frac{4 W_{\ell'}^L C_{\ell'}}{(2\ell +1)(2\ell' +1)}\sum_{ m, m'} \sum_{(\ell m)_{2}} \left(\mathcal G_{m, m', m_2}^{\ell, \ell', \ell_2} \right)^2 \, W_{\ell_2}^L b_{\ell_2, \ell, \ell'}  \\
      & = 4 W_{\ell'}^L C_{\ell'} \sum_{\ell_{2}} \frac{(2\ell_2 +1)}{4\pi} \wtj{\ell_2}{\ell}{\ell'}{0}{0}{0}^2  \, W_{\ell_2}^L b_{\ell_2, \ell, \ell'} \, .
    \end{align}

    Ideally we want to write the sum as an FSB, as both expressions are quite similar. We can add a sum over an additional $\ell$ variable to match the FSB definition (Eq. \ref{eq:fsbest}), but immediately reverse it by adding a Kronecker delta inside the sum: 
    \begin{equation}
      N_{32} (\ell, \ell') = \frac{ 4 W_{\ell'}^L C_{\ell'}}{ 2\ell_3 +1} \sum_{\ell_{2}, \ell_3} \delta^{K}_{\ell \ell_3} \frac{(2\ell_2 +1)(2\ell_3 +1)}{4\pi} \wtj{\ell_2}{\ell_3}{\ell'}{0}{0}{0}^2  \, W_{\ell_2}^L b_{\ell_2, \ell_3, \ell'} \, .
    \end{equation}
    Now we can simply treat this Kronecker delta as a narrow filter; we replace the delta by a filter notation like so: 
    \begin{equation}
      N_{32} (\ell, \ell') = \frac{4 W_{\ell'}^L C_{\ell'}}{2\ell +1} \sum_{\ell_{2}, \ell_3}  \frac{(2\ell_2 +1)(2\ell_3 +1)}{4\pi} \wtj{\ell_2}{\ell_3}{\ell'}{0}{0}{0}^2  \, W_{\ell_2}^L W_{\ell_3}^\ell b_{\ell_2, \ell_3, \ell'}  \, .
    \end{equation}

    A generalised FSB for thin filters then arises in the $N_{32}$ term: 
    \begin{equation}
      N_{32} (\ell, \ell') = \frac{ 4 W_{\ell'}^L C_{\ell'}}{2\ell +1} \Phi_{\ell'}^{L \ell} \, .
    \end{equation}

    An interesting insight we get from this derivation is that the $N_{32}$ term only contributes if $\ell'$ (corresponding to the power spectrum) is evaluated within the FSB filter $L$. In other words, the correlation is higher when the power spectrum shares the long mode of the bispectrum; this result had been previously derived in the literature for the 3D Euclidean case \citep{barreiraSqueezedMatterBispectrum2019, biagettiCovarianceSqueezedBispectrum2022}. 

  \subsection{Summary}
    In this section, we derived the following FSB auto-covariance terms: 
    \begin{align}
      D_{222} (\ell, \ell') & = \frac{ C_{\ell} \delta^K_{\ell \ell'}}{2\pi (2\ell +1)} \sum_{ \ell_{1, 2}}  \left[ (2\ell_1+1) W_{\ell_1}^L W_{\ell_1}^{L'} C_{\ell_1} \right] \left[ (2\ell_2+1) W_{\ell_2}^{L} W_{\ell_2}^{L'} C_{\ell_2} \right] \wtj{\ell}{\ell_1}{\ell_2}{0}{0}{0}^2 , \\
      N_{222} (\ell, \ell') & = \frac{C_{\ell} C_{\ell'} W_{\ell'}^L W_{\ell}^{L'}}{\pi} \sum_{ \ell_{1}} (2\ell_1+1) W_{\ell_1}^L W_{\ell_1}^{L'} C_{\ell_1} \wtj{\ell}{\ell_1}{\ell'}{0}{0}{0}^2 ; 
    \end{align}
    and the following FSB-power spectrum cross-covariance terms:
    \begin{align}
        D_{32} (\ell, \ell') & = 2 \, \Phi_\ell^L \, C_\ell \, \delta^K_{\ell \ell'} , \\ 
        N_{32} (\ell, \ell') & = \frac{4 W_{\ell'}^L C_{\ell'}}{ 2\ell +1} \Phi_{\ell'}^{L \ell} .
    \end{align}
    
\section{Stochastic contributions}\label{app:stochasticity}
  Observationally, we estimate the FSB from pixelised maps of the galaxy overdensity $\delta_p = \frac{N_p}{\bar{N}}-1$, where $N_p$ denotes the number of galaxies in pixel $p$, and $\bar{N}$ the mean number of galaxies per pixel. Assuming galaxies to be Poisson tracers of the underlying density field, the first, second and third moments of the galaxy density field are given by
  \begin{align}
    \langle N_p\rangle_{\mathrm{Pois}}&=\bar{N}_p=\bar{N}(1+\delta_p),\nonumber\\
    \langle N_p N_{p'}\rangle_{\mathrm{Pois}}&=\bar{N}_p\bar{N}_{p'}+\delta_{pp'}^K\bar{N}_p,\label{eq:stochasticity:Poisson}\\
    \langle N_p N_{p'}N_{p''}\rangle_{\mathrm{Pois}}&=\bar{N}_{p}\bar{N}_{p'}\bar{N}_{p''}+\delta_{pp'}^{K}\bar{N}_{p}\bar{N}_{p''}+\delta_{pp''}^{K}\bar{N}_{p}\bar{N}_{p'}+\delta_{p'p''}^{K}\bar{N}_{p}\bar{N}_{p'}+\delta_{pp'}^K\delta_{p'p''}^K\bar{N}_p\nonumber,
  \end{align}
  where $\delta_{pp'}^{K}$ denotes the Kronecker delta and we have only performed the Poisson ensemble average \citep{1980Peebles}.

  In order to obtain an expression for the FSB, we need to compute the expectation value of the product of three overdensities, i.e. 
  \begin{equation}
    \langle\delta_p \delta_{p'}\delta_{p''}\rangle_{\mathrm{Pois}, \delta}=\left\langle\left(\frac{N_p}{\bar{N}}-1\right)\left(\frac{N_{p'}}{\bar{N}}-1\right)\left(\frac{N_{p''}}{\bar{N}}-1\right)\right\rangle_{\mathrm{Pois}, \delta},
  \end{equation}
  over both realisations of the Poisson process and the underlying density field.

  We do this in two steps: first, performing only the Poisson averaging, we get using Eqs. \ref{eq:stochasticity:Poisson}
  \begin{align}
    \left\langle\left(\frac{N_p}{\bar{N}}-1\right)\left(\frac{N_{p'}}{\bar{N}}-1\right)\left(\frac{N_{p''}}{\bar{N}}-1\right)\right\rangle_{\mathrm{Pois}}= & \;\delta^K_{pp'}\bar{N}^{-1}(\delta_{p''}+\delta_{p}\delta_{p''})+\delta^K_{p'p''}\bar{N}^{-1}(\delta_{p}+\delta_{p}\delta_{p'})+\delta^K_{pp''}\bar{N}^{-1}(\delta_{p'}+\delta_{p'}\delta_{p''})\\
    &+\delta_p\delta_{p'}\delta_{p''}+\delta^K_{pp'}\delta^K_{p'p''}\bar{N}^{-2}(1+\delta_p).
  \end{align}
  Now we can also average over the density field realisations to finally obtain
  \begin{align}
    \left\langle\left(\frac{N_p}{\bar{N}}-1\right)\left(\frac{N_{p'}}{\bar{N}}-1\right)\left(\frac{N_{p''}}{\bar{N}}-1\right)\right\rangle = \, \langle \delta_p & \delta_{p'}\delta_{p''}\rangle \, + \, \delta^K_{pp'}\bar{N}^{-1}\langle\delta_{p}\delta_{p''}\rangle \nonumber \\ 
    & \, + \, \delta^K_{p'p''}\bar{N}^{-1}\langle\delta_{p}\delta_{p'}\rangle \, + \, \delta^K_{pp''}\bar{N}^{-1}\langle\delta_{p'}\delta_{p''}\rangle \, + \, \delta^K_{pp'}\delta^K_{p'p''}\bar{N}^{-2} \,,
    \label{eq:stochasticity:bisp}
  \end{align}
  where we have omitted explicitly denoting the averaging procedure for simplicity. With this, we can compute the bispectrum of the galaxy overdensity field for a pixelised map. The spherical harmonic coefficients of the map are given by
  \begin{equation}
    a_{\ell m}=\sum_p\Omega_p\left(\frac{N_p}{\bar{N}}-1\right)Y_{\ell m p}^*,
  \end{equation}
  where we have used the shorthand $Y_{\ell m p}\coloneq Y_{\ell m}(\nv_p)$, and $\Omega_p$ denotes the pixel area. The bispectrum is then given by
  \begin{equation}
    \left\langle a_{\ell m}a_{\ell' m'}a_{\ell'' m''}\right\rangle=\sum_{p, p', p''}\Omega_p^3\left\langle\left(\frac{N_p}{\bar{N}}-1\right)\left(\frac{N_{p'}}{\bar{N}}-1\right)\left(\frac{N_{p''}}{\bar{N}}-1\right)\right\rangle Y_{\ell m p}^*Y_{\ell' m' p'}^*Y_{\ell'' m'' p''}^*.
  \end{equation}
  Inserting the results from Eq.~\ref{eq:stochasticity:bisp}, we get
  \begin{align}
    \left\langle a_{\ell m}a_{\ell' m'}a_{\ell'' m''}\right\rangle=&\sum_{p, p', p''}\Omega_p^3\langle \delta_p\delta_{p'}\delta_{p''}\rangle Y_{\ell m p}^*Y_{\ell' m' p'}^*Y_{\ell'' m'' p''}^*\\
    &+\left(\frac{\Omega_p}{\bar{N}}\right)^2\sum_{p}\Omega_p Y_{\ell m p}^*Y_{\ell' m' p}^*Y_{\ell'' m'' p}^*\\
    &+\left(\frac{\Omega_p}{\bar{N}}\right)\sum_{p, p''}\Omega_p^2 \langle\delta_{p}\delta_{p''}\rangle Y_{\ell m p}^*Y_{\ell' m' p}^*Y_{\ell'' m'' p''}^*
    +2\; \mathrm{perm.}.
  \end{align}
  Using the definition of the Gaunt integral and the reduced bispectrum, the first two terms can be rewritten as
  \begin{align}
    \left\langle a_{\ell m}a_{\ell' m'}a_{\ell'' m''}\right\rangle = \, &\gaunt{\ell}{\ell'}{\ell''}{m}{m'}{m''}\left( b_{\ell\ell'\ell''}+\bar{n}^{-2} \right) \\
    &+ \frac{\Omega_p}{\bar{N}} \sum_{p, p''}\Omega_p^2 \langle\delta_{p}\delta_{p''}\rangle Y_{\ell m p}^*Y_{\ell' m' p}^*Y_{\ell'' m'' p''}^*
    +2\; \mathrm{perm.}.
  \end{align}
  The last term can be simplified by plugging in the definition of the spherical harmonic power spectrum such as
  \begin{align}
    \frac{\Omega_p}{\bar{N}} \sum_{p, p''}\Omega_p^2 \langle\delta_{p}\delta_{p''}\rangle Y_{\ell m p}^*Y_{\ell' m' p}^*Y_{\ell'' m'' p''}^* & = \bar{n}^{-1}\iint\mathrm{d}\nv \mathrm{d}\nv'' Y_{\ell m}^*(\nv)Y_{\ell' m'}^*(\nv)Y_{\ell'' m''}^*(\nv'')\sum_{\ell_1m_1}\sum_{\ell_2m_2}Y_{\ell_1 m_1}^*(\nv)Y_{\ell_2 m_2}(\nv'')C_{\ell_1}\delta_{\ell_1\ell_2}^K\delta_{m_1m_2}^K \nonumber \\
    &=\bar{n}^{-1}\sum_{\ell_1m_1}C_{\ell_1}\int\mathrm{d}\nv  Y_{\ell m}^*(\nv)Y_{\ell' m'}^*(\nv)Y_{\ell_1 m_1}^*(\nv)\int\mathrm{d}\nv''Y_{\ell'' m''}^*(\nv'')Y_{\ell_2 m_2}(\nv'') \nonumber \\
    &=\bar{n}^{-1}C_{\ell''}\int\mathrm{d}\nv Y_{\ell m}^*(\nv)Y_{\ell' m'}^*(\nv)Y_{\ell'' m''}^*(\nv) \nonumber \\
    &=\gaunt{\ell}{\ell'}{\ell''}{m}{m'}{m''}\bar{n}^{-1}C_{\ell''}.
  \end{align}
  Putting everything together, we thus arrive at
  \begin{equation}
    \left\langle a_{\ell m}a_{\ell' m'}a_{\ell'' m''}\right\rangle=\gaunt{\ell}{\ell'}{\ell''}{m}{m'}{m''}\left[b_{\ell\ell'\ell''}+\bar{n}^{-1}(C_{\ell}+C_{\ell'}+C_{\ell''})+\bar{n}^{-2}\right],
  \end{equation}
  which can be plugged into the expression for the FSB to arrive at Eq.~\ref{eq:fsb_stoch}.

\section{2D LPT fields}\label{app:2dlpt}
  This appendix describes the procedure used to generate the fast 2D LPT simulations introduced in Section \ref{sssec:res.sim.2dlpt}. The method is a direct application of the Zel'dovich approximation to 2D fields defined on the sphere, and therefore we start by reviewing this first.

  Under the Zel'dovich approximation \citep{1970A&A.....5...84Z,1401.5466}, the displacement $\mathbf{\Psi}({\bf q},t)$ of a given matter element from its homogeneous initial conditions ${\bf q}$ is the gradient of a displacement potential $\phi$, which itself is related to the linearised overdensity $\Delta^L({\bf q},t)$ through a Poisson-like equation:
  \begin{equation}
    \mathbf{\Psi}=-\nabla\phi,\hspace{12pt}\nabla^2\phi=\Delta^L.
  \end{equation}
  By mass conservation, the overdensity field in Eulerian space is given by the inverse Jacobian of this transformation, $1+\delta=1/{\rm det}(\mathbb{I}+{\sf J})$, where $J_{ij}=\partial_i\Psi_j$ and $\mathbb{I}$ is the identity matrix. A 3D Lagrangian overdensity field can then be easily constructed as follows:
  \begin{itemize}
    \item Generate a Gaussian realisation of $\Delta^L$ from the linear matter power spectrum.
    \item Construct the Jacobian matrix ${\bf J}$ via Fourier-space differentiation:
    \begin{equation}
      J_{ij}\equiv -\partial_i\partial_j\nabla^{-2}\Delta_L=-\int\frac{d^3k}{(2\pi)^3}\frac{k_ik_j}{k^2}\Delta^L_{\bf k}.
    \end{equation}
    \item Generate the Eulerian, non-Gaussian field as $1+\Delta=[{\rm det}(\mathbb{I}+{\sf J})]^{-1}$.
  \end{itemize}

  We construct our 2D LPT simulations following the same steps, replacing partial derivatives in Euclidean space by covariant derivatives on the sphere. Let $\delta^G$ be a Gaussian realisation of an input power spectrum $C_\ell$. The corresponding angular displacement potential $\phi$ can be constructed from $\delta^G$ by inverting the spherical Poisson equation in harmonic space:
  \begin{equation}
    \nabla_\Omega^2\phi=\delta^G
    \hspace{6pt}\longrightarrow\hspace{6pt}
    \phi_{\ell m}=-\frac{\delta_{\ell m}}{\ell(\ell+1)},
  \end{equation}
  where $\nabla_\Omega^2$ denotes the angular Laplacian. Once $\phi$ is constructed, the Jacobian matrix $J_{ab}$ is then given by its second covariant derivatives. As shown in \cite{alonsoRecoveringTidalField2016}, this can be neatly expressed using the differential operator $\eth$, and its conjugate $\bar{\eth}$. The action of these operator on a spin-$s$ field $_sf$ is:
  \begin{equation}
    \eth\,_sf\equiv-\sin^s\theta\left(\partial_\theta+\frac{i\partial_\varphi}{\sin\theta}\right)\sin^{-s}\theta\,_sf,
    \hspace{12pt}
    \bar{\eth}\,_sf\equiv-\sin^{-s}\theta\left(\partial_\theta-\frac{i\partial_\varphi}{\sin\theta}\right)\sin^s\theta\,_sf.
  \end{equation}
  Bearing in mind that $\eth\phi$ is a spin-1 field, the Jacobian is then related to the second derivatives of $\phi$ via
  \begin{equation}
    -\eth^2\phi = (J_{\theta\theta}-J_{\varphi\varphi}) + 2iJ_{\theta\varphi},\hspace{12pt}-\bar{\eth}\eth\phi=J_{\theta\theta}+J_{\varphi\varphi}.
  \end{equation}
  As before, the action of the $\eth^2$ and $\bar{\eth}\eth$ operators can be expressed in harmonic space:
  \begin{equation}
    (\eth^2\phi)_{\ell m}=-\sqrt{\frac{(\ell+2)!}{(\ell-2)!}}\phi_{\ell m},\hspace{12pt}
    (\bar{\eth}\eth\phi)_{\ell m}=-\ell(\ell+1)\phi_{\ell m}=\delta^G_{\ell m}.
  \end{equation}

  Using these results, we can use the following procedure to generate 2D LPT simulations:
  \begin{itemize}
    \item Given an input power spectrum, generate a Gaussian overdensity field $\delta^G(\nv)$.
    \item Construct a spin-2 field $\delta^{G,2}(\nv)\equiv Q(\nv)+iU(\nv)$ with zero $B$-modes and an $E$ mode related to $\delta^G$ in harmonic space via
    \begin{equation}
      E_{\ell m}=\frac{1}{\ell(\ell+1)}\sqrt{\frac{(\ell+2)!}{(\ell-2)!}}\delta^G_{\ell m}.
    \end{equation}
    \item The Jacobian matrix is then constructed from $(\delta^G,Q,U)$ as
    \begin{equation}
      J_{\theta\theta}=-\frac{1}{2}(\delta^G+Q),\hspace{12pt}J_{\varphi\varphi}=-\frac{1}{2}(\delta^G-Q),\hspace{12pt}J_{\theta\varphi}=-\frac{1}{2}U.
    \end{equation}
    \item The non-Gaussian overdensity field is then calculated from the determinant of $\delta_{ab}+J_{ab}$. Expressed in terms of $\delta^G$, $Q$, and $U$, this is simply:
    \begin{equation}\label{eq:2dlpt}
      1+\delta=\left[1-\delta^G+\frac{1}{4}\left((\delta^G)^2-Q^2-U^2\right)\right]^{-1}.
    \end{equation}
  \end{itemize}

  The non-linear and non-local transformation of Eq. \ref{eq:2dlpt} endows the final overdensity field with a non-zero bispectrum that we use in Section \ref{ssec:res.2dlpt} to validate our FSB estimator. Note, however, that due to the connection with LPT, this bispectrum should resemble the shape of the real projected bispectrum to some extent. To see this, consider $\delta^G(\nv)$ to be a projected version of the 3D matter overdensity field $\Delta({\bf x})$ with a particular redshift distribution $p(z)$
  \begin{equation}
    \delta^G(\nv)=\int dz\,p(z)\,\Delta(\chi(z)\nv).
  \end{equation}
  As shown in Section 2.2.2 of \cite{alonsoRecoveringTidalField2016}, the angular displacement field $\psi_a$ associated with the procedure described above is, approximately, related to the underlying 3D displacement field $\Psi_i$ via\footnote{The derivation in \cite{alonsoRecoveringTidalField2016} applies to the projected tidal tensor and its 3D counterpart, but the arguments used are equally valid when applied to the Lagrangian displacement.}
  \begin{equation}
    \psi_a(\nv)=-\nabla^{2D}_a\,\nabla_\Omega^{-2}\int dz\,p(z)\,\Delta(\chi(z)\nv)
    \simeq\int dz\,p(z)\,\frac{1}{\chi(z)}\nabla^{3D}_a\,\nabla^{-2}\Delta=\int dz\,p(z)\,\frac{\Psi_a(\chi(z)\nv)}{\chi(z)},
  \end{equation}
  where $\nabla^{2D}_a$ and $\nabla^{3D}_a$ denote derivatives in the direction $a$ with respect to angular or 3D coordinates, respectively, and the approximation improves for wider redshift distributions. $\psi_a$ is therefore approximately equal to the mean angular displacement of matter elements along the line of sight within the range of redshifts covered by $p(z)$.

\end{document}